\documentclass[longauthor]{aastex62a}
\usepackage{epsfig}
\usepackage{amsmath}
\usepackage{mathtools}
\usepackage{textgreek}

\usepackage{titlesec}
\titleformat{\section}[block]{\large\bfseries\filcenter}{\thesection.}{0.3em}{}

\hypersetup{linkcolor=black,citecolor=black,filecolor=black,urlcolor=black}

\def\cm{{\rm cm}}
\def\oday{{\rm day}}
\def\deg{{\rm deg}}
\def\dof{{\rm dof}}
\def\eff{{\rm eff}}
\def\ej{{\rm ej}}
\def\erg{{\rm erg}}

\def\g{{\rm g}}
\def\GeV{{\rm GeV}}

\def\hr{{\rm hr}}
\def\jpi{{J\textpi}}
\def\K{{\rm K}}
\def\keV{{\rm keV}}
\def\kpc{{\rm kpc}}
\def\MeV{{\rm MeV}}
\def\Mpc{{\rm Mpc}}
\def\mus{{\textmu{\rm s}}}

\def\nonthermal{{\rm nth}}
\def\obs{{\rm obs}}

\def\s{{\rm s}}
\def\thermal{{\rm th}}

\def\yr{{\rm yr}}

\begin{document}

\title{\bf\Large Radioactive ${\mathbf \gamma}$-Ray Emissions from Neutron Star Mergers}


\author{\normalsize Li-Xin Li}
\affiliation{\vspace{0pt}\\
  {\small Kavli Institute for Astronomy and Astrophysics, Peking University, Beijing 100871, P. R. China\\
{\rm Email: lxl@pku.edu.cn}}}

\begin{abstract}
\noindent Gravitational waves and electromagnetic radiations from a neutron star merger were discovered on 17 August 2017. Multiband observations of the optical transient have identified brightness and spectrum features broadly consistent with theoretical predictions. According to the theoretical model, the optical radiation from a neutron star merger originates from the radioactive decay of unstable nuclides freshly synthesized in the merger ejecta. In about a day the ejecta transits from an optically thick state to an optically thin state due to its subrelativistic expansion. Hence, we expect that about a day after the merger, the gamma-ray photons produced by radioactive decays start to escape from the ejecta and make it bright in the MeV band. In this paper, we study the features of the radioactive gamma-ray emission from a neutron star merger, including the brightness and the spectrum, and discuss the observability of the gamma-ray emission. We find that more than $95\%$ of the radiated gamma-ray energy is carried by photons of $0.2$--$4\,\MeV$, with a spectrum shaped by the nucleosynthesis process and the subrelativistic expansion of the ejecta. Under favorable conditions, a prominent pair annihilation line can be present in the gamma-ray spectrum with the energy flux about $3$--$5\%$ of the total. For a merger event similar to GW170817, the gamma-ray emission attains a peak luminosity $\approx 2\times 10^{41}\,\erg\,\s^{-1}$ at $\approx 1.2\,\oday$ after the merger, and fades by a factor of two in about two days. Such a source will be detectable by Satellite-ETCC if it occurs at a distance $\la 12\,\Mpc$.

\vspace{0.3cm}
\end{abstract}


\maketitle

\vspace{0.3cm}

\section{Introduction}
\label{intro}

Mergers of double neutron stars, or a neutron star and a stellar mass black hole, have long been expected to occur in the universe with a rate estimated to be several orders of magnitude lower than the supernova rate \citep{nar91,phi91,van96,blo99}. Three major transient observable phenomena have been predicted to arise from a neutron star merger (a neutron star-neutron star merger, or a neutron star-black hole merger): a gravitational wave signal \citep{cla77,tho87}, a short gamma-ray burst \citep[][and references therein]{goo86,pac86,pac91,eic89,pop99,ber14}, and a UV-optical-NIR (hereafter UVOIR) transient powered by the radioactive decay of unstable heavy elements freshly synthesized in the merger ejecta \citep[][and references therein]{li98,kul05,ros05,met10,rob11,bar13,kas13,tan13,gro14,kas15,met17,ros17,tan18,wol17}. In addition, mergers of neutron stars have been proposed to be a major site for nucleosynthesis of heavy and rare elements in the universe like gold and platinum \citep[][and references therein]{lat74,lat76,lat77,fre99,kor12,baus13,wan14,kas17,thi17,hot18}.

Although the above mentioned three observable phenomena have been firmly predicted for decades and gamma-ray bursts (GRBs) have been observed for more than half a century, mergers of neutron stars have not been directly detected until 17 August 2017 after the joint detection of GW170817 and GRB170817A, and the identification of an optical counterpart SSS17a/AT2017gfo \citep{abb17a,abb17b,cou17,gol17,sav17,sie17,val17}. The gravitational wave signal was consistent with being produced by binary stars with component masses between $0.86$ and $2.26\,M_\odot$, in agreement with the masses of known neutron stars. In the region of GW170817 on the sky ($28\,\deg^2$ jointly determined by Advanced LIGO and Advanced Virgo), a short gamma-ray burst of duration $\approx 2\,\s$, GRB170817A, was detected by {\it Fermi}/GBM and {\it INTEGRAL}/SPI-ACS at $1.7\,\s$ after the coalescence time. About $10.87\,\hr$ later, an optical transient SSS17a/AT2017gfo was detected in the region of GW170817/GRB170817A, which occurred in the outskirts of NGC4993 at about $40\,\Mpc$. This distance agrees with the distance of GW170817 determined by the gravitational wave signal alone, which is $40_{-14}^{+8}\Mpc$.

The possibility of being a supernova or the GRB afterglow for the optical transient was quickly excluded. The UVOIR spectra of SSS17a/AT2017gfo do not have any typical supernova feature. Attempts to spectrally classify the source using the Supernova Identification Code failed to get a good match, even using an expanded template set \citep{tro17}. The luminosity and spectra evolved much faster than those of a supernova. For instance, the $r$-band brightness of the source declined by 1.1\,mag from the peak in one day \citep{val17}. The X-ray and radio emissions were not detected until nine days and two weeks, respectively, after the burst of gravitational waves and are consistent with the GRB afterglow emissions from an off-axis jet \citep{hal17,tro17,ale18,mar18}. The afterglow emissions in the UVOIR range interpolated from the observed X-ray and radio emissions are much fainter than the observed emissions \citep{pia17,sha17,tro17}. The spectra of the transient in the early epoch ($\la 3.5\,\oday$) can be well fitted by blackbodies, while the afterglow spectra of GRBs are usually highly nonthermal.

On the other hand, the observed optical transient has all the features predicted for neutron star mergers: (1)~The emissions are in the UVOIR range, and are characterized by blackbody radiations in the early time; (2)~The peak luminosity is in the supernova range (although in the faint end) and occurs at a time $\sim 1\,\oday$ after the merger; (3)~Both the luminosity and spectra evolve rapidly with time, fading and reddening on a timescale of days. Hence, the optical transit SSS17a/AT2017gfo is clearly identified as the radioactive glow of a neutron star merger, i.e., a kilonova or macronova as often called in the literature. In the early epoch ($\la 2\,\oday$ after the merger), the observed spectra are dominated by strong thermal UV-Optical emissions, with the brightness declining on a timescale of 1--2 days, and the colour reddening on a similar timescale \citep{eva17,mcc17,pia17,buc18}. After a couple of days, the bulk emissions of SSS17a/AT2017gfo shift to the near-infrared range, causing the spectra to redden quickly. This can be interpreted by the variation in the opacity of the merger ejecta, at least in principle.

As pointed out by \citet{kas13} and \citet{tan13}, the opacity of a merger ejecta is very sensitive to the abundance of lanthanide elements. If the mass fraction of lanthanides is $>10^{-2}$, the opacity can be as high as $10\,\cm^2\,\g^{-1}$, due to the bound-bound transition of the $f$-shell electrons of lanthanides. To account for the fact that the spectra of SSS17a/AT2017gfo are dominated by a blue component in the early time and by a red component in the late time, multi-component models of kilonovae have been used to fit the data \citep{cow17,dro17,kil17,pia17,tan17,vil17,wax17}. The presence of multiple components in a merger seems plausible: a dynamical ejecta generated by the tidal and hydrodynamic forces produced by the violent merger process, and a disk-wind ejecta driven by neutrino-antineutrino annihilation following the merger \citep{tho01,pac02}. It is natural to expect that these distinct components have different compositions of heavy elements hence different opacities, and different values of other parameters such as the expansion velocity and mass. However, the later red emissions may also arise from delayed energy injection from a long-lived remnant neutron star at the center \citep{yu17}.

Before the discovery of GW170817, some clues for the existence of kilonovae/macronovae had been found in GRBs 050709, 060614, and 130603B. The very faint near-infrared rebrightening found in their late afterglows was interpreted as the emergence of kilonova/macronova emissions \citep{ber13,tanv13,jin15,jin16,yan15}. GRBs 050709 and 130603B are short bursts with a duration $<2\,\s$. GRB060614 has a duration of $102\,\s$ but is more like a short burst in many other aspects \citep{zha07}. However, all these previous evidences are not strong cases, because of the limit in available data with good qualities. The case of GW170817/GRB170817A and SSS17a/AT2017gfo is a very strong case for the GW-GRB-Kilonova/macronova connection. Without any doubt, GW170717, GRB170817A, and SSS17a/AT2017gfo are different representations at different evolution stages of one physical event: the merger of two neutron stars.

In spite of the successful identification of a kilonova/macronova associated with the GW170817/GRB170817A, a proof of the energy source for powering the UVOIR emission as arising from the decay of radioactive elements in a neutron star merger is not easy. Presumably, the violent merger produces copious radioactive nuclides with different lifetimes and quantum states, whose decay releases energies in the form of neutrinos, gamma-ray photons, and the kinetic energy of electrons, positrons, and other particles. Because the merger ejecta is initially opaque to photons and particles but transparent to neutrinos, only neutrinos can escape freely and the energies carried by photons and particles will be thermalized and eventually escape from the surface of the ejecta in the form of blackbody radiation. Because of the subrelativistic expansion of the ejecta, all emission and absorption lines from the surface of the ejecta are broadened and merged smoothly. As a result, a smooth and almost featureless thermal spectrum is generated \citep[with superposition of smooth undulations that might arise from broad absorptions,][]{tan18}, which is verified by the observations of SSS17a/AT2017gfo \citep{pia17}. The intense near-infrared emissions have sometimes been used to argue for the presence of lanthanides in the merger---presumably produced by the r-process (rapid neutron capture process) in the merger ejecta---but this is very indirect and not conclusive.

The most direct approach for identification of nuclear elements produced during the nucleosynthesis process, and hence the energy mechanism for powering the optical transient from a neutron star merger, would be the direct observation of the gamma-ray photons emitted by the radioactive decay in the merger ejecta. However, this can only be possible after the ejecta becomes transparent to the gamma-ray photons. According to theoretical estimates, for reasonable parameters the ejecta will become optically thin after a day to a few days since the moment of merger. This seems having been confirmed by the optical observations of SSS17a/AT2017gfo. According to the analysis of \citet{pia17}, starting from about three days after the GW170817, the merger ejecta was becoming increasingly transparent to photons and more absorption lines become visible. The analysis by \citet{dro17} also shows that the spectra between 0.5--8.5 days after the merger are broadly consistent with a thermal distribution, then become nonthermal. These conclusions are broadly consistent with the results in other analyses \citep[e.g.,][]{kil17,sha17,tro17,wax17}. If we accept the two-component model for the merger (a blue component plus a red component), we expect that the gamma-ray photons produced by the radioactive decay will start to emerge from about one day after the merger, since the blue component cools very fast. The emerging photons will be in the energy range of $\MeV$ with a peak luminosity of $\sim 10^{41}\,\erg\,\s^{-1}$ (the same order of the optical peak luminosity of SSS17a/AT2017gfo). Since this luminosity is lower than that of the faintest observed GRB by about five orders of magnitude, to observe it requires a very sensitive gamma-ray detector given its distance of $40\,\Mpc$. 

The importance of observations of the gamma-ray emission from Type Ia supernovae (thought to be powered by the decay chain of $\prescript{56}{}{\rm Ni}\rightarrow\prescript{56}{}{\rm Co}\rightarrow\prescript{56}{}{\rm Fe}$) for diagnosing their progenitor and explosion mechanism has been noticed and studied for many decades \citep[][and references therein]{clay69,clay74,sum13,the14}. However, so far only for two supernovae have the gamma-ray emissions produced by radioactive decays been detected. The first detection of gamma-ray emission lines caused by the radioactive decay in a Type Ia supernova was in SN2014J in M82, for which two gamma-ray emission lines of $\prescript{56}{}{\rm Co}$ ($847$ and $1,238\,\keV$, respectively) were detected by {\it INTEGRAL}. From the observed luminosity of the emission lines ($4.7$ and $8.1\times 10^{41}\,\erg\,\s^{-1}$, respectively), it is successfully derived that about $0.6\,M_\odot$ radioactive $\prescript{56}{}{\rm Ni}$ were synthesized during the explosion \citep{chu14}. Before that, the same gamma-ray emission lines were also detected in type II SN1987A (thought to be powered by both radioactive decays and shock waves) with the Solar Max satellite \citep{mat88}. However, the derived mass of $\prescript{56}{}{\rm Co}$ was only a very small fraction ($\approx 1.3\%$) of the total mass of $\prescript{56}{}{\rm Co}$ inferred from the bolometric light curve at a similar time. The rare detection of radioactive gamma-ray lines in supernovae is mainly caused by the fact that we are lacking of gamma-ray detectors with a high enough sensitivity in the $\MeV$ energy range \citep[][and references therein]{tan15}.

Both SN1987A and SN2014J are among the nearest supernovae that have ever been observed, with a distance of $51\,\kpc$ and $3.5\,\Mpc$, respectively. Since the occurrence frequency of neutron star mergers is about 1,000 times smaller than that of supernovae, in principle the closest merger that we have a fair chance to discover would be farther way than the closest supernova by a factor of $\sim 10$. So, for a similar luminosity, we expect that the radioactive gamma-ray emission from neutron star mergers would be more difficult to detect than that from supernovae, since its flux density would be weaker by about two orders of magnitude. However, this does not reduce the importance of observations of the radioactive gamma-ray emission from neutron star mergers. In addition, given the fact that we have discovered SN 1987A although the local rate of type II supernovae is only $\approx 2.5\times 10^{-8}\,\yr^{-1}$ in a spherical volume with a radius of $51\,\kpc$ \citep{li11}, detection of a neutron star merger at a distance $\la 1\,\Mpc$ may not be impossible.

In this paper, we study the gamma-ray emissions due to the radioactive decay of unstable nuclides produced in a neutron star merger. After the merger ejecta becomes transparent a few days after the merger, the gamma-ray photons will escape from the ejecta and become visible. Unlike in the case of supernovae where the dominant gamma-ray emissions come from the decay of a single radioactive nuclear isotope $\prescript{56}{}{\rm Co}$ after the supernova envelope becomes transparent (about 100 days after the explosion), in the case of neutron star mergers the merger ejecta are expected to contain hundreds to thousands of unstable nuclides with a wide distribution in lifetimes. Hence, the gamma-ray emissions from a neutron star merger are expected to contain tons of emission lines with a distribution over the photon energy. The subrelativistic expansion of the ejecta will broaden the emission lines and merge them, resulting in a smooth gamma-ray spectrum in contrast to the case of supernovae where we can see distinct emission lines from one unstable nuclide. In this paper we will calculate the magnitude and the shape of the radioactive gamma-ray spectra of a neutron star merger in its optically thin stage, identify the features in the emission spectrum associated with specific nucleosynthesis processes, and study their dependence on model parameters (expansion velocity, opacity, etc) as well as the observability of the gamma-ray emission.

\citet{hot16} studied the gamma-ray emission resulting from the radioactive decay of r-process elements outside the photosphere in an ejecta of a neutron star merger. They concluded that to observe the emissions, new detectors with a sensitivity higher than current ones by at least a factor of ten are required. Their research was based on a dynamical r-process network. Since in the r-process the dominant nuclear reaction consists of neutron captures, $\beta$-decay, $\alpha$-decay and fissions \citep{mar08}, in the calculation of \citet{hot16} the dominant contribution to the gamma-ray emission comes from the $\beta$-decay of r-nuclides. In our work, without using an r-process network, we assume at some initial time a power-law distribution in the number of radioactive nuclides over their lifetimes, then calculate the energy generation by tracing the decay process of nuclides. The sample of radioactive nuclides is constructed from the NuDat\,2 database at the National Nuclear Data Center according to some selection criteria. For the calculation of the energy generation and the nonthermal gamma-ray spectrum, we make use of the gamma-ray radiation data for each nuclide provided by the NuDat\,2 website. We note that the original work of \citet{li98} was also based on an assumption of power-law distribution of the number of unstable nuclides over their lifetimes, and the luminosity and temperature of blackbody radiations were correctly derived. So, in this work we also take this simple approach. Since our data sample is uniformly extracted from a nuclear database, it includes not only r-nuclides. The sample also includes p-nuclides---proton-rich nuclides, which cannot be produced by the r-process, but the r-nuclides produced during the r-process can serve as the seed for production of p-nuclides if the thermodynamic conditions in the ejecta are appropriate. Inclusion of both r- and p-nuclides in the sample will allow us to identify the specific feature of the gamma-ray emissions produced by each type of nuclides, which is necessary for diagnosing the nucleosynthesis process in the ejecta by observing its gamma-ray emissions. Later in the paper we will also argue that the possibility for the occurrence of p-process---a process for the production of p-nuclides---in a merger ejecta cannot be excluded in principle.

In our model the gamma-ray emission comes from the following five decay processes: $\beta^-$-decay, $\beta^+$-decay, electron capture, $\alpha$-decay, and isomeric transition. The electron capture is a process where a proton-rich nucleus of a neutral or partially-ionized atom absorbs an electron from the K or L shell. It is a process that competes with the $\beta^+$-decay, and has the same effect on the atomic number. The $\beta^-$-decay is a major feature of r-nuclides, through which the unstable and neutron-rich nuclides decay toward the bottom of the valley of nuclear stability in the nuclear chart. The isomeric transition is a process where a long-lived excited nuclear level decays by gamma-ray emissions or internal conversion. We find that $\beta^\pm$-decays and electron captures make the dominant contribution to the gamma-ray energy generation in the merger ejecta. The isomeric transition contributes to the total gamma-ray energy generation with a fraction smaller than that contributed by the electron capture and the $\beta^\pm$-decay, but larger than that contributed by the $\alpha$-decay. We also find that the $\beta^+$-decay and the electron capture produce a gamma-ray spectrum with a feature very different from that generated by the $\beta^-$-decay, including the presence of electron-positron annihilation lines. The $\beta^-$-decay alone cannot produce annihilation lines. This feature, and other differences between the gamma-ray spectrum produced by r-nuclides and that produced by p-nuclides which will be discussed in detail later in the paper, will allow us to distinguish the r-process from the p-process in the merger ejecta through observations of the gamma-ray emissions from a neutron star merger.

In our model we do not include the fission process, since the NuDat\,2 website contains very few radiation data for fissions. However, other works have claimed that the contribution of fissions to the total energy generation is small relative to the $\beta$-decay, though they may make a nonignorable contribution at very late time \citep{met10,hot16}.

The paper is organized as follows. In Section~\ref{outline}, we apply the model of \citet[][with minor modifications]{li98} to fit the UVOIR bolometric luminosity data of SSS17a/AT2017gfo. We derive some critical quantities that will be used as a reference for normalizing the parameters of the model for calculation of the gamma-ray emission. In Section~\ref{model}, we derive the mathematical formulae for calculation of the energy generation by radioactive decays in a merger ejecta, and describe how to calculate the luminosity and spectrum of the gamma-ray emission. In Section~\ref{data}, we construct a sample of radioactive nuclides that will be used in our model, and generate the abundance of each nuclide according to a power-law distribution over their lifetime with a Monte Carlo approach. In Sections~\ref{lum} and \ref{spectra} we present results for the calculation of the energy generation rate, the luminosity and spectrum of the gamma-ray emission, and the efficiency in converting the nuclide mass into nuclear energy by radioactive decays. Section~\ref{dec_chains} contains a discussion on the effect of decay chains on the gamma-ray energy generation.

In Section~\ref{detect}, we take the merger model for GW170817 as an example to calculate the spectra of its gamma-ray emissions, and discuss their observability by comparing the result to the sensitivity of some modern gamma-ray detectors. We argue that the gamma-ray emission from a merger event like GW170817 will be detectable by Satellite-ETCC if it occurs at a distance $\la 12\,\Mpc$. In Section~\ref{sum}, we summarize the result of this work and draw our conclusions. Appendix~\ref{sphere} contains some details not included in Section~\ref{model} on derivation of the formulae for calculation of the energy generation rate and the spectrum of the gamma-ray emission produced by radioactive nuclides in an expanding sphere. Appendix~\ref{tdc} contains the mathematical formulae for the treatment of decay chains.

\section{Model Fitting to the Luminosity Curve of SSS17a/AT2017gfo}
\label{outline}

The model used by \citet{li98} for calculation of the electromagnetic radiation from a merger ejecta in its optically thick phase is simple but robust. The predicted major characters for the electromagnetic radiation produced by a neutron star merger are basically all confirmed (at least qualitatively) by the observations of SSS17a/AT2017gfo: (1)~The early radiation has a thermal spectrum, with the bulk energy in the UV-optical band. Observations of SSS17a/AT2017gfo have shown that this is indeed the case for time $t\la 3.5\,\oday$ after the merger. (2)~The luminosity and spectrum evolve with time rapidly, on a timescale of a few days. The model predicts that the time from the peak luminosity to a luminosity down by a factor of 3 from the peak is about 2 days. The bolometric luminosity of SSS17a/AT2017gfo derived from the observational data drops by a factor of $\sim 3$ from $t=1\,\oday$ to $t=3.5\,\oday$. (3)~The optical transient has a peak luminosity in the supernova range, which is attained at $t\sim 1\,\oday$ after the merger. The peak bolometric luminosity of SSS17a/AT2017gfo is $\approx 8\times 10^{41}\,\erg\,\s^{-1}$, attained at $t\approx 0.6\,\oday$ \citep{wax17}. This peak luminosity is in the range of faint supernovae \citep{fol09,buf14}.

The original model of \citet{li98} contains an $f$ parameter, which roughly represents the efficiency in generation of energy by radioactive decays in the ejecta. The derived peak luminosity of the optical transient $L_m\propto f$, hence the estimated peak luminosity sensitively depends on the value of $f$. In their original work, Li \& Paczy\'nski treated $f$ as a free parameter and took $f=10^{-3}$, $10^{-4}$, and $10^{-5}$ in the presentation of their numerical results. Hence, they got a peak luminosity in the range of $10^{42}$--$10^{44}\,\erg\,\s^{-1}$, i.e., the range of normal to bright supernovae. The precise value of $f$ is hard to determine, since radioactive nuclides have a wide range of efficiency in converting mass to energy, and as a result, the derived value of $f$ sensitively depends on the model. For instance, \citet{met10} derived an effective $f\sim 3\times 10^{-6}$ at $t=1\,\oday$ based on a dynamical r-process network. With a large reaction network, \citet{kor12} derived an analytical heating rate which indicates that $f\sim 0.9\times 10^{-6}$ at $t=1\,\oday$. Basically, the presence of many heavy elements with low radiative efficiency can significantly decrease the derived value of $f$.

Like in the work of \citet{li98}, here we consider a spherical merger ejecta of a constant mass $M_\ej$ and a uniform mass density $\rho$, uniformly expanding with a constant velocity $V$ at its surface. The radius of the expanding sphere is then $R=Vt$, where $t$ is the time since the merger. So we have $M_\ej=(4\pi/3)\rho V^3 t^3$. Assuming that the ejecta material has a constant opacity $\kappa$. Then, the total optical depth of the spherical ejecta is
\begin{eqnarray}
  \tau_s = \frac{3\kappa M_\ej}{4\pi V^2 t^2}= 1.57\left(\frac{\kappa}{0.2\,\cm^2\,\g^{-1}}\right)\left(\frac{M_\ej}{0.01M_\odot}\right)\left(\frac{V}{0.3\,c}\right)^{-2}\left(\frac{t}{1\,\oday}\right)^{-2} \;, \label{tau_s0}
\end{eqnarray}
where $c$ is the speed of light.

Define a critical time $t_c$ by $\tau_s=1$, i.e., the time when the ejecta starts to be transparent to photons. Then by equation (\ref{tau_s0}) we have
\begin{eqnarray}
  t_c = 1.25\,\oday\left(\frac{\kappa}{0.2\,\cm^2\,\g^{-1}}\right)^{1/2}\left(\frac{M_\ej}{0.01M_\odot}\right)^{1/2}\left(\frac{V}{0.3\,c}\right)^{-1} \;. \label{tc}
\end{eqnarray}
In terms of the critical time $t_c$, the total optical depth can be rewritten as
\begin{eqnarray}
  \tau_s=\left(\frac{t}{t_c}\right)^{-2} \;. \label{tau_s}
\end{eqnarray}

We denote by $\epsilon(t)$ the energy generation per unit time and per unit mass by the radioactive decay in the ejecta. Then, the total energy generation per unit time is given by $\dot{E}=\epsilon M_\ej$, where the dot denotes $d/dt$. Among the total energy generated inside the ejecta, a fraction of it is scattered and absorbed by the ejecta matter then re-emitted as thermal photons (i.e., that fraction of the generated energy is thermalized). The rest fraction escapes to infinity in the form of gamma-ray photons. Here we approximate the fraction for thermalization by $1-e^{-\tau_s}$, and the fraction carried away by gamma-ray photons by $e^{-\tau_s}$. That is, $\dot{E}=\dot{E}_\thermal+\dot{E}_\nonthermal$, where
\begin{eqnarray}
  \dot{E}_\thermal = \left(1-e^{-\tau_s}\right)\dot{E}=\left(1-e^{-t_c^2/t^2}\right)\epsilon M_\ej \;, \label{dQ_th}
\end{eqnarray}
and
\begin{eqnarray}
  \dot{E}_\nonthermal = e^{-\tau_s}\dot{E}=e^{-t_c^2/t^2}\epsilon M_\ej \;. \label{dQ_nth}
\end{eqnarray}

When $t\ll t_c$, we have $\tau_s\gg 1$, $\dot{E}_\thermal\approx\dot{E}=\epsilon M_\ej$, and $\dot{E}_\nonthermal\approx 0$. That is, when the ejecta is optically thick, almost all the energy generated inside the ejecta is thermalized. When $t\gg t_c$, we have $\tau_s\ll 1$, $\dot{E}_\thermal\approx \tau_s\dot{E}=\left(t_c^2/t^2\right)\epsilon M_\ej$, and $\dot{E}_\nonthermal\approx\left(1-t_c^2/t^2\right)\epsilon M_\ej\approx\epsilon M_\ej$. That is, when the ejecta is optically thin, almost all the energy generated inside the ejecta escapes to infinity without thermalization. 

After consideration of the effect of optical depth, the equation 9 of \citet{li98} should be modified as
\begin{eqnarray}
  \frac{d}{dt}\left(t^4U\right) = \frac{3M_\ej}{4\pi V^3}\epsilon t\left(1-e^{-t_c^2/t^2}\right)-\frac{\pi Vc}{\kappa M_\ej}\frac{t^5 U}{1+t^2/2t_c^2} \;, \label{dUdt}
\end{eqnarray}
where $U$ is the energy density of radiation. The factor $\left(1+t^2/2t_c^2\right)^{-1}$ in the last term in equation (\ref{dUdt}) comes from the relation $T_{\eff}^4=T^4/(\tau_s+1/2)$, where $T$ is the temperature inside the ejecta, $T_\eff$ is the effective temperature, and inclusion of the number $1/2$ in the denominator is based on the consideration that as $\tau_s\rightarrow 0$ we should have $T_\eff^4\approx 2T^4$ \citep[see, e.g.,][]{car17}.

Equation (\ref{dUdt}) determines the evolution of the temperature inside the ejecta. In \citet{li98} the function $\epsilon$, which is also called the heating rate in the optically thick case, is assumed to be inversely proportional to time $t$. Here, like in other references we take a more general power-law form of $\epsilon(t)$, and write it as
\begin{eqnarray}
  \epsilon =\frac{fc^2}{t_c}\left(\frac{t_c}{t}\right)^{1+\alpha} \;, \label{eps_t}
\end{eqnarray}
where $f$ and $\alpha$ are constant numbers. Of particular interest is in the case of nuclear waste, where we have $\alpha\approx 0.2$ \citep{cot04}. Numerical and analytical works indicate that the value of $\alpha$ should be in the range of $0.1$--$0.4$ \citep{met10,kor12,hot16,hot17}.

The thermal luminosity $L$ is related to the energy density $U$ by the equation 6 of \citet{li98}. With inclusion of the factor $\left(1+t^2/2t_c^2\right)^{-1}$, we get
\begin{eqnarray}
  L=4\pi R^2\sigma T_\eff^4=\left(\frac{4\pi^2V^4c}{3\kappa M}\right)\frac{Ut^4}{1+t^2/2t_c^2} \;.
\end{eqnarray}
Define a dimensionless parameter $\beta$ and dimensionless variables $x$ and $y$ by $\beta\equiv V/c$, $x\equiv t/\tilde{t}_c$ where $\tilde{t}_c\equiv t_c(8\beta/3)^{1/2}$, and
\begin{eqnarray}
  y\equiv\frac{Ut^4}{\tilde{U}_1\tilde{t}_c^4}\left(\frac{8\beta}{3}\right)^{\alpha/2} \;, \hspace{1cm} \tilde{U}_1\equiv\frac{3fMc^2}{4\pi V^3\tilde{t}_c^3} \;.
\end{eqnarray}
Then, substitution of equation (\ref{eps_t}) into equation (\ref{dUdt}) leads to
\begin{eqnarray}
  \frac{dy}{dx}=x^{-\alpha}\left[1-e^{-(3/8\beta) x^{-2}}\right]-\frac{2xy}{1+4\beta x^2/3} \;. \label{dLdt2}
\end{eqnarray}

With a given initial condition for the luminosity, we can solve equation (\ref{dLdt2}) for a solution $y=y(x)$. Then, the luminosity $L$ as a function of time can be calculated by
\begin{eqnarray}
  L=\frac{L_cy}{1+4\beta x^2/3} \;, \hspace{1cm} L_c\equiv\frac{2fMc^2}{\tilde{t_c}}\left(\frac{8\beta}{3}\right)^{-\alpha/2} \;. \label{L_c}
\end{eqnarray}
In practice we can choose $L(t=0)=0$ as an initial condition. Then, solutions satisfying the initial condition exist when $\alpha<1$.

In the limit $x\ll 1$, i.e., $t\ll\tilde{t}_c$, we have the approximate solution $y\approx (1-\alpha)^{-1}x^{1-\alpha}[1-2x^2(3-\alpha)^{-1}]$. In the limit $x\gg(3/8\beta)^{1/2}$, i.e., $t\gg t_c$, we have the asymptotic solution $y\approx x^{-1-\alpha}[4-(8\beta/3)(1+\alpha)]^{-1}$, corresponding to an asymptotic luminosity
\begin{eqnarray}
  L\approx\frac{3L_c}{4\beta[4-(8\beta/3)(1+\alpha)]}\left(\frac{t}{\tilde{t}_c}\right)^{-3-\alpha} \;. \label{L_asy}
\end{eqnarray}

According to the definition, $t_c$ corresponds to the time when the ejecta transits from the optically thick phase to the optically thin phase. When $\alpha=0$, $\tilde{t}_c$ corresponds to the time at the peak of the luminosity, and $L_c$ corresponds to twice the peak luminosity \citep{li98}.

Now let us attempt to apply the above model to fit the bolometric luminosity data of SSS17a/AT2017gfo. Here we use the bolometric luminosity data derived by \citet{wax17}, where three bolometric luminosities were calculated: $L_{\rm int}$ calculated by the trapezoidal integration of multiband photometric data; $L_{\rm bb}$ by fitting a blackbody to the photometric data; and $L_{0.3-2.4\mu{\rm m}}$ by integrating the X-Shooter spectra. They claimed that the $L_{\rm int}$ is more reliable, since it does not depend on modeling of the spectra. In \citet{arc18}, a bolometric luminosity is constructed by dividing the multiband data set into 0.2-day epochs then fitting the data to a blackbody using the Markov Chain Monte Carlo simulation through the {\textsf emcee} package. The derived bolometric luminosity agrees with the $L_{\rm int}$ derived by \citet{wax17} surprisingly well. Therefore, we choose to use the $L_{\rm int}$ for testing the above model.

In \citet{wax17}, fitting errors were only listed for the $L_{\rm bb}$, not for the $L_{\rm int}$. However, the error for $L_{\rm bb}$ can be used as an order of magnitude estimate for the error of $L_{\rm int}$ (E. Ofek \& E. Waxman 2018, private communications). Hence, here we estimate the error of $L_{\rm int}$ by $\delta L_{\rm int}\approx \delta L_{\rm bb} (L_{\rm int}/L_{\rm bb})$.

Integration of equation (\ref{dLdt2}) with the initial condition $y(x=0)=0$ leads to a solution $y=y(x;\alpha,\beta)$. Then we get $L=L_c(t;\tilde{t}_c,\alpha,\beta)$, since $x=t/\tilde{t}_c$. Hence, there are four independent parameters in the calculation of luminosity: $L_c$, $\alpha$, $\beta$, and $\tilde{t}_c$. In the 21 data points of $L_{\rm int}$, the last two data points (at $t=15.5$ and 16.5\,day, respectively) have too large errors. Hence, the last two data points are excluded from our model fitting. If we allow all the four parameters to vary, the least-squares fit leads to a best fit with $\chi^2/\dof=2.19$. But the best fit $\beta$ has a too small value: $\beta=0.0022\pm0.0002$. This small value of $\beta$ is unacceptable, since it is clearly inconsistent with the fact that the observed spectra of SSS17a/AT2017gfo are very smooth in the early time ($t\la 5\,\oday$). The blackbody fit to the multiband photometric data indicates that $\beta\ga 0.2$ at least for the first couple of days \citep{wax17}. If we take a constraint that the value of $\beta$ must be $>0.1$, then we cannot get an acceptable fit to the data with a single set parameters $L_c$, $\alpha$, $\beta$, and $\tilde{t}_c$.

\citet{kru18} also noticed that a single component model cannot fit the data. However, he got a perfect fit to the early six data points (corresponding to $t<1.5\,\oday$) with a single component model. This fact might indicate that a two-component model can fit all the data.

It is easy to see why a one-component model cannot fit the bolometric luminosity data. From the $L_{\rm int}$ derived by \citet{wax17}, the luminosity peaks at a time $t_m<1\,\oday$ (hence $\tilde{t}_c<1\,\oday$), and for $t>1\,\oday$ the luminosity curve clearly has a broken power-law feature: the power-law index jumps from $-0.95\pm 0.06$ for $1\,\oday<t<6.2\pm0.7\,\oday$ to $-2.8\pm0.6$ for $t>6.2\pm 0.7\,\oday$ \citep[Figure~3 and Table~1 in][]{wax17}. The power-law index $-2.8\pm0.6$ is remarkably consistent with our asymptotic solution in equation (\ref{L_asy}), if the value of $\alpha$ is in the range of $-0.8$ to $0.4$. Thus, we can interpret the time $t=6.2\,\oday$ as the time when the ejecta transits from the optically thick phase to the optically thin phase. Hence we should have $t_c\approx 6\,\oday$. We then get $(8\beta/3)^{1/2}=\tilde{t}_c/t_c<1/6$, i.e., $\beta<0.01$. So, to fit the data with a one-component model we have to get a very small expansion velocity.

\begin{figure}[ht!]
\centering{\includegraphics[angle=0,scale=0.65]{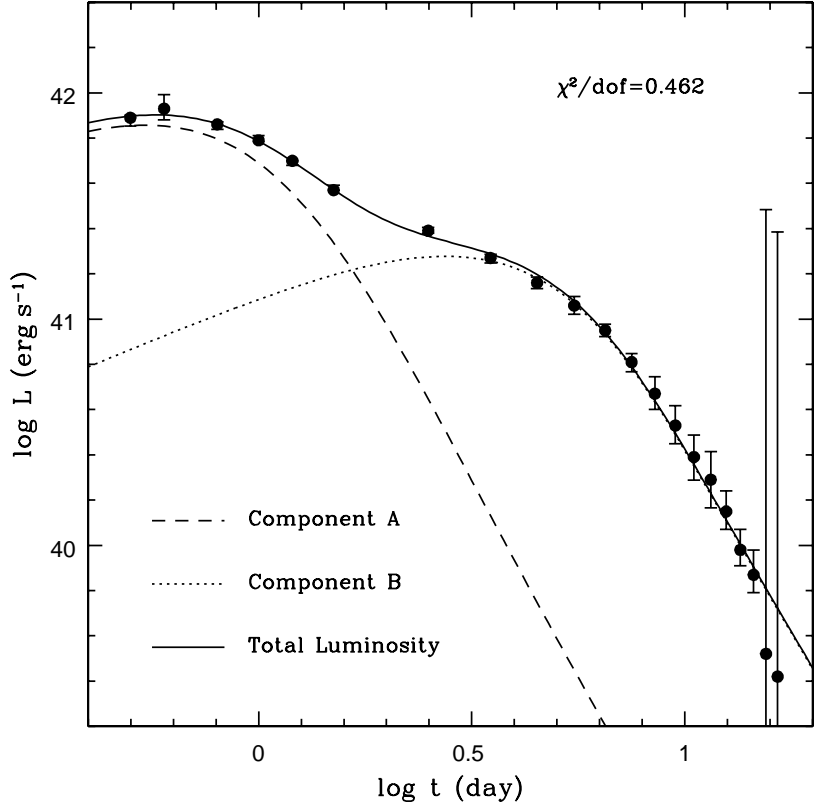}}
\caption{A two-component model fit to the bolometric luminosity data of SSS17a/AT2017gfo. Each component is defined by four independent parameters: the power-law index parameter $\alpha$ in the energy generation rate, the ejecta expansion velocity $V=\beta c$, the critical time $\tilde{t}_c$, and the critical luminosity $L_c$. The $\alpha$ parameter is fixed at 0.2 so that the energy generation rate $\propto t^{-1.2}$. The $\beta$ value is set to 0.3 for component A, and 0.1 for component B. The remaining four parameters ($L_{c,{\rm A}}$, $\tilde{t}_{c,{\rm A}}$, $L_{c,{\rm B}}$, and $\tilde{t}_{c,{\rm B}}$) are determined by the least-squares fit. The bolometric luminosity data are taken from \citet{wax17}, i.e., the $L_{\rm int}$ in their Table~3. The error of $L_{\rm int}$ is estimated by referencing to the error listed in their Table~3 for the blackbody fit luminosity $L_{\rm bb}$ (see the text). The last two data points are excluded from fitting due to their very large errors. The best fitted parameters are listed in Table \ref{parameter}.
\label{lp_fit}}
\end{figure}

Next, we apply a two-component model to fit the data. We assume that the ejecta contains a component A and a component B. For instance, the component A can be a dynamical ejecta, the component B can be a wind ejecta, and {\it vice versa}. The two components can have different values of the parameters $\beta$, $\tilde{t}_c$, and $L_c$, but we assume that they share the same value of $\alpha$. Then, the total luminosity is given by the sum of the luminosity for each component, i.e., $L=L_{\rm A}+L_{\rm B}$, where $L_{\rm A}=L_{c,{\rm A}}y(t;\alpha,\beta_{\rm A},\tilde{t}_{c,{\rm A}})$, and $L_{\rm B}=L_{c,{\rm B}}y(t;\alpha,\beta_{\rm B},\tilde{t}_{c,{\rm B}})$. Each solution of $y$ is determined by equation (\ref{dLdt2}), for given $\alpha$ and $\beta$. Then we have seven independent free parameters: $\alpha$, $L_{c,{\rm A}}$, $\beta_{\rm A}$, $\tilde{t}_{c,{\rm A}}$, $L_{c,{\rm B}}$, $\beta_{\rm B}$, and $\tilde{t}_{c,{\rm B}}$. If we allow all the seven parameters to vary during the fit, we will inevitably get some parameters with too large errors. This is caused by the fact that we have not enough number of data points available (only 19 after removing the data points at $t=15.5$ and $16.5\,\oday$), and that in the optically thick case the luminosity solution does not explicitly depend on the $\beta$ parameter as can be seen from equation (\ref{dLdt2}). So, during the fit we choose to fix the value of $\alpha$ and $\beta$. We choose $\alpha=0.2$ to agree with the value measured in the nuclear waste. For the value of $\beta$, we choose $\beta_{\rm A}=0.3$ and $\beta_{\rm B}=0.1$, to agree with the values obtained by fitting the photometric spectra with blackbody radiation in \citet{wax17}. Then we have four independent parameters to fit: $L_{c,{\rm A}}$, $\tilde{t}_{c,{\rm A}}$, $L_{c,{\rm B}}$, and $\tilde{t}_{c,{\rm B}}$. The number of degrees of freedom for the fitting is then 15. Applying the so-defined two-component model to fit the bolometric luminosity data (with the last two points being excluded, as explained above), we get a best fit with $\chi^2/\dof=0.462$. The fitting results are shown in Figure~\ref{lp_fit}, and the best fit parameters are listed in Table~\ref{parameter}.

\begin{table}[t!]
\centering
\begin{minipage}{174mm}
\caption{Fitted and derived parameters and quantities in a two-component ejecta model, obtained by fitting to the bolometric luminosity data of SSS17a/AT2017gfo. The best fitted $\chi^2/\dof=6.935/15=0.462$. \label{parameter}}
\begin{tabular}{lcccccccc}
\hline\hline
Component$^\dagger$ & $\alpha^{\rm a}$ & ${V}^{\rm b}$ & ${L_c}^{\rm c}$ & ${\tilde{t}_c}^{\;\;{\rm d}}$ & ${t_c}^{\rm e}$  & ${\kappa M_\ej}^{\rm f}$ & ${fM_\ej}^{\rm g}$  & ${\dot{E}(t=1\oday)}^{\rm h}$\\
\hline
A............... & $0.2$ & $0.3c$ & $12.48\pm0.48$ & $0.845\pm0.022$ & $0.944\pm0.025$ & $0.113\pm0.006$ & $0.0249\pm0.0012$ & $5.09\pm0.25$ \\
B............... & $0.2$ & $0.1c$ & $2.97\pm0.10$ & $3.730\pm0.085$ & $7.223\pm0.165$ & $0.737\pm0.034$ & $0.0235\pm0.0010$ & $7.21\pm0.32$ \\
\hline
\end{tabular}\\[1mm]
$^\dagger$Each component is defined by four parameters, $\alpha$, $\beta=V/c$, $\tilde{t}_c$, and $L_c$. The values of $\alpha$ and $\beta$ are fixed for both component A and component B. Only the $\tilde{t}_c$ and $L_c$ for each component are allowed to vary during the fit. The $t_c$, $\kappa M_\ej$, $fM_\ej$, and $\dot{E}(t=1\oday)$ are derived from the fitting result.\\
$^{\rm a}$Parameter in the power-law index in the heating rate in equation~(\ref{eps_t}).\\
$^{\rm b}$Expansion velocity of the merger ejecta, where $c$ is the speed of light.\\
$^{\rm c}$The fitted critical luminosity scale defined by equation (\ref{L_c}), in units of $10^{41}\,\erg\,\s^{-1}$.\\
$^{\rm d}$The fitted critical timescale $\tilde{t}_c$ in units of day, which is related to the $t_c$ in equation (\ref{tc}) by $\tilde{t}_c=t_c(8\beta/3)^{1/2}$.\\
$^{\rm e}$The derived critical timescale $t_c$ defined by equation (\ref{tc}), in units of day. \\
$^{\rm f}$The derived product of the opacity $\kappa$ and the ejecta mass $M_\ej$, in units of $0.01\,M_\odot\,\cm^2\,\g^{-1}$. \\
$^{\rm g}$The derived product of the dimensionless parameter $f$ (see eq.~\ref{eps_t}) and the ejecta mass, in units of $10^{-6}\,M_\odot$.\\
$^{\rm h}$The derived energy generation rate at $t=1\,\oday$, in units of $10^{41}\,\erg\,\s^{-1}$.
\end{minipage}
\end{table}

From equations (\ref{eps_t}) and (\ref{L_c}) we can derive the energy generation rate at $t=1\,\oday$
\begin{eqnarray}
  \dot{E}(t=1\,\oday)=\epsilon(t=1\,\oday)M_\ej=\frac{1}{2}L_c\left(\frac{\tilde{t}_c}{1\,\oday}\right)^{1+\alpha} \;.
\end{eqnarray}
Then, from the best fit values of $L_c$ and $\tilde{t}_c$, we can derive the energy generation rate at $t=1\,\oday$ for each model component. The results are listed in Table~\ref{parameter}.

As can be seen from Figure~\ref{lp_fit} and Table~\ref{parameter}, the two-component model fits the bolometric luminosity data very well. The fit spans the range from $t=0.5\,\oday$ to $t=14.5\,\oday$. The derived values for $t_c$, $\kappa M_\ej$, and $fM_\ej$ for each component are also listed in Table~\ref{parameter}. We see that the derived values for $fM_\ej$ are about the same for both ejecta components. The relation between the parameter $f$ and the average nuclear radiation efficiency $\eta$ in the ejecta will be discussed in the next section. According to equation (\ref{f_fbar}) in Section \ref{model}, we have $f\propto\eta t_c^{-\alpha}$. If the two ejecta components have the same average radiation efficiency (and similar minimum and maximum nuclear lifetime), we should have $f_{\rm B}\approx f_{\rm A}(t_{c,{\rm B}}/t_{c,{\rm A}})^{-\alpha}\approx 0.67f_{\rm A}$. Hence we have $M_{\ej,{\rm A}} \approx 0.0249\pm0.0012\,f_{{\rm A},-6}^{-1}\,M_\odot$ and $M_{\ej,{\rm B}} \approx 0.0353\pm0.0016\,f_{{\rm A},-6}^{-1}\,M_\odot$, where $f_{{\rm A},-6}=f_{\rm A}/10^{-6}$.

From the above fitting results we can get $M_{\ej,{\rm B}}/M_{\ej,{\rm A}}=1.42\pm0.09$. From the derived values for $\kappa M_\ej$ we can get $\kappa_{\rm B}M_{\ej,{\rm B}}/\kappa_{\rm A}M_{\ej,{\rm A}}=6.50\pm0.45$. So we have $\kappa_{\rm B}/\kappa_{\rm A}=4.59\pm0.27$. Hence, the fitting results indicate that the two ejecta components have very different values of the opacity. From the derived values of $\kappa M_\ej$ and $fM_\ej$, we can infer that $\kappa_{\rm A}=0.046\pm0.002\,f_{{\rm A},-6}\,\cm^2\g^{-1}$ and $\kappa_{\rm B}=0.209\pm0.008\,f_{{\rm A},-6}\,\cm^2\g^{-1}$.

Although a red component seems to be present in the UVOIR data of SSS17a/AT2017gfo, the fitting results do not support a very large opacity in the ejecta or outflow. The lanthanide-featured opacity of $\sim10\,\cm^2\,g^{-1}$ as theoretically claimed in some references is not verified, unless the efficiency parameter $f$ is as large as $\sim 4.8\times 10^{-5}$, but then we would get too small ejected masses in both components ($\sim 5.2\times 10^{-4} M_\odot$ and $\sim 7.4\times 10^{-4} M_\odot$, respectively). In other words, our results may indicate that the fraction of lanthanides is $<10^{-4}$ \citep{kas13}. 

The energy generation rate at $t=1\,\oday$, $\dot{E}(t=1\,\oday)$, and the transition time $t_c$, are two important quantities for determining the amplitude and the peak time of the gamma-ray emission to be calculated in the following sections. The values derived here will be used as a reference for input parameters in our modeling for the radioactive gamma-ray emission produced by the merger ejecta.

\section{Theoretical Basis for the Radioactive Gamma-Ray Emission}
\label{model}

In a merger event of neutron stars, a lot of neutron-rich nuclear isotopes are expected to be produced by the complex nucleosynthesis process in the merger ejecta, many of which are unstable. The radioactive decay of the unstable isotopes releases nuclear energy in the form of neutrino energy, gamma-ray photon energy, and the kinetic energy of particle products (electrons, positrons, $\alpha$-particles, etc). The neutrinos escape from the ejecta freely. The positrons will ultimately annihilate with the electrons in the ejecta and produce additional gamma-ray photons. Electrons, $\alpha$-particles, and other charged particles will interact with other charged particles in the ejecta through the Coulomb interaction and be thermalized. The fate of the gamma-ray photons generated during the decay process is determined by the optical thickness of the ejecta. If the ejecta is optically thick, i.e., $\tau_s\gg 1$, the gamma-ray photons will be thermalized inside the ejecta through scattering and absorption by electrons and ions and finally be radiated away with a thermal or quasi-thermal spectrum in the UVOIR band. In the opposite case, if the ejecta is optically thin, i.e., $\tau_s\ll 1$, the gamma-ray photons produced by the radioactive decay will escape from the ejecta freely, without interaction with matter in the ejecta. In this case, the appearing spectrum of the gamma-ray emission is determined by the original energy distribution of the gamma-ray photons produced by nuclear decays, shaped by the subrelativistic expansion of the ejecta through the Doppler effect.\footnote{Photons of energy smaller than a few hundred keV suffer the photoelectric absorption by the atoms in the ejecta. However, as we will see later, more than $94\%$ of the gamma-ray energy generated in the ejecta is carried by photons of energy $>200\,\keV$, for which the photoelectric absorption has ignorable effects.}

To calculate the spectrum of the gamma-ray emission, we must know the species of nuclides inside the merger ejecta and their abundance. The observed spectrum will be given by the superposition of the gamma-ray line spectrum generated by each nuclide, with inclusion of the line broadening effect caused by the subrelativistic expansion of the ejecta. So, our model consists of many different species of unstable nuclides, each nuclide being denoted by a symbol $X_i$ where $i=1$, $2$, ... Let us consider one nuclide, $X_i$, of mass $m_i$ and mean lifetime $\tau_i$. Assuming that at time $t=0$ the total number of $X_i$ is $N_{i,0}$. Because of the radioactive decay, at time $t$ the number of $X_i$ is $N_i=N_{i,0}e^{-t/\tau_i}$. Decay of one $X_i$ releases an energy $\varepsilon_i$. Then, at time $t$, the accumulated energy generated by $X_i$ is $\Delta E_i=\varepsilon_i\Delta N_i=\varepsilon_iN_{i,0}\left(1-e^{-t/\tau_i}\right)$, which leads to a generation rate of the radioactive energy by one species of nuclide
\begin{eqnarray}
  \frac{dE_i}{dt}=\frac{\varepsilon_iN_{i,0}}{\tau_i}e^{-t/\tau_i} \;.
  \label{dEidt}
\end{eqnarray}

The total generation rate of the radioactive energy is given by the sum of the energy generation rate of all nuclides in the ejecta, i.e., 
\begin{eqnarray}
  \frac{dE}{dt}=\sum_i\frac{dE_i}{dt}=\sum_i\frac{\varepsilon_iN_{i,0}}{\tau_i}e^{-t/\tau_i} \;.
  \label{dEdt}
\end{eqnarray}

To convert the sum in equation (\ref{dEdt}) into an integral, let us assume that at time $t=0$ the number of nuclides is given by a distribution over the mean lifetime, so that in an infinitesimal range of the mean lifetime bounded by $\tau$ and $\tau+d\tau$ the number of nuclides is given by $g(\tau)d\tau$. The total number of nuclides at $t=0$ is then given by $\sum_iN_{i,0}=\int g(\tau)d\tau$. Then, the sum in equation (\ref{dEdt}) can be converted to an integral over $\tau$ by
\begin{eqnarray}
  \frac{dE}{dt}=\int\frac{\varepsilon g(\tau)}{\tau}e^{-t/\tau}d\tau \;.
  \label{dEdt2}
\end{eqnarray}

We assume that $g(\tau)$ is a power law of $\tau$, i.e.,
\begin{eqnarray}
  g(\tau)=A\tau^{-1-\alpha} \;, \label{g_tau}
\end{eqnarray}
where $A$ and $\alpha$ are constants. We further assume that $\varepsilon_i$ is not correlated to $\tau_i$, i.e., $\varepsilon$ is not a function of $\tau$. In other words, we take $\varepsilon$ as being an averaged value of $\varepsilon_i$ and hence being independent of $\tau$. Then we have
\begin{eqnarray}
  \frac{dE}{dt}=\varepsilon A\int_{\tau_{\min}}^{\tau_{\max}}\frac{1}{\tau^{2+\alpha}}e^{-t/\tau}d\tau \;,
  \label{dEdt3}
\end{eqnarray}
where $\tau_{\min}$ and $\tau_{\max}$ are minimum and maximum values of $\tau$. The integral can be worked out with the incomplete gamma function. The result is
\begin{eqnarray}
  \frac{dE}{dt}=\frac{\varepsilon A}{t^{1+\alpha}}\zeta(\alpha,t/\tau_{\min},t/\tau_{\max}) \;,
  \label{dEdt4}
\end{eqnarray}
where
\begin{eqnarray}
  \zeta(\alpha,x_1,x_2)\equiv\Gamma(1+\alpha,x_2)-\Gamma(1+\alpha,x_1) \;.
\end{eqnarray}

For time $t$ satisfying the condition $\tau_{\min}\ll t \ll\tau_{\max}$, we have $\zeta(\alpha,t/\tau_{\min},t/\tau_{\max})\approx\Gamma(1+\alpha)$, and
\begin{eqnarray}
  \frac{dE}{dt} \approx \frac{\varepsilon A\Gamma(1+\alpha)}{t^{1+\alpha}} \;.
  \label{dEdt03}
\end{eqnarray}
For $t\ll\tau_{\min}$, we have $dE/dt\approx\varepsilon A(1+\alpha)^{-1}\tau_{\min}^{-1-\alpha}$. For $t\gg\tau_{\max}$, we have $dE/dt \approx \varepsilon A\tau_{\max}^{-\alpha}t^{-1} e^{-t/\tau_{\max}}$.

For any $t$ satisfying the condition $\tau_{\min}\ll t \ll\tau_{\max}$, the dominant contribution to the integral of $dE/dt$ in equation~(\ref{dEdt3}) comes from nuclides with $\tau\sim t$. To see this point, let us define $z=\ln(\tau/t)$ and rewrite equation~(\ref{dEdt3}) as
\begin{eqnarray}
  \frac{dE}{dt}=\frac{\varepsilon A}{t^{1+\alpha}}\int_{-\infty}^{\infty}F(z)dz \;,
\end{eqnarray}
where
\begin{eqnarray}
  F(z)=\exp\left[-(1+\alpha)z-e^{-z}\right] \;. \label{F_z}
\end{eqnarray}
Here we have taken $z_{\min}\equiv\ln(\tau_{\min}/t)=-\infty$, and $z_{\max}\equiv\ln(\tau_{\max}/t)=\infty$. It can be checked that $F(z)\rightarrow 0$ as $z\rightarrow\pm\infty$. Hence, the function $F(z)$ peaks at $z=z_m$, where $z_m$ is determined by $\partial F/\partial z=0$. Since $\partial F/\partial z=-F(z)\left(1+\alpha-e^{-z}\right)$, the solution to $\partial F/\partial z=0$ is $z=z_m\equiv-\ln(1+\alpha)$, i.e., $F(z)$ peaks at
\begin{eqnarray}
   \tau=\tau_m\equiv\frac{t}{1+\alpha} \;.
\end{eqnarray}
At $z=z_m$, we have the peak value of $F(z)$
\begin{eqnarray}
  F_m\equiv F(z_m)=(1+\alpha)^{1+\alpha}e^{-1-\alpha} \;.
\end{eqnarray}

The ``width'' of the integrand function $F(z)$ can be defined by $F\left(z_{\pm1/2}\right)=F_m/2$. For any $\alpha$ in the range of $0\le\alpha\le1$, the solution of $y_{\pm1/2}\equiv \exp(z_{\pm1/2})$ can be approximated by $y_{-1/2} = 0.3734\left(1-0.5369\alpha+0.1867\alpha^2\right)$, and $y_{1/2} = 4.311\left(1-1.843\alpha+2.425\alpha^2-1.912\alpha^3+\right.$ $\left.0.6368\alpha^4\right)$ with a relative error $<1\%$. So, the energy generation at time $t$ mainly comes from nuclides with mean lifetimes in the range of $\sim 0.3t$--$4.5t$.

Integration of equation (\ref{dEdt3}) over time from $t=0$ to $t=\infty$ gives rise to the total energy generated by the radioactive decay
\begin{eqnarray}
  \Delta E = \int_0^\infty\frac{dE}{dt}dt= \varepsilon A\xi\left(\alpha,\tau_{\min},\tau_{\max}\right) \;,
  \label{dE}
\end{eqnarray}
where
\begin{eqnarray}
  \xi\left(\alpha,\tau_{\min},\tau_{\max}\right)\equiv\left\{\begin{array}{ll}
  \ln\left(\tau_{\max}/\tau_{\min}\right) \;, & \quad \alpha=0 \;,\\
  \alpha^{-1}\left(\tau_{\min}^{-\alpha}-\tau_{\max}^{-\alpha}\right) \;, & \quad \alpha>0 \;.
  \end{array} \right.
\end{eqnarray}

Let us denote the total rest mass of the radioactive elements at $t=0$ by $\Delta M_0$ and define an averaged radiation efficiency $\eta$ by $\eta\equiv\Delta E/\Delta M_0 c^2$. Then, from equation (\ref{dE}), we can derive that
\begin{eqnarray}
  \varepsilon A=\eta\xi^{-1}\Delta M_0c^2 \;.
\end{eqnarray}
By equation (\ref{dEdt4}) we have then
\begin{eqnarray}
  \frac{dE}{dt} = \frac{\eta\xi^{-1}\Delta M_0c^2}{t^{1+\alpha}}\zeta(\alpha,t/\tau_{\min},t/\tau_{\max}) \;.
  \label{dEdt5}
\end{eqnarray}
For the case of $\tau_{\min}\ll t \ll\tau_{\max}$, we have
\begin{eqnarray}
  \frac{dE}{dt}\approx\frac{\eta\Delta M_0c^2}{\xi\left(\alpha,\tau_{\min},\tau_{\max}\right)}\frac{\Gamma(1+\alpha)}{t^{1+\alpha}} \;.
  \label{dEdt6}
\end{eqnarray}

Comparison of equations (\ref{eps_t}) and (\ref{dEdt6}) leads to
\begin{eqnarray}
  f=\eta\frac{\Gamma(1+\alpha)}{\xi t_c^\alpha} \;. \label{f_fbar}
\end{eqnarray}
When $\alpha=0$, we have $f=\eta/\ln(\tau_{\max}/\tau_{\min})$. When $\alpha>0$ and $\tau_{\max}\gg\tau_{\min}$, we have $f\approx \eta\alpha\Gamma(1+\alpha)(t_c/\tau_{\min})^{-\alpha}$.

We see that, the parameter $f$ in equation (\ref{eps_t}) is related to the average radiation efficiency $\eta$ of the radioactive decay, but they are not identical. The value of $f$ also depends on a few parameters: the minimum and maximum mean lifetime of nuclides in merger ejecta, and the critical timescale $t_c$ when $\alpha >0$.

To estimate the effect of $\tau_{\min}$, $\tau_{\max}$, and $t_c$ on the value of $f$, let us take $t_{\min}=1\,\s$, $t_{\max}=2.1\times 10^{17}\,\s$ (the mean lifetime of uranium), and $t_c=1\,\oday$. Then we get $f\approx \eta/40$ when $\alpha=0$, and $f\approx \eta/53$ when $\alpha=0.2$. So, it appears that $f$ is smaller than the average radiation efficiency $\eta$ by a significant factor. This is easy to understand. The parameter $f$ describes the strength of the energy generation rate at a given time $t$. According to the above result, at any time $t$ the energy generation is dominantly contributed by nuclides with mean lifetimes comparable to $t$. Hence, an increase in the amount of elements with mean lifetimes much larger or much smaller than $t$ can only increase the total mass of the ejecta but adds little contribution to the total energy released at time $t$, which results in the value of $f$ being significantly reduced. According to equation (\ref{eps_t}), when $\alpha\neq 0$ the strength of the energy generation at time $t$ is described by $ft_c^\alpha$, which explains the appearance of $t_c^{-\alpha}$ in equation (\ref{f_fbar}). 

If we interpret the $\varepsilon_i$ as the gamma-ray energy generated in a radioactive decay, equation (\ref{dEdt}) would give the gamma-ray energy generation rate $dE_\gamma/dt$. In nuclear physics the total energy released in a radioactive decay is usually measured by the $Q$-value, which is defined as the difference in the rest mass energy between the parent nuclide and the daughter nuclide. If in equation (\ref{dEdt}) we substitute $Q_i$ for the $\varepsilon_i$, we would get the total energy generation rate $dE_Q/dt$ which contains the energy released in various forms. According to \citet{met10}, for the $\beta$-decay, which makes the dominant contribution to the total energy generation in their model, fractions of the energy carried by electrons, neutrinos, and gamma-ray photons are respectively: $\epsilon_e\approx\epsilon_\nu\approx 0.25$, and $\epsilon_\gamma\approx 0.5$. However, in our model, as we will see later, the dominant contribution to the gamma-ray energy generation comes from $\beta^+$-decays and electron captures, which is about $65\%$ of the total. The contribution of $\beta^-$-decays to the total gamma-ray generation is about $32\%$, with the remaining $3\%$ contributed by $\alpha$-decays and isomeric transitions. Hence, in our model, the contribution of $\beta$-decay electrons to the heating rate through the thermalization process is about $14\%$.

In the optically thin phase, almost all the gamma-ray energy generated by radioactive decays will escape from the ejecta directly and form the gamma-ray radiation. To take into account the transition from the optically thick phase to the optically thin phase, the gamma-ray energy generation rate should be multiplied by a factor of $e^{-\tau_s}=e^{-t_c^2/t^2}$ to give rise the gamma-ray luminosity. That is, we have
\begin{eqnarray}
  L_\gamma=e^{-t_c^2/t^2}\dot{E}_\gamma =e^{-t_c^2/t^2}f_\gamma\dot{E}_Q \;,  \label{Lgam_Egam}
\end{eqnarray}
where $f_\gamma\sim 0.5$, $\dot{E}_\gamma=dE_\gamma/dt$, and $\dot{E}_Q=dE_Q/dt$. As expected, when $t\gg t_c$ we have $L\approx\dot{E}_\gamma\approx f_\gamma\dot{E}_Q$. Here we have assumed that the critical time $t_c$ is independent of the photon energy, or the $t_c$ can be considered as the value after being averaged over the photon energy (c.f. eq.~\ref{tc_av} in Section~\ref{detect}). In reality, gamma-ray photons of energy $\varepsilon\la$ a few 100 keV seriously suffer the photoelectric absorption by the high-$Z$ nuclei in the ejecta, resulting that $t_c$ increases rapidly with decreasing photon energy for $\varepsilon\la 300\,\keV$. However, as we will see in Section~\ref{spectra}, for the gamma-rays generated by radioactive decays in a merger ejecta, more than $90\%$ of the energy is carried by photons of $\varepsilon>300\,\keV$. Hence, the variation of $t_c$ with photon energy has little influence on the calculation of the gamma-ray luminosity.

To calculate the luminosity and spectrum of the gamma-ray emission produced by radioactive decays from a neutron star merger, we need to consider the radioactive decay process in an expanding medium. Because of the compactness of neutron stars (the radius is on the order of $10\,{\rm km}$ for a neutron star of one solar mass), the merger ejecta can expand with a velocity that is a significant fraction of the speed of light (e.g., $V\sim 0.1$--$0.3c$). The subrelativistic expansion of the merger ejecta causes a number of important effects that must be taken into account in calculation of the luminosity and the observed spectra, including redshift and blueshift of photon energy, relativistic Doppler broadening of emission lines,\footnote{The line broadening due to the thermal motion of atomic nuclei is negligible compared to that caused by the subrelativistic expansion of the ejecta. The thermal velocity of atomic nuclei can be estimated by $V_\thermal\approx 3\times 10^{-5}c\,(T/10^6\,\K)^{1/2}(A/100)^{-1/2}$, which is $\ll V$. Here $T$ is the temperature of the ejecta gas, and $A$ is the average mass number of the nuclei in the ejecta.} and distortion in the spectrum shape and the lightcurve profile. The effect of special relativity must also be taken into account to certain orders. Mathematical details for treatment of the nuclear reaction and energy production in a spherical expanding medium are presented in Appendix \ref{sphere} with the effect of special relativity being properly considered, where the formula for calculation of the energy generation rate and the spectra of gamma-ray emissions as observed by a remote observer are derived.

\section{The Nuclear Data Sample}
\label{data}

We extract from the NuDat\,2 database at the National Nuclear Data Center\footnote{http://www.nndc.bnl.gov/nudat2/} the radioactive decay data for {\it all} nuclides satisfying the following three criteria: (1)~The half-life $t_{1/2}$ of the nuclide satisfies the condition $0.05\,\oday\la t_{1/2}\la 50,000\,\oday$. Note, the half-life is related to the mean lifetime by $t_{1/2}=\tau\ln 2$. (2)~The nuclide and its decay modes have complete information about the energy state and branching ratios. The energy state of a nuclide is specified by the parameter {\jpi}, denoting the angular momentum and the parity of the nuclide. In each {\jpi} state the sum of the branching ratios for all decay modes is close to $100\%$, at least $85\%$. (3)~Each decay mode of a nuclide has available gamma-ray radiation data, although the completeness of the radiation data may be a question for some nuclides.

The condition on the half-life is based on the consideration that we want to calculate the gamma-ray emission in the time interval of $\sim 1$--$100\,\oday$ since the merger time. According to the analysis in Section~\ref{model}, the dominant contribution to the emitted energy at any time $t$ comes from decays with a mean lifetime $\tau\sim t$. So, decay modes in the range of $0.05\,\oday\la t_{1/2}\la 50,000\,\oday$ (corresponding to $0.072\,\oday\la\tau\la 72,000\,\oday$) are enough for our purpose. For instance, at $t=1\,\oday$, the value of $F$ in equation (\ref{F_z}) at $t_{1/2}=0.05\,\oday$ (i.e., at $\tau=0.072\,\oday$) is $\approx 3.6\times 10^{-5}F_m$. At $t=100\,\oday$, the value of $F$ at $t_{1/2}=50,000\,\oday$ (i.e., at $\tau=72,000\,\oday$) is $\approx 3.8\times 10^{-3} F_m$.

Without the information of {\jpi} of a nuclide, it will not be possible to match the radiation data in the radiation database with a given nuclide precisely. For instance, in an isomeric transition we need to know the quantum states of the nuclide before and after the transition. A nuclide in different {\jpi} states can have different decay modes. A nuclide in a given {\jpi} state can have multiple decay modes, each decay mode has a corresponding branching ratio. Obviously, a necessary condition for the data completeness is that the sum of the branching ratios in a given J$\pi$ state for all decay modes should be equal to $100\%$. In practice we require that the sum is at least $>85\%$ so that the data are close to completeness.

From the NuDat\,2 database we extract in total 494 nuclides with 614 total decay modes satisfying the above three criteria, which form the data sample for our investigation. The majority of the nuclides in the sample have a branching ratio sum equal to $100\%$ in a given energy state. In the sample, a nuclide can have multiple {\jpi} states. For instance, $\prescript{101}{}{\rm Rh}$ has two {\jpi} states: $1/2-$ and $9/2+$. A nuclide in a given {\jpi} state can have multiple decay modes and hence multiple branching ratios. For instance, $\prescript{101}{}{\rm Rh}$ in the $\mbox{\jpi}=9/2+$ state has two decay modes: isomeric transition with a branching ratio $7.2\%$, and electron capture with a branching ratio $92.8\%$. In principle, each decay mode has its own half-life. But the half-life listed in the NuDat\,2 database is defined by the total decay constant $\lambda$, i.e., $t_{1/2}=\ln 2/\lambda$. The individual half-life for a particular decay mode is obtained by $t_{1/2,i}=t_{1/2}/B_i$, where $B_i$ is the branching ratio of the $i$-th decay mode.

In the selected data sample, the minimum half-life of nuclides is equal to $0.0475\,\oday$ ($\prescript{174}{}{\rm Ta}$ with $\mbox{\jpi}=3+$, $t_{1/2}=1.14\,\hr$), and the maximum half-life is equal to $51,500\,\oday$ ($\prescript{242}{}{\rm Am}$ with $\mbox{\jpi}=5-$, $t_{1/2}=141\,{\rm yr}$). Each of the 614 decay modes of the 494 nuclides in 537 energy states has its radiation data available in the NuDat\,2 database. The radiation data in the database come from various sources and the completeness of the data is hard to judge, although for most nuclides in the sample the completeness may not be a problem. Although the data completeness can be a big caveat in the current work, we expect that it has little effect on the calculated gamma-ray spectrum which is given by the sum of gamma-ray emissions from all nuclides in the sample. The shape and feature of the spectrum are determined by the collective and statistical properties of the whole radiation data, which will be reasonably precise so long as the sample is uniformly extracted from the database and the radiation data of most nuclides are complete or close to complete. However, data incompleteness may cause an underestimate of the gamma-ray radiation efficiency.

The primary decay modes of the 537 nuclides include five decay types: $\alpha$-decay, $\beta^-$-decay, $\beta^+$-decay, electron capture, and isomeric transition. Obviously, these nuclides uniformly extracted from the NuDat\,2 database include not only r-nuclides, although heavy nuclides with mass number $>60$ occupy $98\%$ of the total mass. In fact, about half of the nuclides in the sample are on the neutron-deficient side of the valley of nuclear stability in the nuclear chart and hence are classified as p-nuclides (proton-rich nuclides). The other half are on the neutron-rich side of the valley and hence are r-nuclides. The unstable r-nuclides are characterized by $\beta^-$-decays to the stable bottom of the valley, while the unstable p-nuclides are characterized by $\beta^+$-decays and electron captures to the stable bottom of the valley. For extremely heavy nuclides, fissions and $\alpha$-decays can happen. As explained in the Introduction, fissions are not included in our data sample due to the lack of radiation data for fissions. Due to the extremely neutron-rich nature of the merger ejecta, it is almost certain that the r-process must occur in the ejecta, as verified by many numerical works. So, inclusion of r-nuclides in the sample is easy to understand. However, in our data sample we also choose to include the p-nuclides, for the following two reasons: (a) The possibility for an effective production of p-nuclides in the ejecta of a neutron star merger cannot be excluded, as will be explained bellow; (b) To identify the unique feature of the gamma-ray spectrum produced by r-nuclides, it is necessary to have the corresponding spectrum produced by p-nuclides to compare.

The p-nuclides can be synthesized from the pre-existing s- and r-nuclide seeds by the following p-processes: $({\rm p},\gamma)$ reactions, $(\gamma,{\rm n})$ photodisintegrations, and capture of neutrinos. Under conditions encountered in astrophysical environments, the formation of p-nuclides through the photodisintegration of s- and r-nuclides (also called the $\gamma$-process) is often more preferable than the capture of protons due to the fact that the Coulomb barrier of a nucleus increases with increasing proton number. To produce p-nuclides efficiently three conditions must be satisfied \citep[][and references therein]{arn03,ili15,lic18}: (1)~There are abundant enough seed nuclei (s- or r-nuclei); (2)~The temperature in the medium is high enough, better in the range of $(1.5$--$3.5)\times 10^9\,\K$; (3)~The time duration of the hot phase is short enough ($\la 1\,\s$) to avoid complete photoerosion of heavy nuclides. Thus, stellar explosions with rapid expansion and cooling of material have been considered as the most plausible site for the production of proton-rich nuclides. So far studies have been focused on type II and type I supernovae as the major site for the formation of p-nuclides \citep[][and references therein]{bur57,woo78,arn03}. To our knowledge, no research has been done yet for the p-process in a neutron star merger. The first condition is clearly satisfied in a merger, where the r-process can produce tons of r-nuclides within a very short timescale, which can be the seeds for the production of p-nuclides. During the rapid expansion of the ejecta, the temperature of the ejecta gas can increase quickly for a short period of time by various heating mechanisms, e.g., by shock wave heating, r-process energy generation, radioactive decays of r-nuclides, or by the energy input from the short GRB central engine. For instance, the research by \citet{kor12} has shown that the energy generated by the r-process can heat the merger gas to a temperature $\sim 10^9\,\K$ for a time period of $\sim 0.1\,\s$. If the short GRB central engine can deposit an energy of $\sim 10^{49}\,\erg$ into the merger ejecta, the temperature can also be easily increased to $>10^9\,\K$ for a time period of $\sim 1\,\s$. Hence, it is reasonable to imagine that the second and the third conditions are also satisfied in a neutron star merger, and that the merger may be another promising site for the production of p-nuclides.

As stated above, unstable p-nuclides are characterized by $\beta^+$-decays and electron captures, and unstable r-nuclides are characterized by $\beta^-$-decays. Hence we expect that the gamma-rays generated by the decays of r- and p-nuclides will have distinguishable spectral features. Among those features the most important one would be the creation of pair annihilation lines. Since the positrons produced by $\beta^+$-decays will annihilate with the electrons in the medium immediately, an annihilation line at $511\,\keV$ is expected to be present in a gamma-ray spectrum of an ejecta dominated by p-nuclides. On the other hand, if the ejecta is dominated by r-nuclides, the annihilation line would be absent. This distinct feature and some other subtle features that will be discussed later make it possible to distinguish the nuclear process and the resulting products in a merger ejecta through observations of the gamma-ray emissions.

\begin{figure}[ht!]
\centering{\includegraphics[angle=0,scale=0.65]{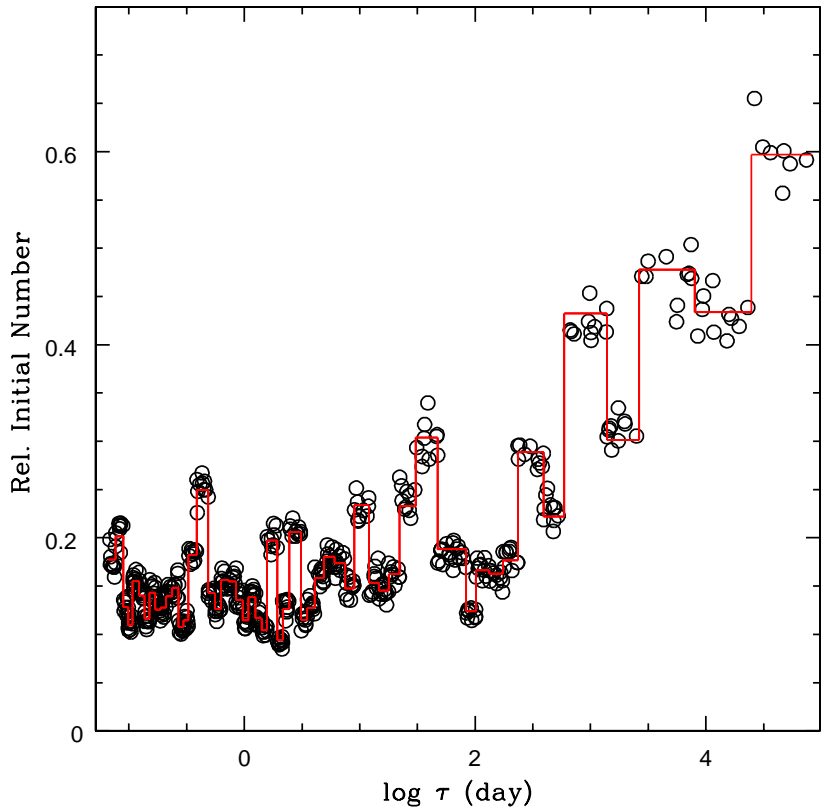}}
\caption{Relative initial number (arbitrarily normalized) of the 537 energy states of the 494 nuclides in the data sample versus the mean lifetime. The initial number of a nuclide is generated randomly as a Gaussian distribution with $5\%$ deviation around a mean abundance (red solid line) defined by the integral of the function $g(\tau)$ (eq.~\ref{g_tau}, $\alpha=0.1$ and $A=1$) over an interval of the mean lifetime $\tau$. The logarithm of $\tau$ is divided into intervals with unequal lengths so that each interval contains a large enough number of nuclide states (see the text for details). 
}
\label{ab_tau}
\end{figure}

Our calculations will be based on all the 614 decay modes of the 494 nuclides selected above. To determine the initial number of each nuclide, we use the function $g(\tau)$ defined by equation (\ref{g_tau}) to model the distribution of the abundance of nuclides over their mean lifetime $\tau$. Since some nuclides have multiple energy states with each state having its own mean lifetime, we treat a nuclide with a given {\jpi} value as an independent nuclide species. Then our sample of nuclides contains 537 independent nuclide species or nuclide elements. The elements are divided into a number of groups according to their mean lifetime in an ascending order. Each group of elements contains 10 elements, except the last group which contains 7 elements. Each group spans an interval in the mean lifetime coordinate, bounded by $\tau_1$ and $\tau_2$. Then, theoretically, the total number of elements in the lifetime interval is given by $\Delta N=\int_{\tau_1}^{\tau_2}g(\tau)d\tau$. Suppose that there are $n$ number of elements in a given group ($n=10$, or $7$ for the last group). The mean number of a nuclide species in that group is then $\overline{N}=\Delta N/n$. For each nuclide element in that group, we generate its initial number or abundance through a Gaussian distribution around $\overline{N}$, with a deviation of $5\%$.

The relative initial number (or, abundance defined in mole fraction) of all nuclide states generated with the above Monte Carlo method is shown in Figure~\ref{ab_tau}, where we have taken the parameter $\alpha=0.1$. We choose this value of $\alpha$ so that the gamma-ray energy generation rate will be $\propto t^{-1.2}$, to be consistent with the result of fitting SSS17a/AT2017gfo in Section \ref{outline} and the value found for the nuclear waste.  Normalization of the number of each nuclide state at the initial time $t=0$ will be determined by scaling the calculated energy generation rate to a given value at some specified time, for instance, to a given energy generation rate at $t=1\,\oday$ after the merger.

The nuclide sample with the relative abundance generated with the above approach would produce an energy generation rate $\dot{E}\sim t^{-1.1}$, according to equation (\ref{dEdt03}). However, the detailed numerical calculation in the next section, done with the sum over nuclide species instead of with the integral, leads to a more accurate gamma-ray energy generation rate $\dot{E}_\gamma\sim t^{-1.2}$. The slight difference in the power-law index of energy generation will be explained in the next section. This value of the time power-law index is in the range of that obtained from numerical simulations based on the r-process network \citep{met10,kor12}. Hence, the nuclide sample that we have constructed should fit the task in this work, at least in principle.

\section{Energetics and the Luminosity}
\label{lum}

With the modeled nuclide abundance, the energy generation rate can be calculated with equation~(\ref{dEdt}), where the sum $\sum_i$ is over all nuclide states and all decay modes in the sample. Each nuclide in a given energy state (specified by the {\jpi} parameter) can have several decay modes. We denote a nuclide state by an index $i$, and a decay mode by an index $j$. Assuming that the $j$-th decay mode of the $i$-th nuclide species has a $Q$-value $Q_{i,j}$ and a branching ratio $B_{i,j}$. Then for the total energy generation by a nuclide we have
\begin{eqnarray}
  \varepsilon_{Q,i}=\sum_jB_{i,j}Q_{i,j} \;.  \label{Q_i}
\end{eqnarray}
and the total energy generation rate is calculated by
\begin{eqnarray}
  \dot{E}_Q=\sum_i\frac{N_{i,0}}{\tau_i}e^{-t/\tau_i}\sum_j B_{i,j}Q_{i,j} \;.
  \label{dEqdt}
\end{eqnarray}

There is no available $Q$-value associated with isomeric transitions, since in these processes parent and daughter nuclides are the same nuclide in different energy levels. For isomeric transitions, we use the parent energy level (defined relative to the daughter energy level) as their $Q$-values. As we have already mentioned, a fraction of the total energy release calculated through the $Q$-value is in the form of gamma-ray photons. A part of the released energy is also in the hard X-ray domain. However, the X-ray radiation only occupies a very small fraction in the total electromagnetic radiation generated by a radioactive decay. In the following part for simplicity we use the term gamma-ray radiation to represent both the gamma-ray and the X-ray radiation. The fraction of the gamma-ray emission in the total released energy can be a function of time. So, the energy generation in gamma-rays should be calculated independently.

To calculate the energy generation rate for the gamma-ray radiation, the $Q_{i,j}$ in equation (\ref{dEqdt}) should be replaced by the energy of the gamma-ray radiation released in a decay. Each decay mode of a nuclide can release multiple photons of different energy with different intensity (probability). We denote each photon energy and the corresponding intensity by an index $k$. Hence, $Q_{i,j}$ should be replaced by $\sum_kh_{i,jk}\varepsilon_{i,jk}$, where $\varepsilon_{i,jk}$ is the energy, $h_{i,jk}$ is the corresponding intensity of the $k$-th photon emitted in the $j$-th decay mode of the $i$-th nuclide state. Then, we have the total gamma-ray energy released by a nuclide species
\begin{eqnarray}
  \varepsilon_{\gamma,i}=\sum_jB_{i,j}\sum_kh_{i,jk}\varepsilon_{i,jk} \;.  \label{vareps_i}
\end{eqnarray}
According to the NuDat\,2, the intensity for the gamma-ray radiation corresponds to the gamma branching ratio for each level, assigning 100 to the strongest gamma-ray.

\begin{figure}[ht!]
\centering{\includegraphics[angle=0,scale=0.65]{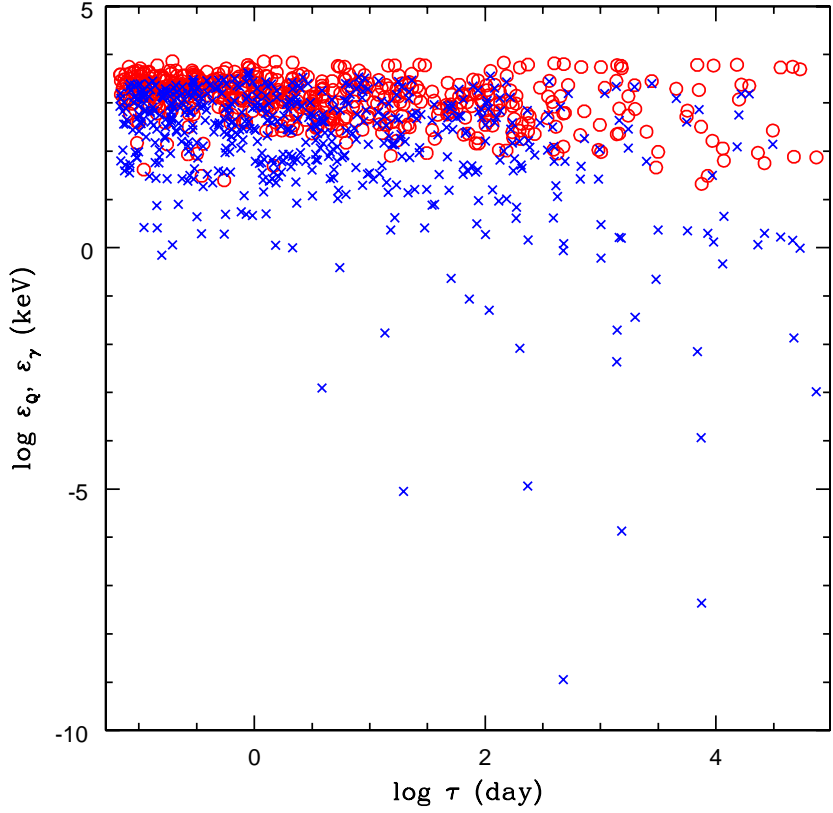}}
\caption{The total energy and the gamma-ray energy emitted by a radioactive nuclide versus its mean lifetime. The total energy $\varepsilon_Q$ for each nuclide state is calculated by equation (\ref{Q_i}) and shown with red circles. The gamma-ray energy $\varepsilon_\gamma$ is calculated by equation~(\ref{vareps_i}) and shown with blue crosses.
}
\label{ener_tau}
\end{figure}

The $\varepsilon_{Q,i}$ and $\varepsilon_{\gamma,i}$ defined by equations (\ref{Q_i}) and (\ref{vareps_i}) are calculated for all nuclides in the sample. The results are shown in Figure~\ref{ener_tau}, from which we see a weak anticorrelation between the released energy and the mean lifetime of nuclides, especially for the gamma-ray energy $\varepsilon_{\gamma,i}$.

\begin{figure}[ht!]
\centering{\includegraphics[angle=0,scale=0.65]{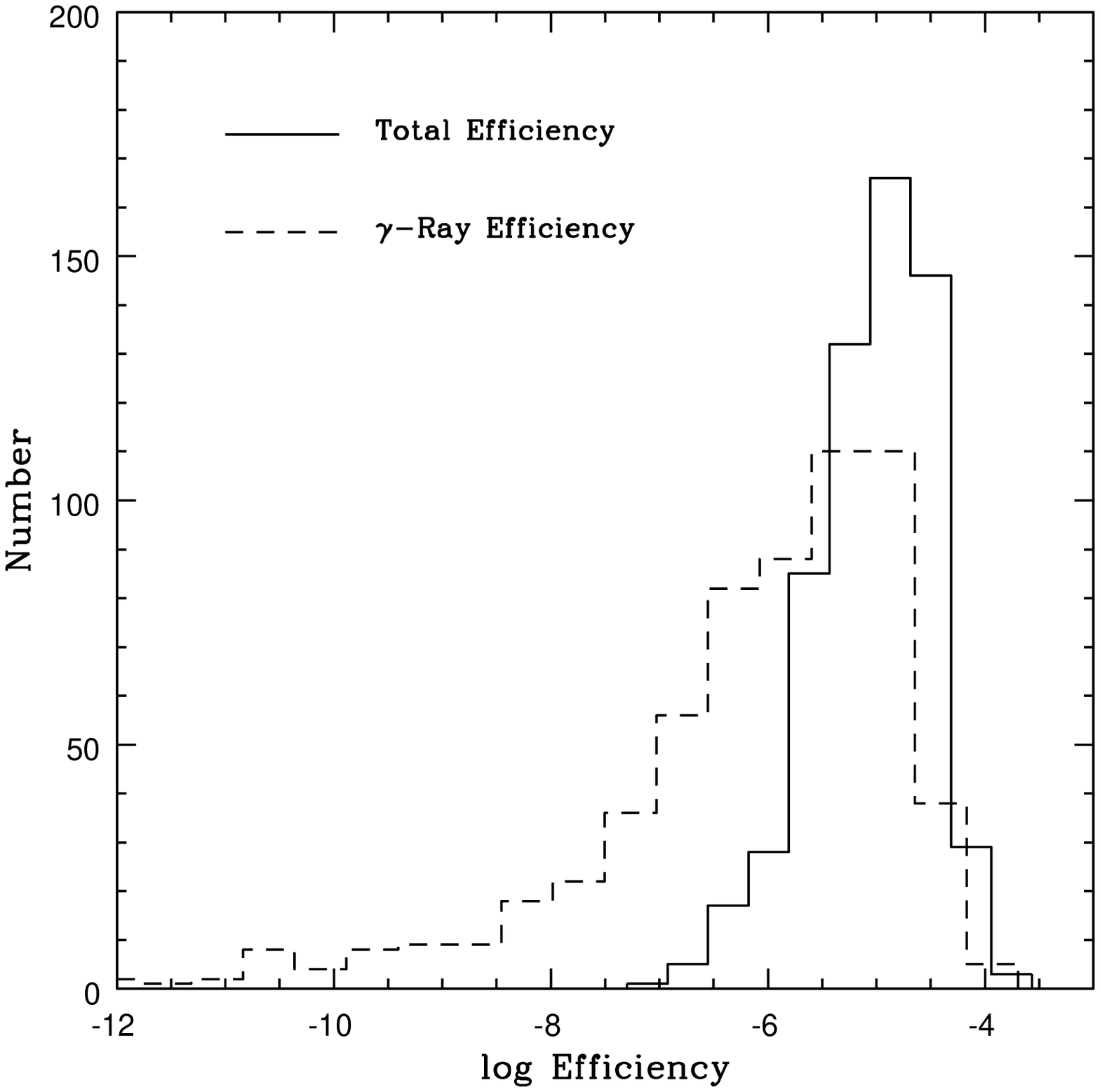}}
\caption{Distribution of the radiation efficiency for the 612 decay modes of the nuclides in the sample, where two decay modes with negative $Q$-values are excluded. The solid line histogram shows the efficiency calculated with the $Q$-value, i.e., the efficiency for the total energy released in various forms during a radioactive decay. The dashed line histogram shows the efficiency for the gamma-ray energy emission.
}
\label{eff}
\end{figure}

For each decay mode of a nuclide, we can define a total radiation efficiency by $\eta_{Q,ij}=Q_{i,j}/m_ic^2$, and a gamma-ray radiation efficiency by $\eta_{\gamma,ij}=\sum_kh_{i,jk}\varepsilon_{i,jk}/m_ic^2$. In Figure~\ref{eff} we show the histogram distribution of $\eta_{Q,ij}$ and $\eta_{\gamma,ij}$ for the 612 decay modes with positive $Q$-values. The two decay modes with negative $Q$-values ($\prescript{87}{}{\rm Sr}$ with $Q=-282.2\,\keV$ for electron capture, and $\prescript{180}{}{\rm Hf}$ with $Q=-846\,\keV$ for $\beta^-$-decay) are excluded from the data shown in Figure~\ref{eff}. From the data for the 612 decay modes, we derive that the mean of the total radiation efficiency is $\approx 1.69\times 10^{-5}$, and the mean of the gamma-ray radiation efficiency is $\approx 6.45\times 10^{-6}$. We see that the gamma-ray radiation efficiency has an extremely wide distribution. If we exclude efficiency bins with number of nuclides smaller than 20 to reduce statistical errors, we find that the gamma-ray radiation efficiency is distributed in the range $\sim 10^{-8}$--$10^{-4}$, over four orders of magnitude.

To calculate the luminosity of the gamma-ray emission, the energy generation rate defined in the rest frame of ejecta must be converted to the energy rate in the observer frame. After taking into account the subrelativistic expansion of the ejecta, for a single nuclide species the gamma-ray energy generation rate defined in the observer frame is given by equation~(\ref{dEidt_obs2a}) in Appendix~\ref{sphere}. After summation over all nuclide species, decay modes, and emission lines, we get the total gamma-ray energy generation rate as measured by the observer
\begin{eqnarray}
  \dot{E}_{\gamma,\obs} = \sum_i\frac{N_{i,0}}{\tau_i}I_1(\alpha^\prime_i,\beta)\sum_jB_{i,j}\sum_kh_{i,jk}\varepsilon_{i,jk} \;, 
  \label{dEgdt_obs} 
\end{eqnarray}
where $\alpha^\prime_i\equiv t/\tau_i$, $y_-\equiv (1+\beta)^{-1}$, $y_+\equiv (1-\beta)^{-1}$, and $I_1$ is defined by equation (\ref{I1}). Here $\beta\equiv V/c$, $V$ is the expansion velocity at the surface of the ejecta. Then, by equation~(\ref{Lgam_Egam}), we get
\begin{eqnarray}
  L_\gamma = e^{-t_c^2/t^2}\sum_i\frac{N_{i,0}}{\tau_i}I_1(\alpha^\prime_i,\beta)\sum_jB_{i,j}\sum_kh_{i,jk}\varepsilon_{i,jk} \;,
  \label{Lg} 
\end{eqnarray}
after considering the effect of optical depth.

\begin{figure}[ht!]
\centering{\includegraphics[angle=0,scale=0.65]{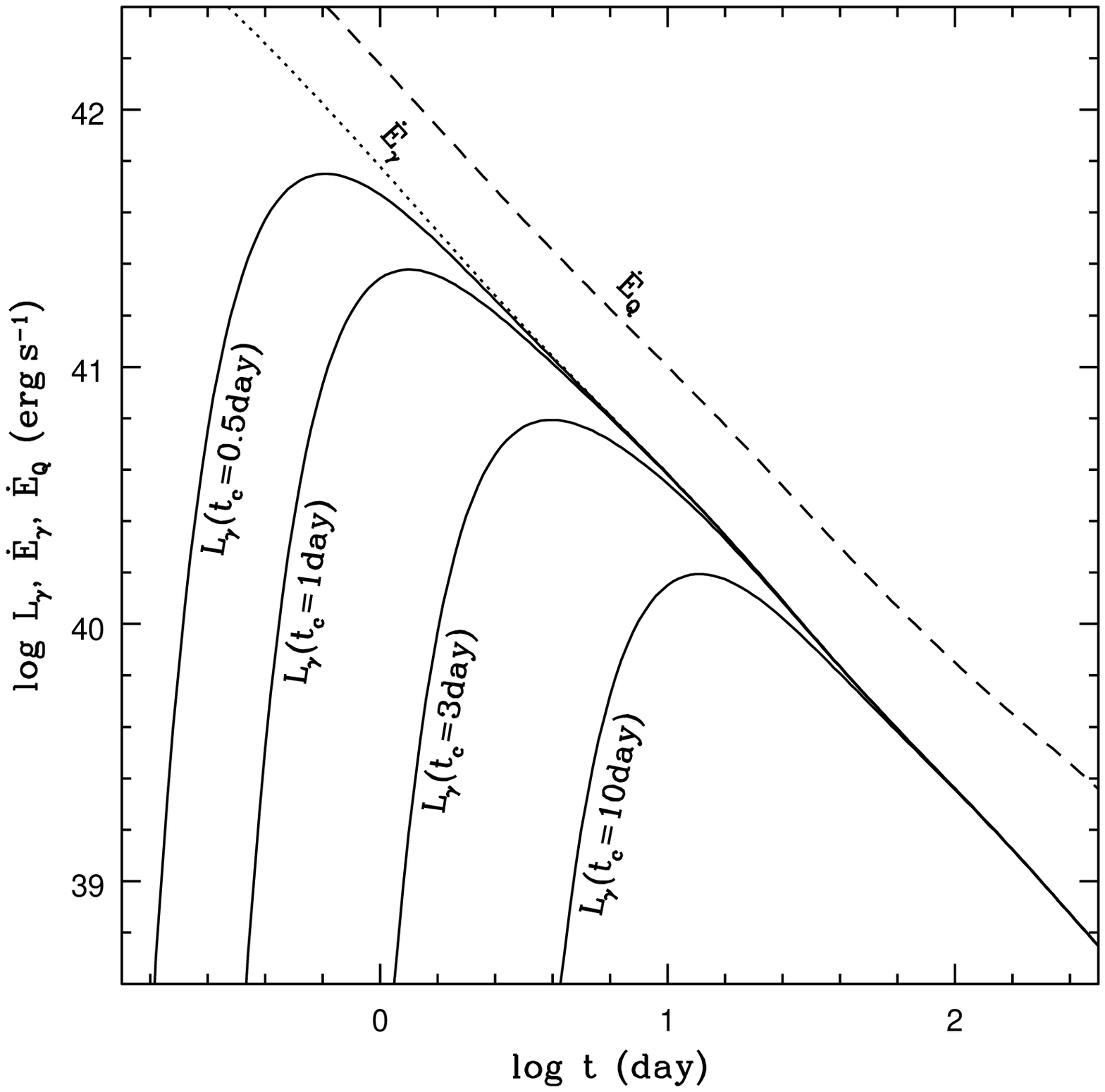}}
\caption{The gamma-ray luminosity $L_\gamma$ (solid line), the gamma-ray energy generation rate $\dot{E}_\gamma$ (dotted line), and the total energy generation rate  $\dot{E}_Q$ (dashed line). Different luminosity curves correspond to different values of the critical time, from $t_c=0.5\,\oday$ to $t_c= 10\,\oday$. For the ejecta expansion velocity we have adopted $V=0.3c$. As explained in the text, the luminosity and the energy generation rate depend weakly on the value of $V$ in the subrelativistic situation. The energy generation is normalized so that $\dot{E}_\gamma=6\times 10^{41}\,\erg\,\s^{-1}$ at $t=1\,\oday$.
}
\label{power}
\end{figure}

The gamma-ray luminosity $L_\gamma$, the gamma-ray energy generation rate $\dot{E}_\gamma=\dot{E}_{\gamma,\obs}$, and the total energy generation rate $\dot{E}_Q$ calculated with the above formulae, are shown in Figure~\ref{power}. In calculation of the luminosity we have taken a number of values for the critical time, from $t_c=0.5\,\oday$ to $t_c=10\,\oday$. To determine the absolute number of each nuclide species in the ejecta, we have adopted the following normalization condition: $\dot{E}_\gamma=6\times 10^{41}\,\erg\,\s^{-1}$ at $t=1\,\oday$ (c.f. the case of SSS17a/AT2017gfo in Table~\ref{parameter}). Our calculation results indicate that $\dot{E}_\gamma\approx 0.4\dot{E}_Q$ at $t=1\,\oday$. Asymptotically, we have $L_\gamma=\dot{E}_\gamma\propto t^{-1.21}$ and $\dot{E}_Q\propto t^{-1.17}$, which are also in agreement with the fitting results for SSS17a/AT2017gfo. Both power-law indices slightly differ from the theoretical index $-1.1$, as inferred from the power-law distribution function $g(\tau)\propto \tau^{-1.1}$ used in generating the abundance of the nuclide species. This is caused by a slight statistical anticorrelation between the energy generated by radioactive decays and the mean lifetime of nuclides, as can be seen in Figure~\ref{ener_tau}. A straight line fit to the data in Figure~\ref{ener_tau} leads to $\varepsilon_Q\propto\tau^{-0.1}$, and $\varepsilon_\gamma\propto\tau^{-0.1}$ if data points with $\varepsilon_\gamma<3\,\keV$ are excluded from the fit.

\begin{figure}[ht!]
\centering{\includegraphics[angle=0,scale=0.65]{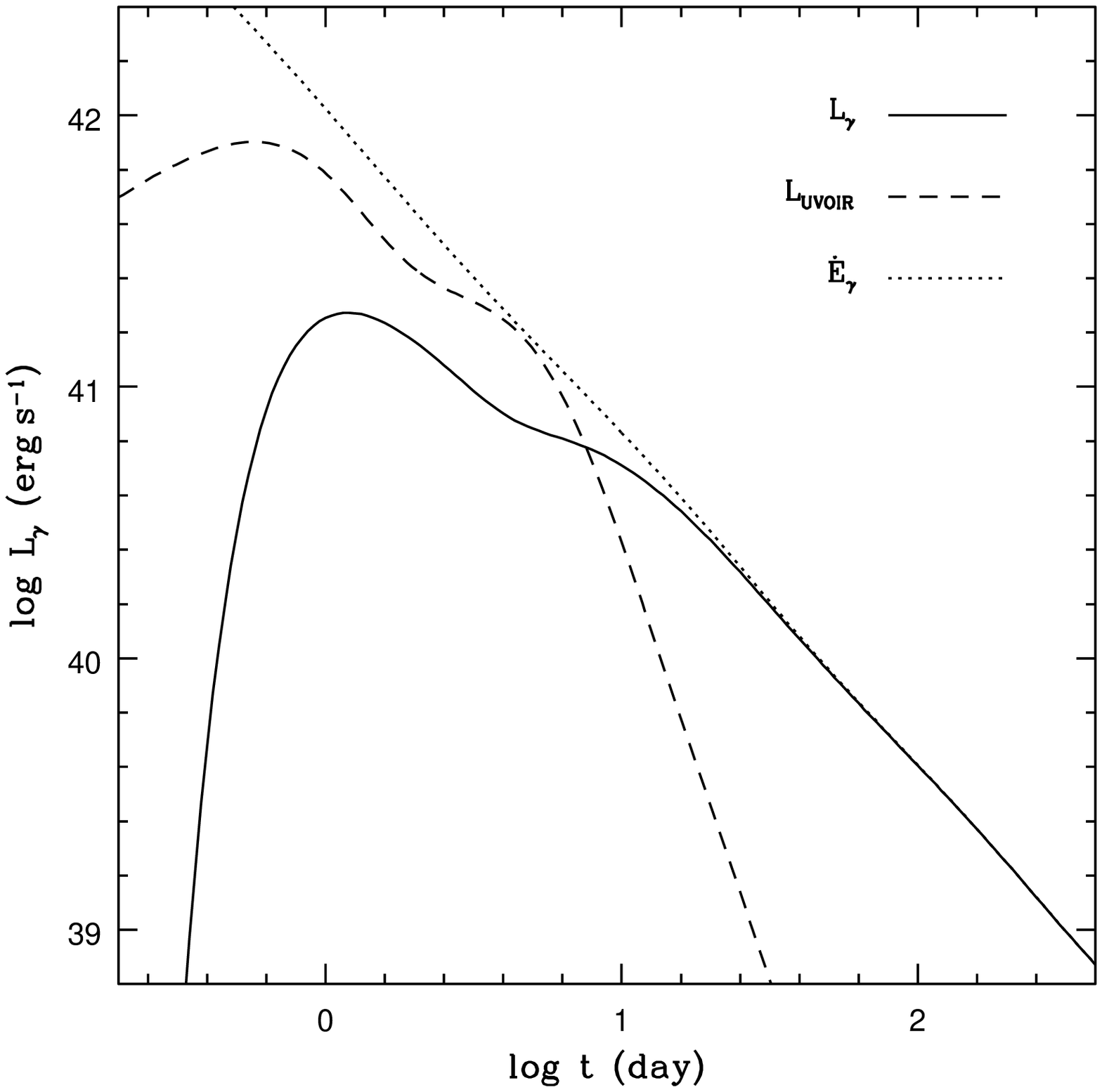}}
\caption{The gamma-ray luminosity as a function of time for the two-component model used to fit the UVOIR bolometric light curve of SSS17a/AT2017gfo. For comparison, the UVOIR bolometric light curve is shown with a dashed curve (i.e., the solid curve in Figure~\ref{lp_fit}). The dotted curve is the total gamma-ray energy generation rate in the ejecta.
}
\label{power2}
\end{figure}

In Figure~\ref{power2} we show the luminosity calculated for the gamma-ray emission in the two-component model used to fit the UVOIR bolometric light curve of SSS17a/AT2017gfo. For comparison, the UVOIR bolometric light curve and the total gamma-ray energy generation rate  are also shown. As we claimed in Section~\ref{model}, $\beta^-$-decay electrons contribute about $14\%$ to the heating rate through the thermalization process, so the gamma-ray energy generation rate $\dot{E}_\gamma$ is related to the heating rate $\epsilon$ by $\dot{E}_\gamma=0.86\epsilon M_\ej=0.86\dot{E}$, where the values of $\dot{E}$ at $t=1\,\oday$ are given in Table~\ref{parameter}. The UVOIR luminosity includes the contribution of $\beta^-$-decay electrons, but the gamma-ray luminosity does not since $\beta^-$-decay electrons do not contribute to the gamma-ray emission. In calculation of the gamma-ray luminosity this correction has been included. The gamma-ray light curve peaks at $t\approx 1.2\,\oday$ after merger, with a peak luminosity $\approx 1.9\times 10^{41}\,\erg\,\s^{-1}$. The UVOIR light curve peaks at $t\approx 0.58\,\oday$, with a peak luminosity $\approx 8.0\times 10^{41}\,\erg\,\s^{-1}$.

From $L_\gamma=e^{-t_c^2/t^2}\dot{E}_\gamma\propto e^{-t_c^2/t^2}t^{-1-\alpha}$, we can derive the time at the peak of $L_\gamma$
\begin{eqnarray}
  t_{\rm p}=t_c\left(\frac{2}{1+\alpha}\right)^{1/2} \;.  \label{tp_gam}
\end{eqnarray}
The peak gamma-ray luminosity, $L_{\gamma,{\rm p}}$, is related to the gamma-ray energy generation rate at $t=1\,\oday$, $\dot{E}_{\gamma,1}$, by
\begin{eqnarray}
  L_{\gamma,{\rm p}} = \dot{E}_{\gamma,1}e^{-(1+\alpha)/2}\left(\frac{t_{\rm p}}{1\,\oday}\right)^{-1-\alpha} = \dot{E}_{\gamma,1}\left(\frac{2e}{1+\alpha}\right)^{-(1+\alpha)/2}\left(\frac{t_c}{1\,\oday}\right)^{-1-\alpha} \;. \label{Lp_gam}
\end{eqnarray}
For $\alpha=0.2$, $t_c=0.944\,\oday$, and $\dot{E}_{\gamma,1}=0.86\dot{E}_1=4.38\times 10^{41}\,\erg\,\s^{-1}$ (the parameters for component A, see Table~\ref{parameter}), we get $t_{\rm p}=1.29t_c=1.22\,\oday$ and $L_{\gamma,{\rm p}}=1.9\,\times 10^{41}\erg\,\s^{-1}$, consistent with the numerical result.

The integration of equation~(\ref{dEgdt_obs}) over $t$ from $t=0$ to $t=\infty$ leads to the total observed energy contained in the gamma-ray emission
\begin{eqnarray}
  E_{\gamma,\obs} = \sum_i\frac{3N_{i,0}\varepsilon_{\gamma,i}}{2\beta^3}\left[\beta-\frac{1}{2}\left(1-\beta^2\right)\ln\frac{1+\beta}{1-\beta}\right]= \sum_iN_{i,0}\varepsilon_{\gamma,i}\left[1+\frac{1}{5}\beta^2+{\cal O}\left(\beta^4\right)\right] \;,
\end{eqnarray}
where $\varepsilon_{\gamma,i}$ is defined by equation (\ref{vareps_i}). Therefore, the expansion of the ejecta affects the energy generation rate only on the order of $\beta^2$.

\section{Spectrum of the Gamma-Ray Emission}
\label{spectra}

For a single nuclide species in a given decay mode, the photon number rate in a range of photon energy from $\varepsilon=\varepsilon_1$ to $\varepsilon=\varepsilon_2>\varepsilon_1$ defined in the observer frame is calculated by equation~(\ref{dcNidt_obs3}). After summation over all nuclide species, decay modes, and gamma-ray emission lines, we get the total observed photon number rate in a bin of photon energy defined by $(\varepsilon_1,\,\varepsilon_2)$
\begin{eqnarray}
  \Delta\dot{\cal N} = \sum_i\frac{N_{i,0}}{\tau_i}\sum_jB_{i,j}\sum_kh_{i,jk}{\cal I}_2\left(\alpha^\prime_i,\beta,y_1,y_2\right) \;, 
  \label{dcNidt_obs6} 
\end{eqnarray}
where
\begin{eqnarray}
  {\cal I}_2\left(\alpha_i,\beta,y_1,y_2\right)=
  \left\{\begin{array}{ll}
  I_2(y_1,y_2) \;, & \quad y_-<y_1<y_2<y_+ \;, \\
  I_2(y_-,y_2) \;, & \quad y_1<y_-<y_2<y_+ \;, \\
  I_2(y_1,y_+) \;, & \quad y_-<y_1<y_+<y_2 \;, \\
  I_2(y_-,y_+) \;, & \quad y_1<y_-<y_+<y_2 \;, \\
  0 \;, & \quad \mbox{else} \;,
  \end{array}\right.
  \label{cal_I_eval2} 
\end{eqnarray}
$y_1=\varepsilon_1/\varepsilon_{i,jk}$, $y_2=\varepsilon_2/\varepsilon_{i,jk}$, $I_2(a,b)\equiv I_2(b)-I_2(a)$, and $I_2(y)$ is defined by equation~(\ref{I_int}).

We choose to calculate the photon number rate in a photon energy bin defined by $(\varepsilon_1,\varepsilon_2)$ instead of the specific photon number rate $\dot{\cal N}_\varepsilon$ at any photon energy $\varepsilon$ (eq.~\ref{dcNeidt_obs}) because of the following considerations. First, since we are calculating the observed spectrum arising from many individual emission lines, when the emission lines are very narrow and sharp, some lines can easily be missed as we sample the photon energy numerically if we choose to calculate $\dot{\cal N}_\varepsilon$ at a given photon energy. This problem can be avoided if we choose to calculate the $\Delta\dot{\cal N}$ for an interval of photon energy. Second, if the size of the photon energy bin, $\Delta\varepsilon$, is sufficiently small, after we get the $\Delta\dot{\cal N}$ for each energy bin we can easily derive the specific photon number rate $\dot{\cal N}_\varepsilon$ and the specific photon energy rate $L_\varepsilon$ through the following relations
\begin{eqnarray}
  \dot{\cal N}_\varepsilon=\frac{\Delta\dot{\cal N}}{\Delta\varepsilon} \;, \hspace{1cm}
  L_\varepsilon=\varepsilon\dot{\cal N}_\varepsilon=\varepsilon\frac{\Delta\dot{\cal N}}{\Delta\varepsilon} \;. \label{dotNe}
\end{eqnarray}

In the data sample, the minimum of the photon energy is $0.34\,\keV$, and the maximum is $5\,\MeV$. The range of photon energy spans about four orders of magnitude. Hence, for calculation of the spectrum of the gamma-ray emission, we choose to divide the $\log\varepsilon$ uniformly, where the photon energy $\varepsilon$ is in keV. Considering the Doppler effect caused by the expansion of the spherical ejecta, in our calculation we take $\log\varepsilon_{\min}=-0.6$ and $\log\varepsilon_{\max}=3.9$ and uniformly divide the total range of $\log\varepsilon$ into 600 bins. Each bin of $\log\varepsilon$ has then a size of $\delta=0.0075$, corresponding to a nonuniform division of the photon energy with $\Delta\varepsilon=(\ln 10)\varepsilon\delta=0.01727\varepsilon$. In each bin of the photon energy, the observed photon number rate is calculated by equations~(\ref{dcNidt_obs6}) and (\ref{cal_I_eval2}). Then, by equation (\ref{dotNe}) we get $\dot{\cal N}_\varepsilon=(\ln 10)^{-1}\Delta\dot{\cal N}/(\varepsilon\delta)$, and $\varepsilon\dot{\cal N}_\varepsilon=L_\varepsilon=(\ln 10)^{-1}\Delta\dot{\cal N}/\delta$.

\begin{figure}[ht!]
\centering{\includegraphics[angle=0,scale=0.65]{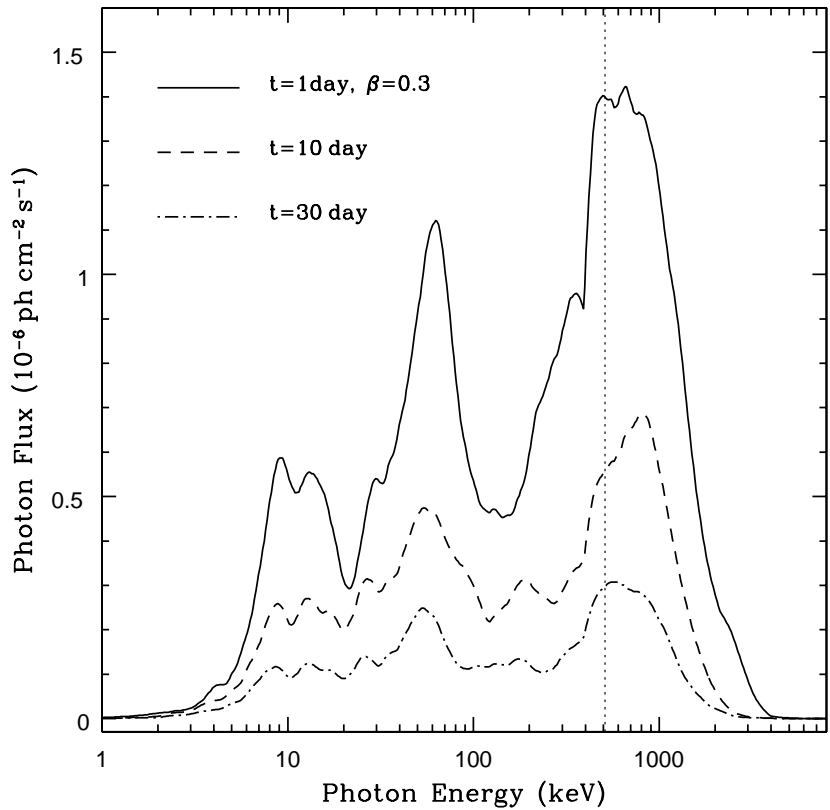}}
\caption{The intrinsic photon flux spectrum of the gamma-ray emission produced by radioactive decays in a merger at a distance $D=40\,\Mpc$ and time $t=1\,\oday$, $10\,\oday$, and $30\,\oday$. The energy generation is normalized so that $\dot{E}_\gamma=6\times 10^{41}\,\erg\,\s^{-1}$ at $t=1\,\oday$. The merger ejecta is assumed to have an expansion velocity $V=0.3c$. For better visibility, the photon fluxes at $t=10\,\oday$ and $t=30\,\oday$ have been multiplied by a factor of 6 and 10, respectively. The vertical dotted line denotes the annihilation line energy of electrons and positrons, which is $511\,\keV$.
}
\label{spec1}
\end{figure}

\begin{figure}[ht!]
\centering{\includegraphics[angle=0,scale=0.65]{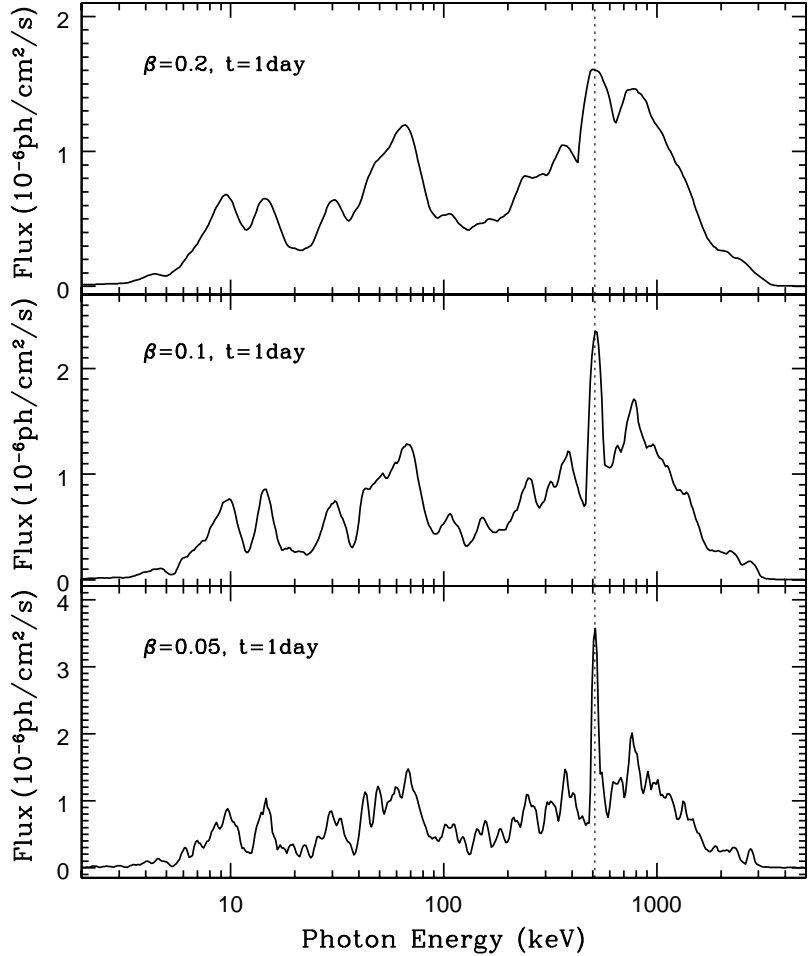}}
\caption{Similar to Figure~\ref{spec1} but for $V=0.2c$ (top panel), $V=0.1c$ (middle panel), and $V=0.05c$ (bottom panel) at $t=1\,\oday$.
}
\label{spec2}
\end{figure}

The quantity $\varepsilon\dot{\cal N}_\varepsilon$ describes the photon number rate in a photon energy band. It is easier to display the pattern in the shape of the spectrum with $\varepsilon\dot{\cal N}_\varepsilon$ than with $\dot{\cal N}_\varepsilon$. In Figures~\ref{spec1} and \ref{spec2} we show the photon flux calculated from $\varepsilon\dot{\cal N}_\varepsilon$ for a number of models. The photon flux is defined by
\begin{eqnarray}
  \mbox{Photon Flux}=\frac{\varepsilon\dot{\cal N}_\varepsilon}{4\pi D^2} \;, \label{flux}
\end{eqnarray}
where $D$ is the distance from the merger to the observer. In the calculation we take $D=40\,\Mpc$, the distance to the host galaxy of GW170817. Like in Figure~\ref{power}, we have normalized the gamma-ray energy generation so that $\dot{E}_\gamma=6\times 10^{41}\,\erg\,\s^{-1}$ at $t=1\,\oday$.

In Figure~\ref{spec1}, we show the cases of $\beta=0.3$ at $t=1\,\oday$, $10\,\oday$, and $30\,\oday$, respectively. In Figure~\ref{spec2}, we show the cases of $\beta=0.2$, $0.1$, and $0.05$, at time $t=1\,\oday$ since the merger. From these figures of spectra, we see that the emitted photons are roughly clustered in three groups in terms of photon energy: a group with the strongest emissions in $150$--$3,000\,\keV$, a group with intermediate strong emissions in $20$-$150\,\keV$, and a group with the weakest emissions in $3$--$20\,\keV$. We also see that in the case of $\beta\la 0.2$ strong annihilation lines of electrons and positrons (with $\varepsilon=511\,\keV$) are present in the spectrum. In the case of $\beta=0.3$, the strong line broadening effect arising from the expansion of the ejecta causes the pair annihilation lines smeared (but still visible).

Figure~\ref{spec1} also shows some subtle difference in the spectrum patterns at different times, which is caused by the fact that at different times the dominant radiation comes from different groups of radioactive nuclides (see Section~\ref{model}). However, there is no obvious evolution in the photon energy like that seen in the UVOIR spectra, if $\beta$ remains constant during the expansion. The subtle difference between the spectra at different times can tell us important information about the element composition and radioactive process inside the ejecta. For instance, from Figure~\ref{spec1} we see that the pair annihilation line at $t=10\,\oday$ and $30\,\oday$ is weaker than that at $t=1\,\oday$, which indicates that the number of $\beta^+$-decay nuclides with mean lifetime $\sim 10\,\oday$ and $\sim 30\,\oday$ is smaller than that with mean lifetime $\sim 1\,\oday$. This inference is verified by checking the data of the nuclides in the sample.

The flux defined by equation (\ref{flux}) and the spectra shown in Figures~\ref{spec1} and \ref{spec2} are ``intrinsic'' or ``naked'' quantities (i.e., not the observable quantities), since the effect of optical depth has not been included yet. If the opacity in the ejecta is a constant as we  have assumed so far, the optical depth does not depend on the photon energy and is a function of time only. In this simple case, the observed flux is simply equal to the intrinsic flux multiplied by a factor $e^{-t_c^2/t^2}$ according to equation (\ref{dQ_nth}) and hence the observed spectrum has the same shape as the intrinsic spectrum. In reality, the opacity and hence the optical depth can be a function of the photon energy. For a merger ejecta composed of heavy elements, for photons $\la 300\,\keV$ the opacity is dominated by the contribution from the photoelectric absorption and increases quickly with decreasing photon energy. As a result, the low energy part of the spectra shown in Figures~\ref{spec1} and \ref{spec2} with the photon energy $\la 300\,\keV$ will be absorbed by the ejecta and hence will not be visible in the observed spectra, unless at the very late time when the ejecta becomes optically transparent to low energy photons also. This effect will be discussed in detail in Section~\ref{detect} when we investigate the observability of the gamma-ray emission from a neutron star merger.

\begin{figure}[ht!]
\centering{\includegraphics[angle=0,scale=0.7]{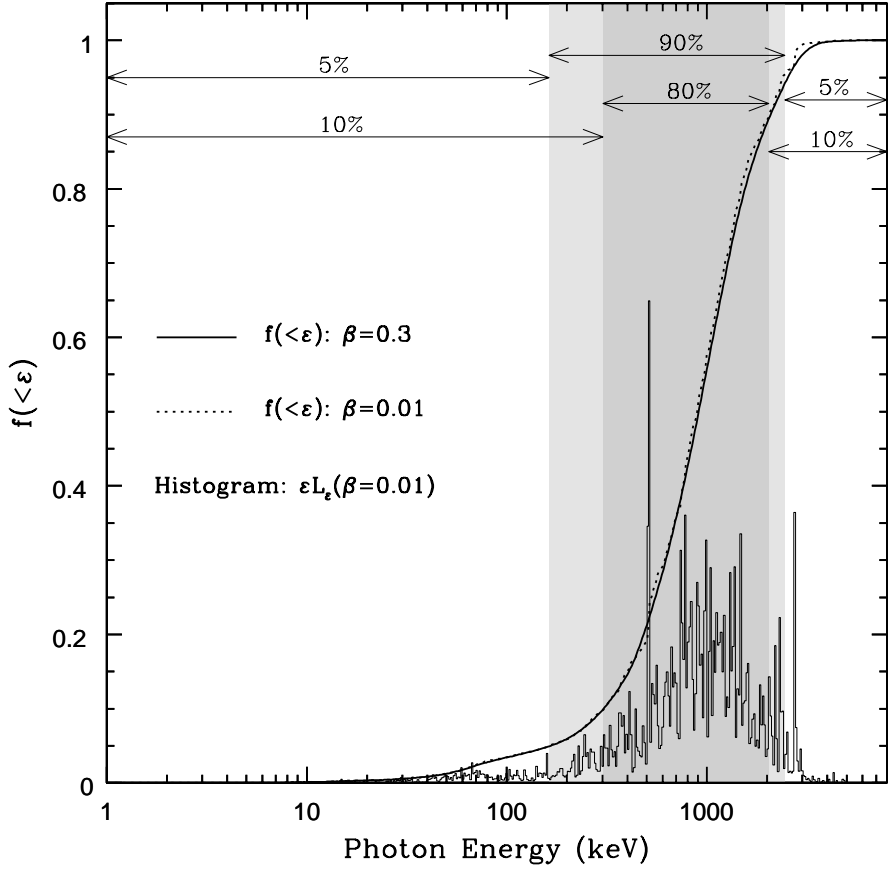}}
\caption{Fraction of the gamma-ray energy rate contained in the photon energy range $(0,\varepsilon)$ in the total gamma-ray energy rate, $f(<\varepsilon)$ (eq.~\ref{fe}), versus the photon energy at $t=1\,\oday$. Two models with different expansion velocity are shown: $V=0.3c$ (solid curve) and $0.01c$ (dotted curve). The light shaded region covers the energy range containing $90\%$ of the total integrated flux, with $5\%$ on the left and $5\%$ on the right. The dark shaded region covers the energy range containing $80\%$ of the total integrated flux, with $10\%$ on the left and $10\%$ on the right. The histogram shows the $\varepsilon L_\varepsilon$ (in arbitrary units) versus $\varepsilon$ for the case of $V=0.01c$, which roughly represents the energy rate in an energy band around a given photon energy $\varepsilon$.
}
\label{specInt}
\end{figure}

In Figure~\ref{specInt} we show the fraction of the gamma-ray energy rate defined in the photon energy range $(0,\varepsilon)$ in the total gamma-ray energy rate, i.e.,
\begin{eqnarray}
  f(<\varepsilon)=\frac{\int_0^\varepsilon L_\varepsilon d\varepsilon}{\int_0^\infty L_\varepsilon d\varepsilon} \;, \label{fe}
\end{eqnarray}
for the cases of $\beta=0.01$ and $0.3$ at $t=1\,\oday$. The choice of $\beta=0.01$ is for showing the case with the minimum line broadening effect. The figure shows that the photon energy of the radioactive emission is distributed in a relative narrow range. About $90\%$ of the total emitted gamma-ray energy is carried by photons with energy in the range of $160$--$2,500\,\keV$ (with $5\%$ energy by photons with $\varepsilon<160\,\keV$, and the remaining $5\%$ by photons with $\varepsilon>2,500\,\keV$). About $80\%$ of the total emitted gamma-ray energy is carried by photons with energy in the range of $300$--$2,000\,\keV$ (with $10\%$ energy by photons with $\varepsilon<300\,\keV$, and the remaining $10\%$ by photons with $\varepsilon>2,000\,\keV$). The energy of annihilation lines at $511\,\keV$ contributes about $3$--$5\%$ to the total gamma-ray energy flux.

Therefore, $90\%$ of the gamma-ray energy emitted by radioactive nuclides is carried by photons of energy $>300\,\keV$, only $10\%$ is carried by photons of energy $<300\,\keV$. Although the photoelectric absorption has a significant effect on the low energy part of the observed photon flux spectrum, its influence on the calculation of the observed gamma-ray luminosity is minor.

\begin{figure}[ht!]
\centering{\includegraphics[angle=0,scale=0.65]{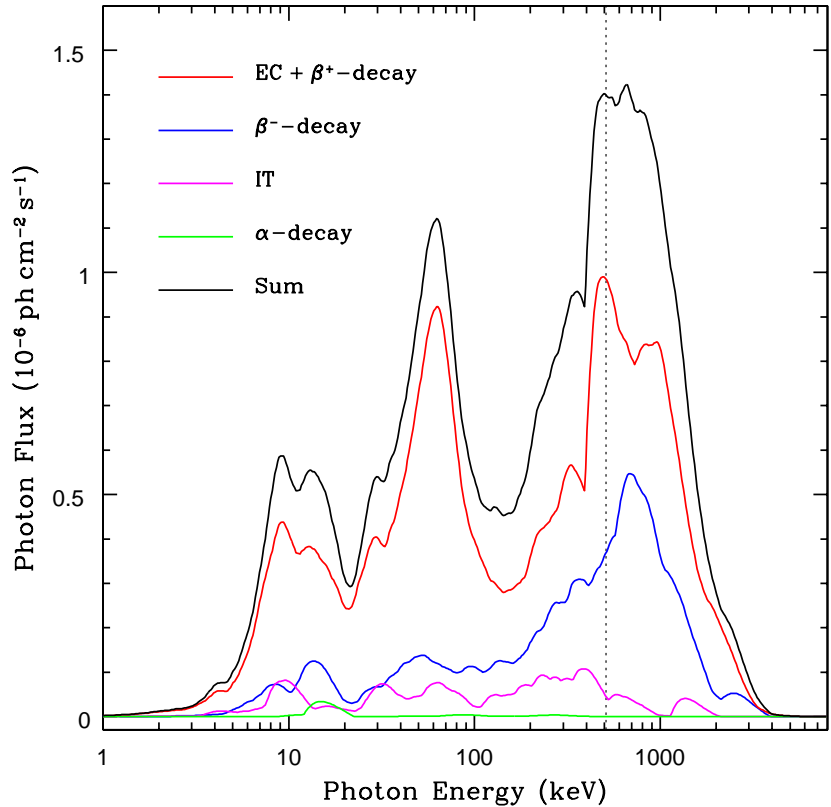}}
\caption{Photon flux of the gamma-ray emission produced by radioactive nuclides in each decay mode, versus the photon energy. The model is the same as that in Figure~\ref{spec1}, shown for the case at $t=1\,\oday$. EC=electron capture, IT=isomeric transition, Sum=the sum of fluxes in all decay modes. The electron capture and the $\beta^+$-decay are shown together since they often occur competitively for the same nucleus. The vertical dotted line denotes the energy of pair annihilation lines, $511\,\keV$.
}
\label{spec_mode}
\end{figure}

To see the contribution of the five decay modes ($\beta^-$-decay, $\beta^+$-decay, electron capture, isomeric transition, and $\alpha$-decay) to the gamma-ray emission, in Figure~\ref{spec_mode} we plot separately the photon flux versus the photon energy for photons associated with each decay mode, for the same model in Figure~\ref{spec1} (at $t=1\,\oday$). We note that, if the energy difference between parent and daughter atoms is larger than $1.022\,\MeV$, positron emission is allowed then the $\beta^+$-decay can compete and accompany the electron capture, and {\it vice versa}. In our data sample, about half of the electron captures are accompanied by $\beta^+$-decays and almost all the $\beta^+$-decays are accompanied by electron captures, for which the contribution of $\beta^+$-decays and electron captures to the photon flux is hard to distinguish. Hence, in Figure~\ref{spec_mode}, photon fluxes generated by the $\beta^+$-decay and the electron capture are shown with one curve (the red curve), which represents the sum of their contributions. We see that, $\beta^+$-decays and electron captures (ECs) make the biggest contribution to the gamma-ray emission. In terms of the gamma-ray energy power obtained by integration of the energy flux over the photon energy, $\beta^+$-decays and ECs contribute $66.17\%$ to the total. The next dominant contribution comes from $\beta^-$-decays, which contribute $30.09\%$ to the total power. Next, isomeric transitions (ITs) contribute $3.69\%$, and $\alpha$-decays contribute the least: only $0.05\%$.

From Figure \ref{spec_mode} we also see that the spectra of photons generated in different decay modes have very different features. The spectrum generated by the EC and $\beta^+$-decay has a remarkable triple-finger shape, i.e., it has three distinct peaks around $600\,\keV$, $60\,\keV$, and $9\,\keV$, respectively. During an EC process, besides the gamma-rays generated when the daughter nucleus is in an excited state, characteristic X-rays can also be produced when an outer electron fills an inner hole of the atom left by the capture of a K or L electron. A major feature of the photon emission produced by $\beta^+$-decays is the presence of electron-positron annihilation lines of $511\,\keV$, which contribute $3$-$5\%$ to the total energy flux. The spectrum generated by the $\beta^-$ decay has a single prominent peak around $700\,\keV$, and two small bumps around $50\,\keV$ and $14\,\keV$. The IT contributes a spectrum that is relatively flat from $8\,\keV$ to $700\,\keV$. Emissions produced by the $\alpha$-decay are dominantly around $15\,\keV$, which makes a negligible contribution to the total gamma-ray energy generation. Since in our model $\beta^+$-decays and ECs make the biggest contribution to the total spectrum, the shape of the total spectrum is closer to that of the EC+$\beta^+$-decay spectrum. This may not be the case in other models. Hence, observation of the gamma-ray spectrum can in principle provide important information about the contribution of each decay mode to the energy generation, put constraints on the nucleosynthesis process in the merger ejecta, and test different theoretical models.

For each type of radioactive decay, different nuclides emit gamma-ray photons with spectra broadly similar in shapes. However, the hardness ratio of photon flux---i.e., the ratio of the flux of high energy photons to that of low energy photons---varies from nuclide to nuclide. Hence, we expect that at a given time, each peak or bump in the emission spectrum is produced by many unstable nuclides with similar mean lifetime, with no sharp change in the fraction of the contribution of each nuclide in the total flux around that peak. To identify the unstable nuclides that make the dominant contribution to the photon flux around a peak in the gamma-ray spectrum, we have calculated the contribution of each nuclide in the sample to the photon flux at a given time in a given decay channel, then sorted the fluxes of all nuclides in a descending order. The top six nuclides responsible for a spectral peak are listed in Table~\ref{ele_bpec} for the case of $\beta^+$-decay/electron capture (red curve in Figure~\ref{spec_mode}), and in Table~\ref{ele_bm} for the case of $\beta^-$-decay (blue curve in Figure~\ref{spec_mode}), at time $t=1\,\oday$ and $10\,\oday$, respectively. Fractions of the contribution of the nuclides in the flux around a spectral peak are also listed. For instance, for the peak near $600\,\keV$ in the spectrum of $\beta^+$-decay/electron capture shown in Figure~\ref{spec_mode} there are in total 34 unstable nuclides with their contribution $>1\%$ in the total flux in the range of $390$--$1,340\,\keV$, but only six nuclides with contribution $>2.5\%$ are listed. While for the peak near $700\,\keV$ in the spectrum of $\beta^-$-decay in Figure~\ref{spec_mode}, there are in total 28 unstable nuclides with their contribution $>1\%$ in the total flux in the range of $410$--$1,200\,\keV$, but only six nuclides with contribution $>4.2\%$ are listed. 

\begin{table}[ht!]
\centering
\begin{minipage}{173mm}
\caption{Nuclides with dominant contribution to the gamma-ray spectrum of $\beta^+$-decay/electron capture
  \label{ele_bpec}}
\begin{tabular}{lll}
\hline\hline
Peak Region$^{\rm a}$ & Time$^{\rm b}$ & Nuclides with dominant flux contribution$^{\rm c}$ \\
\hline
$390$--$1,340\,\keV$, & $t=1\,\oday$:..........~~ & $\prescript{82}{}{\rm Rb} (4.6\%)$, $\prescript{86}{}{\rm Y} (4.2\%)$, $\prescript{204}{}{\rm Bi} (3.9\%)$, $\prescript{76}{}{\rm Br} (2.9\%)$, $\prescript{55}{}{\rm Co} (2.9\%)$, $\prescript{90}{}{\rm Nb} (2.7\%)$ \\
& $t=10\,\oday$:........ & $\prescript{106}{}{\rm Ag} (8.8\%)$, $\prescript{206}{}{\rm Bi} (8.4\%)$, $\prescript{48}{}{\rm V} (5.9\%)$, $\prescript{52}{}{\rm Mn} (5.2\%)$, $\prescript{96}{}{\rm Tc} (4.8\%)$, $\prescript{146}{}{\rm Eu} (4.7\%)$ \\
\hline
$36$--$96\,\keV$, & $t=1\,\oday$:.......... & $\prescript{165}{}{\rm Tm} (3.3\%)$, $\prescript{183}{}{\rm Os} (3.1\%)$, $\prescript{182}{}{\rm Re} (2.8\%)$, $\prescript{175}{}{\rm Ta} (2.7\%)$, $\prescript{166}{}{\rm Tm} (2.7\%)$, $\prescript{189}{}{\rm Pt} (2.6\%)$  \\
& $t=10\,\oday$:........ & $\prescript{169}{}{\rm Yb} (7.0\%)$, $\prescript{171}{}{\rm Lu} (6.3\%)$, $\prescript{172}{}{\rm Lu} (5.4\%)$, $\prescript{206}{}{\rm Bi} (4.6\%)$, $\prescript{155}{}{\rm Tb} (4.2\%)$, $\prescript{156}{}{\rm Tb} (3.4\%)$ \\
\hline
$7$--$20\,\keV$, & $t=1\,\oday$:.......... & $\prescript{239}{}{\rm Am} (3.3\%)$, $\prescript{86}{}{\rm Zr} (3.2\%)$, $\prescript{246}{}{\rm Bk} (3.2\%)$, $\prescript{240}{}{\rm Am} (3.1\%)$, $\prescript{204}{}{\rm Bi} (2.5\%)$, $\prescript{76}{}{\rm Kr} (2.0\%)$ \\
& $t=10\,\oday$:........ & $\prescript{231}{}{\rm U} (4.8\%)$, $\prescript{245}{}{\rm Bk} (4.0\%)$, $\prescript{241}{}{\rm Cm} (3.8\%)$, $\prescript{171}{}{\rm Lu} (3.4\%)$, $\prescript{82}{}{\rm Sr} (3.4\%)$, $\prescript{206}{}{\rm Bi} (3.4\%)$ \\
\hline
\end{tabular}\\[1mm]
$^{\rm a}$The (somewhat arbitrarily chosen) range of photon energy enclosing the spectral peaks near $600\,\keV$, $60\,\keV$, and $9\,\keV$ shown in the red curve in Figure~\ref{spec_mode}.\\
$^{\rm b}$The time since the merger of neutron stars.\\
$^{\rm c}$The number in parenthesis after a nuclide is the fraction of the photon flux generated by that nuclide in the total photon flux defined in the given photon energy range.
\end{minipage}
\end{table}

\begin{table}[ht!]
\centering
\begin{minipage}{175mm}
\caption{Nuclides with dominant contribution to the gamma-ray spectrum of $\beta^-$-decay
  \label{ele_bm}}
\begin{tabular}{lll}
\hline\hline
Peak Region$^{\rm a}$ & Time$^{\rm b}$ & Nuclides with dominant flux contribution$^{\rm c}$ \\
\hline
$410$--$1,200\,\keV$, & $t=1\,\oday$:..........~~ & $\prescript{130}{}{\rm I} (10.1\%)$, $\prescript{128}{}{\rm Sb} (9.6\%)$, $\prescript{48}{}{\rm Sc} (6.6\%)$, $\prescript{96}{}{\rm Nb} (6.4\%)$, $\prescript{72}{}{\rm Ga} (4.5\%)$, $\prescript{82}{}{\rm Br} (4.3\%)$ \\
& $t=10\,\oday$:........ & $\prescript{126}{}{\rm Sb} (23.2\%)$, $\prescript{136}{}{\rm Cs} (9.8\%)$, $\prescript{148}{}{\rm Pm} (8.0\%)$, $\prescript{127}{}{\rm Sb} (5.3\%)$, $\prescript{48}{}{\rm Sc} (4.7\%)$, $\prescript{95}{}{\rm Nb} (3.4\%)$ \\
\hline
$36$--$80\,\keV$, & $t=1\,\oday$:.......... & $\prescript{157}{}{\rm Eu} (10.7\%)$, $\prescript{153}{}{\rm Sm} (8.1\%)$, $\prescript{171}{}{\rm Er} (7.9\%)$, $\prescript{183}{}{\rm Ta} (5.6\%)$, $\prescript{172}{}{\rm Er} (5.3\%)$, $\prescript{151}{}{\rm Pm} (4.4\%)$  \\
& $t=10\,\oday$:........ & $\prescript{183}{}{\rm Ta} (16.3\%)$, $\prescript{161}{}{\rm Tb} (11.0\%)$, $\prescript{191}{}{\rm Os} (10.0\%)$, $\prescript{237}{}{\rm U} (8.4\%)$, $\prescript{166}{}{\rm Dy} (7.0\%)$, $\prescript{147}{}{\rm Nd} (5.5\%)$ \\
\hline
$10$--$19\,\keV$, & $t=1\,\oday$:.......... & $\prescript{244}{}{\rm Am} (15.6\%)$, $\prescript{232}{}{\rm Pa} (14.6\%)$, $\prescript{231}{}{\rm Th} (9.8\%)$, $\prescript{240}{}{\rm U} (6.5\%)$, $\prescript{72}{}{\rm Zn} (5.8\%)$, $\prescript{237}{}{\rm U} (5.6\%)$ \\
& $t=10\,\oday$:........ & $\prescript{237}{}{\rm U} (33.1\%)$, $\prescript{233}{}{\rm Pa} (19.9\%)$, $\prescript{246}{}{\rm Pu} (13.9\%)$, $\prescript{191}{}{\rm Os} (7.0\%)$, $\prescript{239}{}{\rm Np} (5.0\%)$, $\prescript{225}{}{\rm Ra} (4.6\%)$ \\
\hline
\end{tabular}\\[1mm]
$^{\rm a}$The (somewhat arbitrarily chosen) range of photon energy enclosing the spectral peaks near $700\,\keV$, $50\,\keV$, and $14\,\keV$ shown in the blue curve in Figure~\ref{spec_mode}.\\
$^{\rm b}$The time since the merger of neutron stars.\\
$^{\rm c}$The number in parenthesis after a nuclide is the fraction of the photon flux generated by that nuclide in the total photon flux defined in the given photon energy range.
\end{minipage}
\end{table}

As we have derived in Section~\ref{model}, at any time $t$ the dominant contribution to the radioactive gamma-ray emission comes from unstable nuclides with their mean lifetime comparable to $t$, i.e., with $\tau$ in the range of $\sim 0.3t$--$4.5t$. Therefore, we expect that the member of nuclides that make the dominant contribution to the photon flux around a peak in the gamma-ray spectrum evolves with time. This point is confirmed by the data in Tables~\ref{ele_bpec} and \ref{ele_bm}. For a given spectral peak, the top six nuclides making the dominant contribution to the photon flux clearly differ at different time. From the data we can also see the following interesting effect: for a given spectral peak, the fraction of the photon flux generated by a dominant unstable nuclide increases with time. For instance, for the peak around $600\,\keV$ in the spectrum of $\beta^+$-decay/electron capture, at $t=1\,\oday$ the top six nuclides contribute in total $21\%$ of the photon flux. At $t=10\,\oday$, the top six nuclides contribute in total $38\%$ of the photon flux. For the peak around $700\,\keV$ in the spectrum of $\beta^-$-decay, at $t=1\,\oday$ the top six nuclides contribute in total $42\%$ of the photon flux. At $t=10\,\oday$, the top six nuclides contribute in total $54\%$ of the photon flux. This effect arises from the fact that the number of nuclide species decreases with increasing mean lifetime.

\section{On the Effect of Decay Chains}
\label{dec_chains}

So far, in our calculation of the radioactive decays of nuclides we have assumed that all the nuclides in the sample undergo one-step decays, i.e., parent nuclides directly decay to stable daughter nuclides. This is true for most of the nuclides in the sample. If we treat all nuclides with half-life greater than $50,000\,\oday$ as stable since they make a negligible contribution to the radiation power, we find that among the $614$ daughter nuclides produced by the $614$ decay modes in the sample, $383$ of them are stable, and the remaining $231$ are unstable. So, $231$ of the daughter nuclides will continue to decay, until at some step stable nuclides are produced. In this section we discuss the effect of these decay chains on the generation of gamma-ray energy in the merger ejecta.

Of the 231 decay chains, 17 of them bifurcate at some intermediate decay stage. For instance, $\prescript{212}{}{\rm Pb}$ ($\mbox{\jpi}=0+$) decays to $\prescript{212}{}{\rm Bi}$ ($1-$) through the $\beta^-$-decay with $t_{1/2}=10.64\,\hr$. The $\prescript{212}{}{\rm Bi}$ ($1-$) is unstable and has $t_{1/2}=1.01\,\hr$. Then, the $\prescript{212}{}{\rm Bi}$ ($1-$) decays to $\prescript{212}{}{\rm Po}$ ($0+$) through the $\beta^-$-decay with a branching ratio $64.06\%$, and to $\prescript{208}{}{\rm Tl}$ ($5+$) through the $\alpha$-decay with a branching ratio $35.94\%$. Both the $\prescript{212}{}{\rm Po}$ ($0+$) and $\prescript{208}{}{\rm Tl}$ ($5+$) are unstable, with $t_{1/2}=0.299\,\mbox{\mus}$ and $3.053\,{\rm min}$, respectively. The $\prescript{212}{}{\rm Po}$ ($0+$) decays to the stable $\prescript{208}{}{\rm Pb}$ ($0+$) through the $\alpha$-decay, and the $\prescript{208}{}{\rm Tl}$ ($5+$) decays to the stable $\prescript{208}{}{\rm Pb}$ ($0+$) through the $\beta^-$-decay. Hence, the decay chain of $\prescript{212}{}{\rm Pb}$ ($0+$) ends at the stable daughter nuclide $\prescript{208}{}{\rm Pb}$ ($0+$).

There are in total 404 decay modes for daughter decays contained in the 231 decay chains. The number distribution of the 614 parent decay modes, and of the 231 daughter decays, are listed in Table~\ref{chain1}. Note, here the $\beta^+$-decay and the electron capture are not strictly distinguished. As explained in the previous section, $\beta^+$-decays and electron captures always occur competitively for atoms with available energy larger than $1.022\,\MeV$. As a result, many radiation data for electron captures listed on the webpage of NuDat\,2 contain radiations from $\beta^+$-decays, and {\it vice versa}. The classifications listed in Table~\ref{chain1} are according to the classification given by NuDat\,2. From Table~\ref{chain1} we see an interesting fact that the daughter decays contain about three times more $\alpha$-decays than the parent decays (136 vs 40). Since $\alpha$-decays are not efficient in producing gamma-ray photons (Figure~\ref{spec_mode}), and only about one-third of the parent decays have chain decays, we expect that decay chains will only moderately affect the gamma-ray energy generation in the ejecta. The calculation of the gamma-ray energy generation presented below will confirm this inference.

\begin{table}[ht!]
\centering
\begin{minipage}{153mm}
\caption{Count of the number of decay modes
  \label{chain1}}
\begin{tabular}{llllllr}
\hline\hline
Decay Modes:~~~~~~~~~~~~~~~~~~ & EC~~~~~~~~~~~~~ & B$-$~~~~~~~~~~~~~ & B$+$~~~~~~~~~~~~ & IT~~~~~~~~~~~~~ & A~~~~~~~~~~~ & Total\\
\hline
Number A............ & $300$ & $199$ & $8$ & $67$ &  $40$ & $614$  \\
Number B............ & $144$ & $121$ & $2$ &  $1$ & $136$ & $404$  \\
\hline
\end{tabular}\\
\tablecomments{This table lists the number of decay modes associated with the $537$ radioactive nuclides (Number A), and the number of decay modes associated with their unstable daughters (Number B). EC = Electron Capture, B$-$ = $\beta^-$-decay, B$+$ = $\beta^+$-decay, IT = Isomeric Transition, A = $\alpha$-decay. The last column lists the total number of decay modes associated with the parent nuclides and their unstable daughters.}
\end{minipage}
\end{table}

In Table~\ref{chain2} we list the number distribution of the length of the decay chains contained in the sample. The length of a decay chain is defined as the sum of the decay steps contained in the chain. For instance, if $X_0$ decays to a stable $X_1$, the length of the decay chain is equal to one. If $X_0$ decays to $X_1$ then $X_1$ decays to a stable $X_2$, the length of the decay chain is equal to two, etc. When a bifurcation occurs on a decay path, the length counts the total decay steps on both bifurcation branches. However, the decay of the nuclide at the place where a bifurcation occurs is counted only once. For instance, for the decay chain of $\prescript{212}{}{\rm Pb}$ ($0+$) cited above, the length of the chain is equal to four for both branches. From Table~\ref{chain2} we see that, most of the decay chains in the sample have their length equal to one or two. The total number of decay chains with length equal to one and two is 559, which is $91\%$ of the total number 614. The total number of decay chains with length equal to one, two, and three is 588, which is $96\%$ of the total number. There are only $4\%$ of the decay chains that have length larger than three.

\begin{table}[ht!]
\centering
\begin{minipage}{156mm}
\caption{Number distribution of the length of the decay chains contained in the sample.}
\label{chain2}
\begin{tabular}{lrrrrrrrrrrrr}
\hline\hline
Length$^{\rm a}$............. & ~~~1~~~ & ~~~2~~~ & ~~~3~~~ & ~~~4~~~ & ~~~5~~~ & ~~~6~~~ & ~~~7~~~ & ~~~8~~~ & ~~~9~~~ & ~~~10~~~ & ~~~11~~~ & ~~~12 \\
\hline
Number$^{\rm b}$...........   & 383~~~ & 176~~~ & 29~~~ & 1~~~ & 4~~~ & 6~~~ & 3~~~ & 4~~~ & 2~~~ & 3~~~ & 3~~~ & ~~1 \\
\hline
\end{tabular}\\[1mm]
$^{\rm a}$The length of a decay chain, defined as the sum of decay steps contained in the chain.\\
$^{\rm b}$The number of decay chains with a given length. The total number of decay chains is 614.
\end{minipage}
\end{table}

\begin{figure}[ht!]
\centering{\includegraphics[angle=0,scale=0.65]{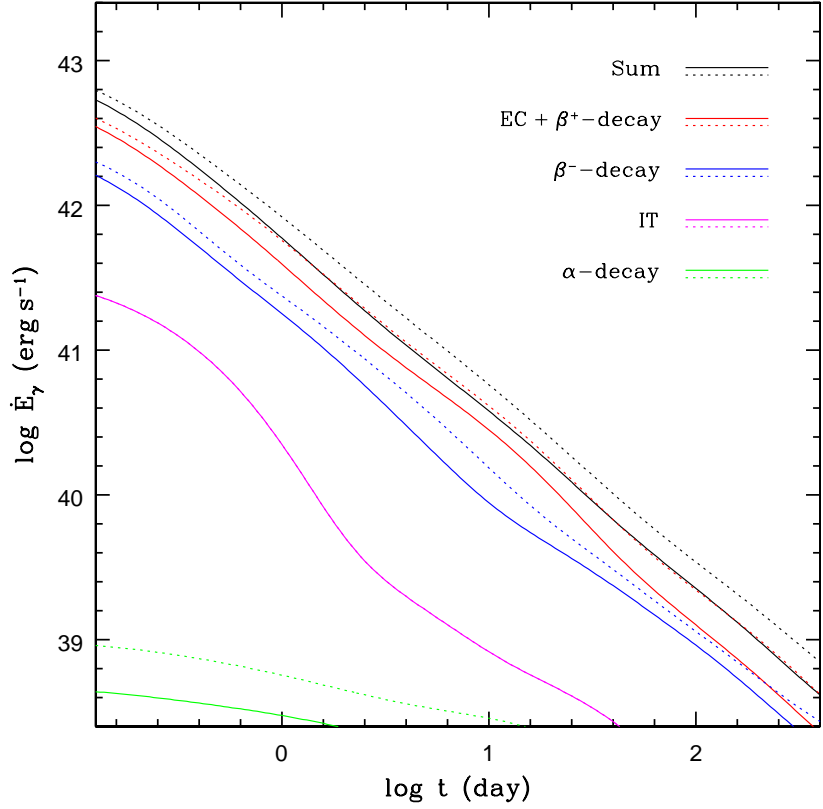}}
\caption{The gamma-ray energy generation rate in each decay mode produced by radioactive decays in a merger ejecta. The solid curves show the case when the effect of decay chains is ignored. The dotted curves show the case when the effect of decay chains is included. The total energy generation rate in the case without decay chains (solid black curve) is normalized in the same way as that in Figure~\ref{power}, i.e., $\dot{E}_\gamma=6\times 10^{41}\,\erg\,\s^{-1}$ at $t=1\,\oday$. The energy generation rates in the IT mode with and without decay chains are almost identical so the magenta solid and dotted curves visually appear indistinguishable. As in Figure~\ref{spec_mode} the electron capture and the $\beta^+$-decay are shown together with red curves.
}
\label{power_q2}
\end{figure}

The mathematical formalism for the calculation of decay chains is presented in Appendix~\ref{tdc}. With the formalism we have calculated the gamma-ray energy generation rate in the ejecta in each of the five decay modes (EC, $\beta^-$-decay, $\beta^+$-decay, IT, and $\alpha$-decay), with the effect of decay chains being included and not being included. The results are shown in Figure~\ref{power_q2}. The results without decay chains are shown with solid lines, while the results with decay chains are shown with dotted lines. We see that, for the total energy generation rate (black solid and dotted lines), the successive chain decays of parent nuclides can enhance the gamma-ray energy production by a factor of $\sim 1.5$. In agreement with the result in Figure~\ref{spec_mode}, the electron capture and the $\beta^\pm$-decay make the dominant contribution to the total gamma-ray energy generation. Effects of decay chains on the gamma-ray energy generation by the electron capture and the $\beta^\pm$-decay are similar, to enhance the corresponding energy generation by a factor of $\sim 1.5$. This agrees with the number count of decay modes in Table~\ref{chain1}: the EC and the $\beta^-$ decay modes contained in daughter decays are roughly $50\%$ of that contained in parent decays (144 vs 300, and 121 vs 199, respectively). In our model, decay chains have little effect on the generation of the gamma-ray energy by the isomeric transition, which is caused by the fact that daughter decays contain very few isomeric transitions relative to parent decays (1 vs 67). Decay chains have the largest effect on the generation of gamma-ray energy by the $\alpha$-decay, with an enhancement factor of $\sim 3$, in agreement with the number count for $\alpha$-decays in Table~\ref{chain1} (136 vs 40).

The integrated gamma-ray energy generated in each decay mode can be calculated with equation (\ref{delE_chain}). The results are shown in Table~\ref{chain3}, which confirm the previous conclusion that the dominant contribution to the gamma-ray energy generation comes from the electron capture and the $\beta^\pm$-decay, which in combination contribute about $97\%$ to the total gamma-ray energy generation. When the effect of decay chains is included, the integrated gamma-ray energy generation is increased by a factor of $1.42$, in agreement with the result in Figure~\ref{power_q2}. The result in Table~\ref{chain3} also confirms that the largest effect of decay chains on the gamma-ray energy generation is on that produced by the $\alpha$-decay, which is increased by a factor of three. However, since the $\alpha$-decay makes a negligible contribution to the gamma-ray energy generation, the large enhancement in the number of $\alpha$-decays has little influence on the total gamma-ray energy generation in the ejecta.

\begin{table}[ht!]
\centering
\begin{minipage}{169mm}
\caption{Integrated gamma-ray energy generation in each decay mode and the percentage in the total gamma-ray energy generation for models with and without decay chains. \label{chain3}}
\begin{tabular}{lccccc}
\hline\hline
Decay$^{\rm a}$~~~~~~~~~~~~~~ & ~~~~~~~~${\Delta E_\gamma}^{\rm b}$~~~~~~~~ & ~~~~~~~~${\xi_r(\%)}^{\rm c}$~~~~~~~~ & ~~~~~~~~${\Delta E^\prime_\gamma}^{\rm d}$~~~~~~~~ & ~~~~~~~~${\xi_r^\prime(\%)}^{\rm e}$~~~~~~~~ & ~~~~~~${\Delta E^\prime_\gamma/\Delta E_\gamma}^{\rm f}$\\
\hline
EC.............  & $2.63$ & $60.48$ & $3.74$ & $60.58$ & ~~~~~$1.42$ \\
B$-$............ & $1.37$ & $31.61$ & $2.01$ & $32.58$ & ~~~~~$1.46$ \\
B$+$............ & $0.19$ & $4.42$  & $0.25$ & $4.04$  & ~~~~~$1.30$ \\
IT.............. & $0.14$ & $3.25$  & $0.14$ & $2.29$  & ~~~~~$1.00$ \\
A............... & $0.01$ & $0.24$  & $0.03$ & $0.51$  & ~~~~~$3.09$ \\
\hline
Sum$^\dagger$     & $4.34$ & $100$ & $6.17$ & $100$ & ~~~~$1.42$ \\
\hline
\end{tabular}\\[1mm]
$^{\rm a}$Decay modes: EC = Electron Capture, B$-$ = $\beta^-$-decay, B$+$ = $\beta^+$-decay, IT = Isomeric Transition, A = $\alpha$-decay.\\
$^{\rm b}$Integrated gamma-ray energy generation in each decay mode in the model without decay chains, in units of $10^{47}\,\erg$.\\
$^{\rm c}$Percentage of the gamma-ray energy generation in each decay mode in the total gamma-ray energy generation, for the model without decay chains.\\
$^{\rm d}$Integrated gamma-ray energy generation in each decay mode in the model with decay chains, in units of $10^{47}\,\erg$.\\
$^{\rm e}$Percentage of the gamma-ray energy generation in each decay mode in the total gamma-ray energy generation, for the model with decay chains.\\
$^{\rm f}$Ratio between the integrated gamma-ray energy generation in the two models in each decay mode.\\
$^\dagger$Sum of the quantity in each column, except the last which is the ratio between the summed gamma-ray energy in the two models.
\end{minipage}
\end{table}

\begin{figure}
\centering{\includegraphics[angle=0,scale=0.65]{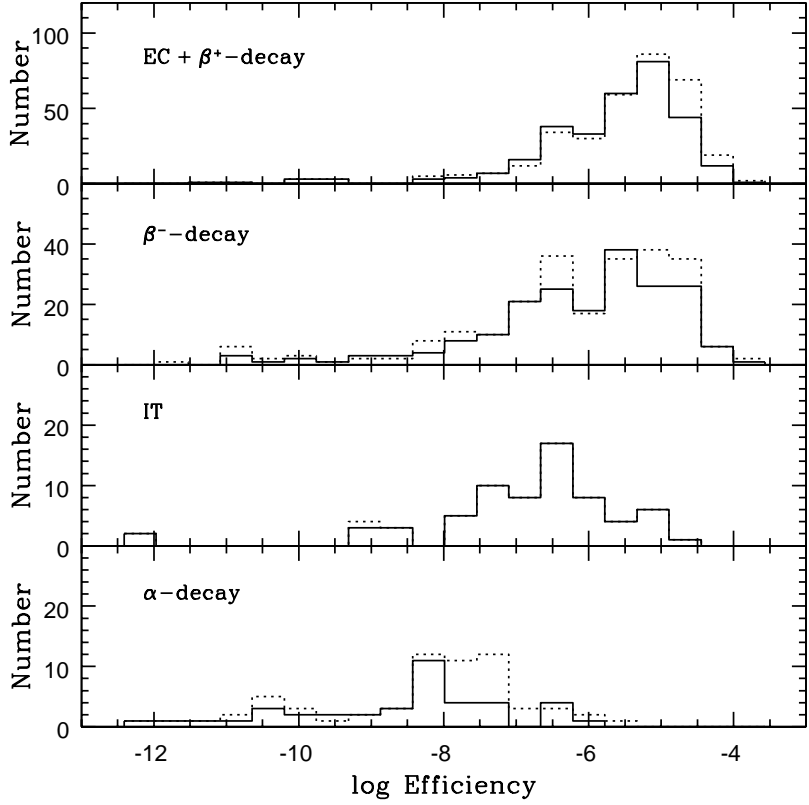}}
\caption{Distribution of the efficiency in gamma-ray energy generation in each decay mode. The solid histograms show the case when the effect of decay chains is ignored. The dotted histograms show the case when the effect is included.
}
\label{eff_g_dc}
\end{figure}

In Figure~\ref{eff_g_dc} we show the distribution of the efficiency in gamma-ray energy generation in each decay mode. The electron capture and the $\beta^\pm$-decay have a similar efficiency distributions. Similar to Figures~\ref{spec_mode} and \ref{power_q2}, the efficiency distribution of the electron capture and the $\beta^+$-decay are shown together, since in the data sample the distinction between the two decay modes is not strict. The isomeric transition has a smaller mean efficiency in generating the gamma-ray energy than the electron capture and the $\beta^\pm$-decay. Not surprisingly, the $\alpha$-decay has the lowest efficiency in generating the gamma-ray energy.

From the result in this section we see that the chain decay has a moderate effect on the efficiency of radioactive nuclides in producing the gamma-ray energy. Without consideration of decay chains, the average efficiency in producing the gamma-ray energy, defined by the ratio of the integrated gamma-ray energy production to the total mass energy of the radioactive nuclides, is $\eta_0=4.39\times 10^{-6}$. When the effect of decay chains is included in the calculation, the average efficiency becomes $\eta=6.23\times 10^{-6}$, which is 1.42 times larger than the $\eta_0$. However, chain decays are not expected to affect the profile of the gamma-ray light curve and the shape of the continuous gamma-ray spectrum, which are in principle determined by the collective and statistical properties of all the radioactive decays contained in the ejecta. Hence, we expect that the gamma-ray energy generation rate, the luminosity, and the spectrum of the gamma-ray emission calculated in the previous sections are not significantly affected by the presence of decay chains, since these quantities are normalized by the UVOIR peak luminosity of SSS17a/AT2017gfo.

\section{Are the Gamma-Ray Emissions Detectable?}
\label{detect}

According to the results in Section~\ref{spectra}, about $95\%$ of the gamma-ray energy emitted by the radioactive decay in a merger ejecta is carried by photons in the energy range of $160\,\keV$--$4\,\MeV$ (Figure~\ref{specInt}). However, plenty number of photons are emitted in the range of $3$--$160\,\keV$ (Figures~\ref{spec1} and \ref{spec2}), which occupy $44\%$ of the total photon number but contribute only $5\%$ to the total gamma-ray energy. These low energy photons seriously suffer the photoelectric absorption by heavy atoms in the ejecta. Hence, when we observe the gamma-ray emission from a neutron star merger, we expect that in the early time the spectrum is dominated by MeV photons, while in the late time photons of hundred keV will start to be present in the spectrum. In this section we investigate the observability of the gamma-ray emission from a neutron star merger event similar to that associated with GW170817, which we will call a ``typical'' merger.

Below a few $100\,\keV$, the interaction of gamma-ray photons with matter is dominated by the photoelectric absorption. For high-$Z$ elements, the opacity of the photoelectric absorption can be larger than that of the Compton scattering by orders of magnitude for photons of energy $\la 100\,\keV$. From a few $100\,\keV$ to about $5\,\MeV$, the opacity is dominated by the Compton scattering. Beyond $5\,\MeV$, the opacity is dominated by pair production in the nuclear field. The total opacity, given by the sum of the opacities for the photoelectric absorption, the Compton scattering, and the pair production, varies relatively slowly with the photon energy for photons of energy $\varepsilon\ga 300\,\keV$. We write the total opacity in the ejecta as $\kappa=\kappa_{\rm pe}+\kappa_{\rm C-pp}$, where $\kappa_{\rm pe}$ is the opacity arising from the photoelectric absorption, and $\kappa_{\rm C-pp}=\kappa_{\rm C}+\kappa_{\rm pp}$ is the sum of the opacities arising from the Compton scattering and pair production. Following \citet{hot16}, we take $\kappa_{\rm pe}\approx 2.5\,\cm^2\,\g^{-1}(\varepsilon/100\,\keV)^{-1.8}$ for $\varepsilon<100\,\keV$, and $\kappa_{\rm pe}\approx 2.5\,\cm^2\,\g^{-1}(\varepsilon/100\,\keV)^{-2.7}$ for $\varepsilon>100\,\keV$. For $\kappa_{\rm C-pp}$, we take the following approximate formula
\begin{eqnarray}
  y &=& -2.44604 + 1.74242 x - 1.49915 x^2 + 1.91335 x^3 - 1.71039 x^4 + 0.875052 x^5 - 0.259065 x^6 \nonumber\\
  && + 0.041322 x^7 - 0.002722 x^8 \;, \label{y_x}
\end{eqnarray}
where $x=\log \varepsilon (\keV)$ and $y=\log\kappa_{\rm C-pp} (\cm^2\,\g^{-1})$. The formula is obtained by polynomial fitting to the numerical opacities for lead atoms evaluated with XCOM.\footnote{XCOM: Photon Cross Sections Database (NIST), https://www.nist.gov/pml/xcom-photon-cross-sections-database} The $\kappa_{\rm C-pp}$ calculated with equation (\ref{y_x}) has a relative error $\la 2\%$ for $1\,\keV\le\varepsilon\le 10\,\MeV$ (i.e., $0\le x\le 4$).

\begin{figure}[ht!]
\centering{\includegraphics[angle=0,scale=0.65]{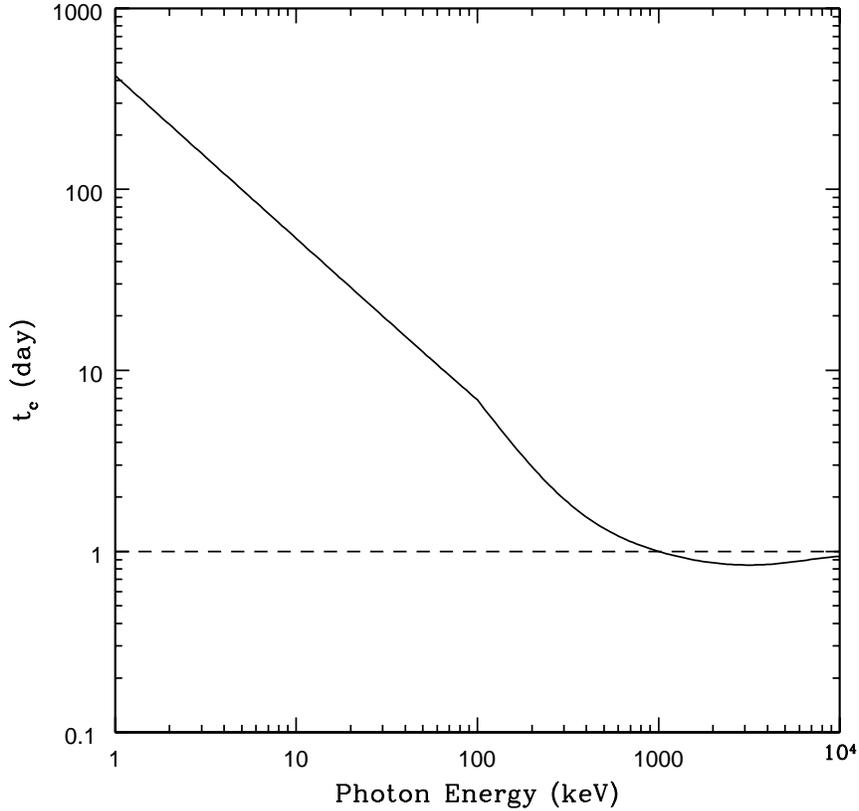}}
\caption{The critical time $t_c$ as a function of the photon energy. The optical opacity is defined by $\kappa=\kappa_{\rm pe}+\kappa_{\rm C}+\kappa_{\rm pp}$, where $\kappa_{\rm pe}$ is from the contribution of the photoelectric absorption, $\kappa_{\rm C}$ from the Compton scattering, and $\kappa_{\rm pp}$ from the pair production (see the text). The critical time is related to the opacity by $t_c\propto\kappa^{1/2}$. Here $t_c$ is set to be equal to one day at $1\,\MeV$.
}
\label{tc_e}
\end{figure}

\begin{figure}[ht!]
\centering{\includegraphics[angle=0,scale=0.65]{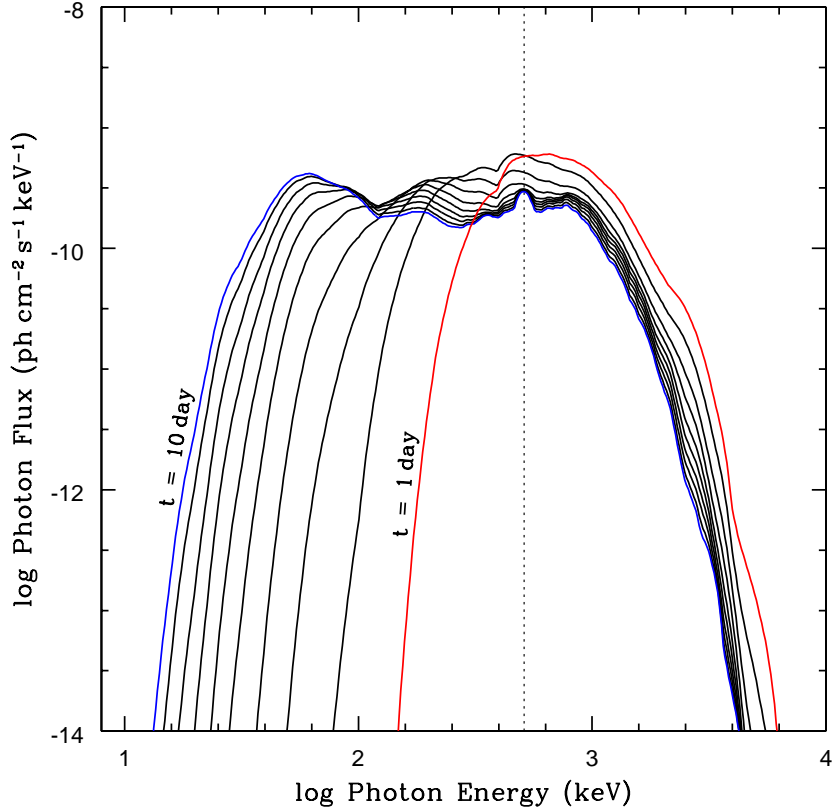}}
\caption{The observed photon flux spectra of the gamma-ray emission produced by radioactive nuclides in a ``typical model'' for neutron star mergers. The model is a copy of that used to fit the UVOIR data of SSS17a/AT2017gfo (Section~\ref{outline}), which contains two ejecta components (A and B) with parameters listed in Table~\ref{parameter}. For the critical time, we have included the contribution of photoelectric absorption to the opacity (Figure~\ref{tc_e}), and required that the energy flux averaged critical time $\langle t_c\rangle$ to be equal to the values listed in Table~\ref{parameter} (see the text). Component A is defined by $V=0.3 c$, $\langle t_c\rangle=0.944\,\oday$, and $\dot{E}_{\gamma,1}=0.86\dot{E}(t=1\,\oday)=4.38\times 10^{41}\,\erg\,\s^{-1}$. Component B is defined by $V=0.1 c$, $\langle t_c\rangle =7.223\,\oday$, and $\dot{E}_{\gamma,1}=6.20\times 10^{41}\,\erg\,\s^{-1}$. The spectra shown in the figure correspond to the time $t=1$, 2, ... $10\,\oday$ after the merger, from right to left as indicated in the graph. The merger is assumed to occur at a distance $D=40\,\Mpc$. The vertical dotted line denotes the electron-positron annihilation energy, which is $511\,\keV$.
}
\label{spec_lp}
\end{figure}

Then, by $t_c\propto \kappa^{1/2}$, we can calculate the variation of the critical timescale $t_c$ versus the photon energy. The result is shown in Figure~\ref{tc_e}, where we have assumed that $t_c=1\,\oday$ at $\varepsilon=1\,\MeV$. We see that for photons of energy $\la$ a few $100\,\keV$, the critical timescale for the ejecta to become transparent to gamma-ray photons increases quickly with decreasing photon energy. For instance, if $t_c\approx 1\,\oday$ for a $1\,\MeV$ photon, we would have $t_c\approx 7\,\oday$ for a $100\,\keV$ photon, and $t_c\approx 50\,\oday$ for a $10\,\keV$ photon. This indicates that, in the energy range $\la$ a few $100\,\keV$, low energy photons appear later than high energy photons. As a result, we expect that in the early spectrum of the gamma-ray emission we would see significant absorption of photons of energy below several $100\,\keV$. As time goes on low energy photons start to emerge from the ejecta, hence we expect to see more low energy photons in later spectra. In other words, the observed gamma-ray spectra would appear to evolve with time with a feature that the spectrum broadens toward the low energy end as time goes on, while the high energy part of the spectrum has a shape that remains almost invariant with time. This feature is shown in Figure~\ref{spec_lp}, where the observed spectra are calculated for the ``typical'' merger model, i.e., for the same two-component model adopted in Figure~\ref{power2}.

In the model, for the components A and B we adopt the parameters ($\beta$, $t_c$, and $\dot{E}$ at $t=1\,\oday$) listed in Table~\ref{parameter}, which are obtained by fitting the UVOIR data of SSS17a/AT2017gfo. For the energy-dependent critical time $t_c$, we determine its normalization by requiring that the energy flux averaged critical time calculated at a time close to the peak of the gamma-ray emission for each component is equal to the value listed in Table~\ref{parameter}. The energy flux averaged critical time $\langle t_c\rangle$ is defined by
\begin{eqnarray}
  e^{-\langle t_c\rangle^2/t^2}=\frac{\int \exp\left(-t_c^2/t^2\right)F_\varepsilon d\varepsilon}{\int F_\varepsilon d\varepsilon} \;. \label{tc_av}
\end{eqnarray}
For component A we take $t=1\,\oday$ and require that $\langle t_c\rangle=0.944\,\oday$. For component B we take $t=10\,\oday$ and require that $\langle t_c\rangle=7.223\,\oday$. With this normalization of $t_c$ we can reproduce the luminosity curve in Figure~\ref{power2}.

From the spectra in Figure~\ref{spec_lp} we see that, the peak in the energy range of $20$--$150\,\keV$ present in the intrinsic gamma-ray spectrum (Figures~\ref{spec1}, \ref{spec2}, and \ref{spec_mode}) starts to be seen only after $t\approx 5\,\oday$. For $t\la 5\,\oday$, we can only see the peak in the energy range of $150$--$3,000\,\keV$, since photons of low energy are seriously absorbed due to the photoelectric effect. A fraction of the energy absorbed by the ejecta matter may be re-emitted as fluorescence line emissions in the X-ray domain, which are not shown in the figure since they only contribute a very small fraction to the total energy emission. We also see that, for $t\ga 5\,\oday$, a broadened electron-positron annihilation line is clearly seen in the spectrum. This is caused by the fact that after $t\approx 5\,\oday$, the emission from the ejecta component B starts to dominate, and the component B has a slower expansion velocity ($V=0.1c$) than the component A ($V=0.3c$). However, for $t\la 5\,\oday$, a bump of annihilation lines around $511\,\keV$ is also clearly visible.

\begin{figure}[ht!]
\centering{\includegraphics[angle=0,scale=0.65]{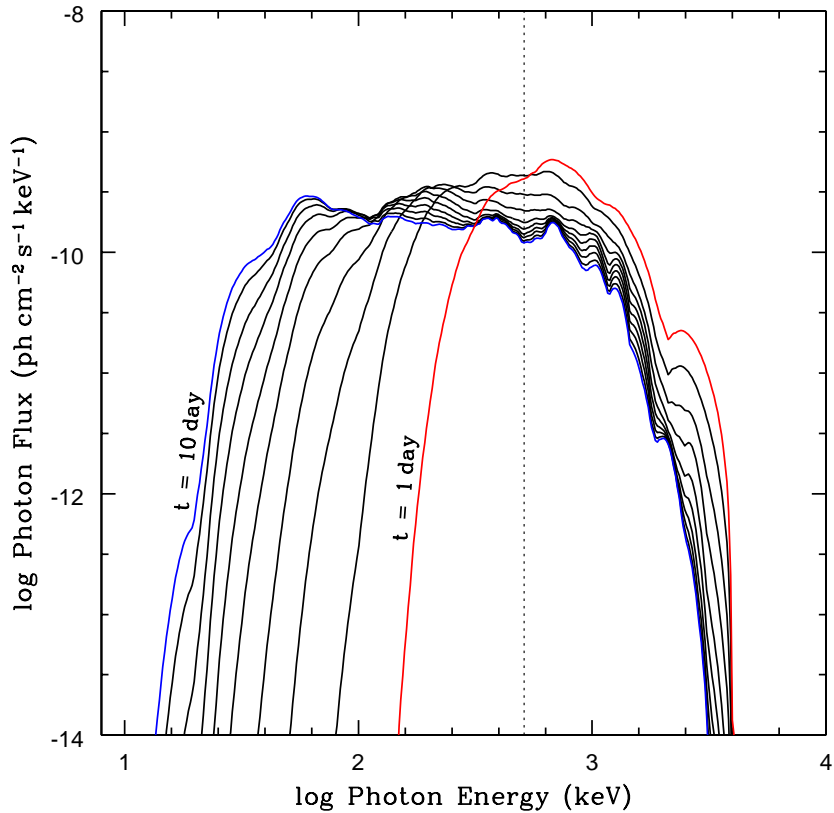}}
\caption{Same as Figure~\ref{spec_lp} but for a sample of nuclides without electron captures and $\beta^+$-decays, i.e., the observed photon flux spectra of the gamma-ray emission produced by r-nuclides alone. The flux is normalized so that the component A attains $\dot{E}_{\gamma}=0.69\dot{E}(t=1\,\oday)=3.51\times 10^{41}\,\erg\,\s^{-1}$ at $t=1\,\oday$ (see the text).
}
\label{spec_lp2}
\end{figure}

To see the effect of the nucleosynthesis process in the merger ejecta on the observed gamma-ray spectrum, in Figure~\ref{spec_lp2} we show the photon flux spectra generated by nuclides without electron captures and $\beta^+$-decays. As we stated previously, electron captures and $\beta^+$-decays are the major feature of p-nuclides. So, with exclusion of electron captures and $\beta^+$-decays, the obtained spectra are in principle close to that produced by the r-nuclides alone. In this case we should have $\dot{E}_{\gamma,1}=0.69\dot{E}(t=1\,\oday)$, since $\beta^-$-decay electrons contribute $31\%$ to the heating rate but zero to the gamma-ray emission. Hence, for the same heating rate, the brightness of the gamma-ray radiation produced by an ejecta with r-nuclides alone is about $0.8$ times that produced by an ejecta with both r- and p-nuclides. 

Comparison of Figure~\ref{spec_lp2} to Figure~\ref{spec_lp} can give us some indications about the difference in the spectra generated by r-nuclides and that generated by p-nuclides, since in Figure~\ref{spec_lp} electron captures and $\beta^+$-decays contribute $65\%$ to the total generation of gamma-ray energy and in Figure~\ref{spec_lp2} $\beta^-$-decays contribute $90\%$ to the total energy generation. Identification of the difference in the spectra will allow us to diagnose the details of the nucleosynthesis process in the merger, e.g., whether it is r-process dominated or a combination of the r-process and the p-process. By comparing Figure~\ref{spec_lp2} to Figure~\ref{spec_lp}, we can see the major differences in the spectra: (1)~For a merger ejecta dominated by p-nuclides, the gamma-ray spectrum has a prominent annihilation line (i.e., the bump around $511\,\keV$ in Figure~\ref{spec_lp}). This feature is not seen in the spectrum of a merger ejecta dominated by r-nuclides (Figure~\ref{spec_lp2}). (2)~The spectrum of the gamma-rays generated by the r-nuclides is cut-off at the high energy end ($\ga 4\,\MeV$) more abruptly than that generated by p-nuclides, and there are very few photons beyond $4\,\MeV$ in the spectrum shown in Figure~\ref{spec_lp2}. (3)~The gamma-ray emissions of r-nuclides show a more ``wavier'' character. For instance, the early time spectra of the r-nuclide emissions have a shoulder around $2.4\,\MeV$, where the spectra of the p-nuclide emissions are smooth.

We find that, despite of some critical differences in the details, including the presence and absence of annihilation lines, the gamma-ray spectra generated by an r-nucide dominated ejecta and a p-nuclide dominated ejecta have very similar global shapes. After consideration of the absorption of low energy photons by the photoelectric effect, in both cases more than $95\%$ of the gamma-ray energy is carried by photons of energy in the range of $0.2$--$4\,\MeV$. This fact indicates that the main calculation results in this work are not sensitive to the nuclear elements used in our data sample. Only some fine features of the gamma-ray spectra are affected by the types of nuclear elements. 

In observational gamma-ray astronomy, the photon energy range of $0.2$--$100\,\MeV$ is a field that is largely unexplored, due to huge backgrounds and the big difficulty in building detectors with good sensitivities in this energy range \citep{nak14,tat16}. For example, above $100\,\MeV$ over 3,000 steady sources have been discovered by {\it Fermi}/LAT \citep{ace15}, and in the range of $14$--$195\,\keV$ over 1,000 sources have been detected by {\it Swift}/BAT \citep{bau13}. But in the range of $0.2$--$100\,\MeV$ only several tens of steady sources have been detected so far by {\it CGRO}/COMPTEL \citep{sch00}. However, the astrophysics in the photon energy range $0.2$--$100\,\MeV$ is very rich, including GRBs, blazars, neutron stars, supernovae, etc; particularly the radioactive decay emissions from various sources. To study the spectacular astrophysics in the MeV gamma-ray range, a few missions and telescopes have been launched or proposed, for instance, the Satellite-ETCC \citep[Electron-Tracking Compton Experiments,][]{tan15}, the e-ASTROGAM space mission \citep{tat16}, and the AMEGO \citep[All-sky Medium Energy Gamma-ray Observatory,][]{moi17,ran17}.

\begin{figure}[ht!]
\centering{\includegraphics[angle=0,scale=0.7]{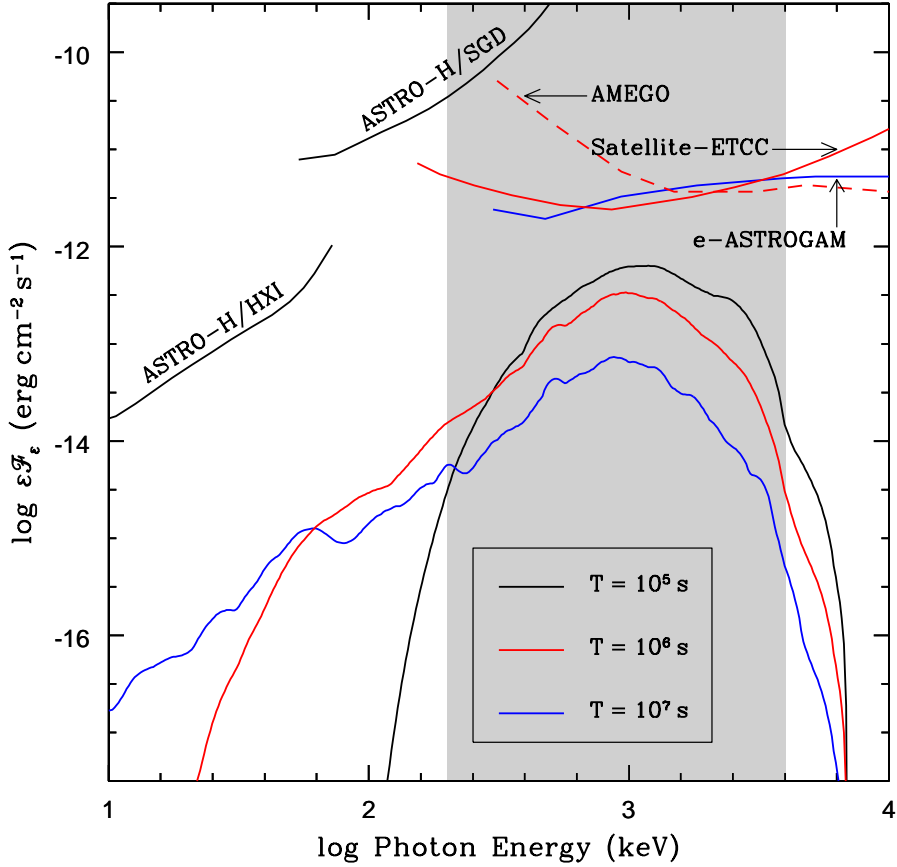}}
\caption{Averaged energy flux spectra of the gamma-ray emission produced by radioactive nuclides in a ``typical model'' for neutron star mergers. The energy flux spectra are averaged over an observation time of $10^5\,\s$ (starting from $t=0.6\,\oday$, black curve), $10^6\,\s$ (from $t=0.5\,\oday$, red curve), and $10^7\,\s$ (starting from $t=0.5\,\oday$, blue curve). Sensitivity curves of some detectors (with 3-$\sigma$ detection) are shown, including {\it ASTRO-H}/(HXI,SGD) ($10^5\,\s$, black line), Satellite-ETCC ($10^6\,\s$, red line), e-ASTROGAM ($1\,\yr$, blue line), and AMEGO ($10^6\,\s$, red dashed line). The grey shaded region encloses the photon energy range of $0.2$--$4\,\MeV$, which contains more than $95\%$ of the observed gamma-ray energy.
}
\label{spec_ener}
\end{figure}

To explore the testability of the gamma-ray emission from radioactive decays of the unstable nuclides synthesized in a neutron star merger, in Figure~\ref{spec_ener} we show the energy flux spectra calculated for the same model, and the sensitivity curves of some gamma-ray detectors. The merger is assumed to occur at the same distance as GW170817, i.e., at $D=40\,\Mpc$. To compare with the sensitivity of detectors, the energy flux shown in Figure~\ref{spec_ener} has been averaged over a period of time according to
\begin{eqnarray}
  {\cal F}_\varepsilon(t_0,T)\equiv\frac{1}{T}\int_{t_0}^{t_0+T}F_\varepsilon(t) dt \;, \label{F_av}
\end{eqnarray}
where $t_0$ is the start time of observation, and $T$ is the total observation time (i.e., the total exposure time). The sensitivity curves of detectors are calculated in an effective exposure time $10^5$--$10^7\,\s$ \citep{tak12,tan15,tat16,moi17}. Therefore, in Figure~\ref{spec_ener} we show the energy flux spectra averaged for three different values of the exposure time: $T=10^5\,\s$ ($t_0=0.6\,\oday$), $T=10^6\,\s$ ($t_0=0.5\,\oday$), and $T=10^7\,\s$ ($t_0=0.5\,\oday$).\footnote{Due to the rapid fading of the gamma-ray emission from a merger event, an exposure time of $10^7\,\s$ may not be very appropriate since for $t>10^6\,\s$ the gamma-ray emission would be too faint. Here we show a spectrum curve with a $10^7\,\s$ exposure time to match the sensitivity curve of  e-ASTROGAM which has an exposure time of $1\,\yr$.}

The gamma-ray detectors shown in Figure~\ref{spec_ener} (with 3-$\sigma$ detection) include {\it ASTRO-H}/HXI ($5$--$80\,\keV, 10^5\,\s$), {\it ASTRO-H}/SGD ($40$--$600\,\keV, 10^5\,\s$), Satellite-ETCC ($0.15$--$20\,\MeV$, $10^6\,\s$), e-ASTROGAM ($0.3\,\MeV$--$2.9\,\GeV$, $10^7\,\s$), and AMEGO ($0.3\,\MeV$--$10\,\GeV$, $10^6\,\s$). The sensitivity curves of {\it ASTRO-H}/(HXI,SGD) and e-ASTROGAM are taken from \citet{tat16} and \citet{tak12}. The sensitivity curves of Satellite-ETCC and AMEGO are taken from \citet{tan15} and \citet{moi17}, respectively.

From Figure~\ref{spec_ener} we see that, for a ``typical'' merger event like the GW170817, the gamma-ray emissions are about one to two orders of magnitude fainter than the sensitivities of the current gamma-ray detectors. For instance, the spectrum curve shown in red colour has been averaged over an observation time of $10^6\,\s$, consistent with the observation time for the sensitivity curve of Satellite-ETCC. If the gamma-ray emission were brighter by a factor of 10, it would be detectable by Satellite-ETCC (with 3-$\sigma$ detection). If the gamma-ray emission were brighter by a factor of 20, it would be detectable by AMEGO. The sensitivity curve of e-ASTROGAM has a longer observation time of $10^7\,\s$. If it is converted to a $10^6\,\s$ observation time, the sensitivity curve of the e-ASTROGAM should be moved upward by a factor $\sim 3$, assuming that the detector's flux sensitivity is $\propto T^{-1/2}$. Then we get that the gamma-ray emission would be detectable by e-ASTROGAM if it were brighter by a factor of 40. The same conclusion can also be obtained by comparing the $10^7\,\s$ spectrum curve (in blue colour) to the sensitivity curve of e-ASTROGAM.

The spectrum with a $10^5\,\s$ exposure time starting from $t=0.6\,\oday$ (black curve) shows a serious absorption feature for photon energy $\la 300\,\keV$. It is below the sensitivity curve of {\it ASTRO-H} (now called {\it Hitomi}; the sensitivity curve has also an exposure time of $10^5\,\s$) by three orders of magnitude. If it is converted to a $10^6\,\s$ observation time, the sensitivity curves of {\it ASTRO-H} should be moved downward by a factor $\sim 3$. Even with this correction, the $10^6\,\s$ spectrum curve (red curve) is still under the sensitivity curves of {\it ASTRO-H} by about two orders of magnitude. From the figure we see that to detect the gamma-ray emission from a merger effectively, a detector covering the $0.2$--$4\,\MeV$ energy range and having an energy flux threshold $\la 4\times 10^{-13}\,\erg\,\cm^{-2}\,\s^{-1}$ in a $10^6\,\s$ exposure time would be most desired.

The result shown in Figure~\ref{spec_ener} indicates that, to detect the gamma-ray emission from a neutron star merger like the one associated with GW170817 and GRB170817A, the sensitivity of current detectors should be improved by at least one order of magnitude. It is unclear whether the merger event associated with GW170817 indeed represents a typical merger event, i.e., if it is not on the faint end or the bright end of the merger luminosity function. The GRB170817A associated with GW170817 is extremely faint compared to other short GRBs at cosmological distances, most likely indicating that the GRB emission arises from an off-axis jet \citep{gol17,ale18,zha18}. Compared to normal short GRBs at cosmological distances, the X-ray and radio emissions of GRB170817A are fainter by a factor of 3,000 and $\ga 10,000$ respectively in terms of isotropic luminosities \citep{fon17}. However, after the off-axis effect is taken into account, the derived jet energy and the ambient particle density are remarkably consistent with those derived for on-axis short GRBs \citep{ale17,fon17,mar17}, suggesting that GRB170817A is not intrinsically faint. However, compared to previously claimed kilonovae following short GRBs, the UVOIR luminosity of the kilonovae associated with GW170817/GRB170817A is fainter by a factor of $\approx 3$--$5$, suggesting the existence of a broad range of kilonova luminosities, colours, and timescales \citep{fon17}.

The occurrence rate density of neutron star mergers is estimated to be $\approx 10^{-7}$--$10^{-6}\,\Mpc^{-3}\,\yr^{-1}$ \citep{abb17a,jin17,chr18}. The occurrence rate within a spherical volume of radius $40\,\Mpc$ is then $\approx 0.03$--$0.3\,\yr^{-1}$. So, the detection of GW170817 is already very lucky, to some extent. The merger rate within a volume of radius $10\,\Mpc$ would be $\approx 0.0005$--$0.005\,\yr^{-1}$, so the chance for detection of a merger event with a distance $\la 10\,\Mpc$ would be very low. However, this does not necessarily mean that discovery of a merger event at a very close distance is not possible. For instance, the local rate density of type II supernovae is only $4.45\times 10^{-5}\,\Mpc^{-3}\,\yr^{-1}$ \citep{li11}, which indicates that the rate of type II supernovae in a volume of radius $51\,\kpc$ is only $2.5\times 10^{-8}\,\yr^{-1}$. But we have discovered the SN1987A at a distance of $51\,\kpc$. So, discovery of a neutron star merger at a distance $\la 10\,\Mpc$ may not be completely impossible. According to Figure~\ref{spec_ener}, the gamma-ray emissions from a merger event like GW170817 would be detectable with Satellite-ETCC if it occurs at a distance $\la 12\,\Mpc$, and would be detectable with e-ASTROGAM if it occurs at a distance $\la 6\,\Mpc$.

\section{Summary and Discussion}
\label{sum}

A neutron star merger is expected to produce a subrelativistic ejecta with heavy and unstable nuclei arising from the complex nucleosynthesis process in the rapidly decompressed nuclear density matter. The radioactive decay of the unstable nuclei provides a long-term energy source for the expanding ejecta. During the initial optically thick stage, an optical transient is produced with a fast evolving brightness and spectrum. The existence of such a transient was solidly verified on 17 August 2017 by the discovery of the optical source SSS17a/AT2017gfo associated with GW170817/GRB170817A. The comprehensive multiband observation of SSS17a/AT2017gfo revealed that the dominant radiation covers the UV, optical, and near-IR bands, with a peak bolometric luminosity $\approx 8\times 10^{41}\,\erg\,\s^{-1}$ at $t\approx 0.6\,\oday$ after the merger. All the observed features agree nicely with theoretical predictions, including the fast evolution with time of the luminosity and the spectrum, and the thermal feature in the early emission.

Due to the subrelativistic expansion, the optical depth of the merger ejecta decreases quickly with time. After about a day to a few days from the time of merger, the ejecta is expected to become optically thin to MeV photons. Then the gamma-ray photons generated inside the ejecta by the continuing radioactive decay process will start to escape without interaction with the ejecta material. Detection and observation of these radioactive gamma-ray photons would be the best approach for directly probing the physical conditions and the nuclear reaction process inside the merger ejecta, understanding the physics of the merger process, and be the most robust test for the hypothesis of neutron star mergers as a major site for the formation of heavy and rare elements in the universe.

In this work we have calculated the luminosity and the spectrum of the gamma-ray emission produced by the radioactive decay of the unstable nuclides freshly synthesized in a neutron star merger ejecta. The calculation is based on a model constructed as follows: we extract from the NuDat\,2 database at the National Nuclear Data Center a sample of radioactive nuclides which have their half-lives in the range of $0.05$--$50,000\,\oday$ and satisfy some other conditions related to data completeness. We get in total 494 isotopes in 537 energy states that satisfy the conditions. An isotope in a given energy state (determined by a given {\jpi} value) is treated as an independent nuclide species. Then we have a sample of 537 nuclide species, each of which has available gamma-ray radiation data. We assume that the nuclides are uniformly distributed in the ejecta, with their relative abundances at time $t=0$ determined by a distribution over the mean lifetime according to a power-law with a Gaussian deviation. Then, by tracing the decay process of the nuclides, we can calculate the gamma-ray energy generation rate in the ejecta, the luminosity, and the spectrum of the emerging gamma-ray emission.

Assuming that the number of nuclide species in a given interval of lifetime $\tau$ is $\propto\tau^{-1.1}$, we get a gamma-ray energy generation rate that is approximately $\propto t^{-1.2}$ (Figure~\ref{power}). This result agrees with that obtained by numerical simulations based on the r-process network \citep{met10,kor12}, and is consistent with the result of model fitting to the UVOIR data of SSS17a/AT2017gfo obtained in Section~\ref{outline}. Therefore, the model that we have constructed is suitable for calculation of the luminosity and spectrum of the radioactive gamma-ray emission produced by a neutron star merger. In our calculations, we determine the absolute magnitude of the abundance of each nuclide species by normalizing the calculated gamma-ray energy generation rate at $t=1\,\oday$ to a reference value obtained by the model fitting to SSS17a/AT2017gfo (Table~\ref{parameter}), with the fraction of the gamma-ray energy generation in the total heating rate being properly taken into account. Then the luminosity and the spectrum of the gamma-ray emission are calculated, with the result displayed in Figures~\ref{power}--\ref{eff_g_dc} and \ref{spec_lp}--\ref{spec_ener}.

After fixing the power-law index in the distribution of nuclide abundances over their lifetime, the model contains three independent parameters: the expansion velocity of the ejecta $V$; the critical time $t_c$, corresponding to the time when the ejecta becomes optically thin; and the normalization of the gamma-ray energy generation rate, $\dot{E}_{\gamma,1}=\dot{E}_\gamma(t=1\,\oday)$. The $\dot{E}_{\gamma,1}$ simply affects the amplitudes of the luminosity and the spectrum of the gamma-ray emission. If the value of $\dot{E}_{\gamma,1}$ is boosted by a factor of three but the values of $V$ and $t_c$ are fixed, for instance, the amplitudes of the luminosity and the spectrum are also boosted by a factor of three. The expansion velocity $V$ has a very small effect on the luminosity, only to the order of $V^2/c^2$ if the other two parameters are fixed. However, the value of $V$ can significantly affect the shape of the gamma-ray spectrum through the line broadening effect, which can be clearly seen by comparing Figure~\ref{spec1} to Figure~\ref{spec2}. For instance, when $V\la 0.2c$, a pair annihilation line of $511\,\keV$ superposed on the continuous spectrum is clearly seen (Figure~\ref{spec2}). When $V\ga 0.3c$, the annihilation line is significantly smeared by the line broadening effect and hence becomes hard to identify, resulting in a more smoother spectrum (Figure~\ref{spec1}).

In a realistic model the opacity in the ejecta can be a function of the photon energy. For photon energy $\varepsilon\la 300\,\keV$, the opacity increases quickly with decreasing photon energy, caused by the strong photoelectric absorption of photons by the heavy elements inside the ejecta. For $\varepsilon\ga 300\,\keV$, the opacity varies very slowly with the photon energy and is dominantly contributed by the Compton scattering and the pair production in the nuclear field. As a result, the critical time $t_c$ is also a function of the photon energy, implying that low energy photons emerge from the ejecta later than high energy photons. This effect causes the observed spectrum of the gamma-ray emission to have a ``width'' broadening toward the low energy end as time goes on (Figures~\ref{spec_lp}--\ref{spec_ener}). In our calculations, we require that the energy flux averaged critical time, $\langle t_c\rangle$, is equal to the value obtained by fitting the UVOIR data of SSS17a/AT2017gfo where a constant opacity has been assumed. For the gamma-ray emission produced by radioactive decays, about $90\%$ of the total emitted energy is carried by photons with $\varepsilon\ga 300\,\keV$ which are not affected by the photoelectric absorption (Figure~\ref{specInt}). Hence, the absorption of low energy photons has little effect on the calculation of the energy generation in the merger ejecta and the luminosity of the gamma-ray emission. However, as we have stated, the photon absorption can have important effects on the observed spectrum of the gamma-ray emission, causing that the low energy part of the gamma-ray spectrum is seriously cut off in the early epoch. 

From the calculated intrinsic photon spectra, i.e., the spectra without consideration of the effect of optical depth (Figures~\ref{spec1} and \ref{spec2}), we see that the emitted photons are clustered in three distinct energy groups: a group with the strongest emission in the range of $150$--$3,000\,\keV$, a group with the intermediate strong emission in the range of $20$--$150\,\keV$, and a group with the weakest emission in the range of $3$--$20\,\keV$. This is a feature of the emission produced by $\beta^\pm$-decays and electron captures, which make the dominant contribution to the gamma-ray emission in the ejecta. Calculations of the radiation spectrum in each decay mode (Figure~\ref{spec_mode}) reveal that in our model, the electron capture and the $\beta^+$-decay contribute about $65\%$ to the total energy of the gamma-ray emission, and the $\beta^-$-decay contributes about $32\%$ (see Table~\ref{chain3}). The remaining $3\%$ radiation comes from the contribution of the $\alpha$-decay and the isomeric transition. The spectra of the radiation generated by different decay modes have some subtle differences in their shapes. Since $\beta^\pm$-decays and electron captures make the dominant contribution to the gamma-ray energy generation, the shape of the emerging gamma-ray spectrum of the ejecta is dominantly determined by the radiation produced by the $\beta^\pm$-decay and the electron capture. The nuclide species with dominant contribution to the spectral peaks evolve with time, as indicated by Tables~\ref{ele_bpec} and \ref{ele_bm}.

After taking into account the effect of decay chains, we get an averaged gamma-ray radiation efficiency for the dadioactive decay: $\eta\approx 6.23\times 10^{-6}$. This number may have been somewhat underestimated considering the fact that the radiation data in the sample may not be complete. However, the possible incompleteness in the radiation data should not have affected the calculation of the luminosity and the spectrum seriously. The profile and shape of the luminosity and the spectrum are determined by the collective and statistical properties of the gamma-ray radiation by all radioactive nuclides in the sample, which are not seriously affected by the slight data incompleteness. The magnitudes of the luminosity and the spectrum, on the other hand, are normalized by referencing to the corresponding values obtained by fitting the UVOIR data of SSS17a/AT2017gfo.

Inclusion of the $\beta^+$-decay and the electron capture in the calculation of the gamma-ray energy generation in a merger ejecta is a major feature distinguishing our model from other existing models based on the r-process network. The $\beta^+$-decay and the electron capture can arise from proton-rich nuclides in the ejecta. The p-nuclides can, under favorable conditions, be synthesized from the abundant r-nuclides produced by the r-process. As we have argued in Section~\ref{data}, these favorable conditions can be satisfied in a merger ejecta, at least in principle. Hence, it appears that the presence of p-nuclides in a merger ejecta cannot be excluded {\it a priori}. In our model, the contribution of the $\beta^+$-decay and the electron capture to the total gamma-ray energy generation is about twice the contribution by the $\beta^-$-decay. As a result, a prominent electron-positron annihilation line at $511\,\keV$ can be created in the observed gamma-ray spectrum, which is a critical feature that other models do not have. Detection of strong pair annihilation lines in a neutron star merger will be a solid proof of our model.

The results obtained in this work are generally consistent with that obtained by \citet{hot16}, except that proton-rich nuclides are not included in their model based on an r-process network and hence pair annihilation lines are not present in their spectra. In particular, our calculations give rise to a specific gamma-ray energy generation rate $\epsilon_\gamma\approx 7.7\times 10^9\,\erg\,\s^{-1}\,\g^{-1}(t/1\,\oday)^{-1.2}$, which is in agreement with their $\epsilon_\gamma\approx 8\times 10^9\,\erg\,\s^{-1}\,\g^{-1}(t/1\,\oday)^{-1.3}$. Although observed spectra are not presented by \citet{hot16}, their intrinsic gamma-ray spectra are broadly consistent with what we have got, in particular if only r-nuclides are included in our data sample. This is not surprising, since all r-nuclides produce gamma-ray emissions with similar spectra. Inclusion of p-nuclides in our model allows us to compare the gamma-ray spectra produced by r-nuclides to that produced by p-nuclides, and to identify the presence of pair annihilation lines in a p-nuclide dominant ejecta. Since we have treated the opacity in the merger ejecta in a similar way to that taken by \citet{hot16}, the gamma-ray luminosity and the observed spectra calculated in both works should agree in principle, except for some specific features for the gamma-ray emission by p-nuclides.

To study the detectability of the gamma-ray emission from neutron star mergers, we have calculated the gamma-ray radiation for a two-component model corresponding to the case of GW170817/GRB170817A. The model contains an ejecta component A and an ejecta component B, with the parameters for each component given in Table~\ref{parameter} where the $t_c$ is understood as the energy flux averaged critical time. The calculated gamma-ray luminosity curve for this model is shown in Figure~\ref{power2}. The peak of the gamma-ray luminosity, where the major contribution to the emission comes from component A, occurs at $t\approx 1.2\,\oday$ after the merger. The peak gamma-ray luminosity is $\approx 2\times 10^{41}\,\erg\,\s^{-1}$. The contribution of the component B to the luminosity starts to be seen at $t\approx 5\,\oday$ and dominates in later times. The observable spectra of the gamma-ray emission are calculated and shown in Figures~\ref{spec_lp}--\ref{spec_ener}. More than $95\%$ of the radiated gamma-ray energy is carried by photons in the energy range of $0.2$--$4\,\MeV$. The cut-off arising from the photoelectric absorption for photons of energy $\la 300\,\keV$ is clearly seen in the very early spectra. 

To detect such a ``typical'' merger event at $D=40\,\Mpc$, we need a detector with an energy flux threshold $\la 4\times 10^{-13}\,\erg\,\cm^{-2}\,\s^{-1}$ in the photon energy range of $0.2$--$4\,\MeV$, with an exposure time of $10^{6}\,\s$. The modern advanced gamma-ray detectors, such as Satellite-ETCC and e-ASTROGAM, cover this photon energy range but have sensitivities above the required energy flux threshold by a factor of 10 and 40, respectively. The proposed AMEGO also covers this photon energy range, but has an energy flux sensitivity above the required threshold by a factor of 20. However, if the merger event occurs at a distance $\la 12\,\Mpc$, it would be detectable with Satellite-ETCC. If the merger event occurs at a distance $\la 6\,\Mpc$, it would also be detectable with e-ASTROGAM. The probability for detection of a neutron star merger event at such a near distance is very small, but it may not be completely impossible. Of course, the detection probability can be significantly larger for a much brighter merger event (e.g., brighter than SSS17a/AT2017gfo by a factor of 10), whose existence in nature cannot be excluded.

Finally, we remark that the major results in this paper regarding the gamma-ray emission from a neutron star merger do not depend on the details of the nuclear ingredients contained in the nuclear data sample. These results include the brightness and the peak time of the gamma-ray emission, the rate of the brightness declining with time, and the energy range of the radiated gamma-ray photons. From the UVOIR light curve of SSS17a/AT2017gfo we can derive the heating rate during the optically thick phase, from which we can get the gamma-ray energy generation rate as a function of time with an assumption about the fraction of the gamma-ray energy rate in the total heating rate. Then, with the theoretically estimated and observationally determined critical time for the transition from the optically thick stage to the optically thin stage, we can get the gamma-ray luminosity and its peak time. From nuclear physics it is well known that the gamma-rays emitted by the decay of radioactive nuclei are typically in the MeV range. Hence, the results mentioned above are general and robust, no matter whether the merger ejecta is p-nuclide dominated or r-nuclide dominated. The only critical difference between the gamma-ray spectrum generated by a p-nuclide dominated ejecta and that by an r-nuclide dominated ejecta is in the presence and absence of pair annihilation lines at $511\,\keV$. Observations of the annihilation line and other subtle spectral features as discussed in Section~\ref{detect} can be used to determine the element composition of the merger ejecta and diagnose the relevant nucleosynthesis process.

\acknowledgments

The author acknowledges Eli Waxman and Eran Ofek for helpful communications about their work. He also thanks Alejandro Sonzogni for kind help in understanding the data in the NuDat\,2 database at the National Nuclear Data Center, and an anonymous referee for a very constructive and enlightening report. This work was supported by the National Basic Research Program (973 Program) of China (Grant No. 2014CB845800) and the National Natural Science Foundation of China (Grant Nos. 11373012 and 11721303).

\appendix

\section{The Case of a Uniformly Expanding Sphere}
\label{sphere}

Assuming that a sphere uniformly expands with a subrelativistic surface speed $V$. The speed $V$ can be a fraction of the light speed $c$ (e.g., $V=0.3c$), but the corresponding Lorentz factor is always $\sim 1$. The radius of the sphere surface is $R=Vt$. Each spherical shell with a radius $r<R$ expands with a speed $v=Vr/R$. Hence we have $r=vt=\beta ct$, where $\beta\equiv v/c<1$. Each spherical shell can be specified by a ``comoving coordinate'' $\beta$, $0\le\beta\le\beta_0\equiv V/c$. At any time $t$, a volume element at radius $r$ is $dV=2\pi r^2\sin\theta dr d\theta=2\pi c^3t^3\beta^2\sin\theta d\beta d\theta$, where $0\le\theta<\pi$.

For a nuclide $X_i$ uniformly distributed in the sphere, its initial total number is $N_{i,0}$. Then, it can be derived that in the volume element $dV$ the initial number of the $X_i$ is
\begin{eqnarray}
  dN_{i,0}=\frac{3N_{i,0}}{2\beta_0^3}\beta^2d\beta\sin\theta d\theta \;.
\end{eqnarray}
By equation (\ref{dEidt}), the energy generation rate of the $i$-th nuclide in the volume element $dV$ is
\begin{eqnarray}
  d\left(\frac{dE_i}{dt}\right) = \frac{3\varepsilon_i N_{i,0}}{2\beta_0^3\tau_i}e^{-t/\tau_i}\beta^2d\beta\sin\theta d\theta \;. \label{dEidtx}
\end{eqnarray}

The above energy generation rate is defined in the rest frame of $dV$. The time $t$ at the volume element is related to the observer time $t_\obs$ by $t=t_\obs-D/c+t\beta\cos\theta$, where $D$ is the distance from an observer at $\theta=0$ to the sphere center. Then we have
\begin{eqnarray}
  t=\frac{1}{1-\beta\cos\theta}\left(t_\obs-\frac{D}{c}\right) \;, \label{t_tobs}
\end{eqnarray}
and
\begin{eqnarray}
  dt_\obs=dt(1-\beta\cos\theta) \;. \label{dt_dtobs}
\end{eqnarray}

Let us assume that, at some moment, the nuclide emits a photon of energy $\varepsilon_i$ in the rest frame of the nuclide. As the photon arrives at the observer, the observer detects it with an energy $\varepsilon_{i,\obs}=\Gamma^{-1}\varepsilon_i/(1-\beta\cos\theta)$, due to the relativistic Doppler effect. Here $\Gamma=\left(1-\beta^2\right)^{-1/2}$ is the Lorentz factor of the volume element. In our calculations we keep the linear effect of the velocity but ignore second and higher order effects. Then we have $\Gamma\approx 1$ and
\begin{eqnarray}
  \varepsilon_{i,\obs}\approx\frac{\varepsilon_i}{1-\beta\cos\theta} \;. \label{Doppler}
\end{eqnarray}
From equations (\ref{dt_dtobs}) we can derive that $dE_{i,\obs}/dt_\obs=(dE_i/dt)/(1-\beta\cos\theta)^2$. Then, by equation (\ref{dEidtx}) we have
\begin{eqnarray}
  d\left(\frac{dE_i}{dt}\right)_\obs = \frac{3\varepsilon_i N_{i,0}}{2\beta_0^3\tau_i}\exp\left[-\frac{t_\obs-D/c}{\tau_i(1-\beta\cos\theta)}\right]\frac{\beta^2d\beta\sin\theta d\theta}{(1-\beta\cos\theta)^2} \;.
\end{eqnarray}

Because of the relative motion of the emitter and the observer, special relativity has two effects here. One is the Doppler effect, i.e., the energy of a photon as measured by the remote observer differs from the energy measured at the rest frame of the emitter by a redshift/blueshift factor, as given by equation (\ref{Doppler}). The other is the distortion of time given by equations (\ref{t_tobs}) and (\ref{dt_dtobs}), which causes the following outcome: for an element moving toward the observer emitting a photon to the observer, the photon arrives at the observer earlier by an amount of time
\begin{eqnarray}
  \Delta t=t\beta\cos\theta\approx\left(t_\obs-\frac{D}{c}\right)\beta\cos\theta \;,
\end{eqnarray}
than a photon emitted at the sphere center at the same time. In addition, because of equation (\ref{dt_dtobs}), for an emitter moving toward the observer the photon emission rate is amplified by a factor $(1-\beta\cos\theta)^{-1}\approx 1+\beta\cos\theta$.

Hence, for a photon emitter moving toward the observer, the emitted photon is blueshifted and the emission rate is amplified. For an emitter moving away from the observer, the emitted photon is redshifted and the emission rate is reduced. This can cause a distortion to the observed spectra and the luminosity lightcurve, in addition to the broadening of emission lines.

Defining a variable $x\equiv\beta\cos\theta$ (then $dx=-\beta\sin\theta d\theta$), we have
\begin{eqnarray}
  \varepsilon_{i,\obs}=\frac{\varepsilon_i}{1-x} \;, \hspace{1cm} d\varepsilon_{i,\obs}=\frac{\varepsilon_i dx}{(1-x)^2}, \label{epsi_x}
\end{eqnarray}
and
\begin{eqnarray}
  d\left(\frac{dE_i}{dt}\right)_\obs = \frac{3\varepsilon_i N_{i,0}}{2\beta_0^3\tau_i}\exp\left[-\alpha^\prime_i(1-x)^{-1}\right](1-x)^{-2}\beta d\beta dx \;,
  \label{dEidt_obs1}
\end{eqnarray}
where $\alpha^\prime_i\equiv(t_\obs-D/c)/\tau_i$. Note that, in equation (\ref{dEidt_obs1}) we have dropped a minus sign since after variable change we take $x$ to vary from $-\beta$ to $\beta$, rather than from $+\beta$ to $-\beta$.

Equation (\ref{epsi_x}) indicates that $x$ is related to the observed photon energy, so equation (\ref{dEidt_obs1}) essentially describes the observed spectrum of the photons emitted by the expanding sphere. Submitting equation (\ref{epsi_x}) into equation (\ref{dEidt_obs1}), we get
\begin{eqnarray}
  d\left(\frac{dE_i}{dt}\right)_\obs = \frac{3N_{i,0}}{2\beta_0^3\tau_i}\exp\left(-\alpha^\prime_i\varepsilon_{i,\obs}/\varepsilon_i\right)d\varepsilon_{i,\obs}\beta d\beta \;,
  \label{dEidt_obs1a}
\end{eqnarray}
where
\begin{eqnarray}
  \frac{\varepsilon_i}{1+\beta}\le\varepsilon_{i,\obs}\le\frac{\varepsilon_i}{1-\beta} \;, \hspace{1cm}
  0\le\beta\le\beta_0 \;. \label{vareps_lim}
\end{eqnarray}
For the photon number rate measured by the remote observer, we have
\begin{eqnarray}
  d\left(\frac{d{\cal N}_i}{dt}\right)_\obs = \frac{3N_{i,0}}{2\beta_0^3\tau_i}\exp\left(-\alpha^\prime_i\varepsilon_{i,\obs}/\varepsilon_i\right)\frac{d\varepsilon_{i,\obs}}{\varepsilon_{i,\obs}}\beta d\beta \;.
  \label{dcNidt_obs1a}
\end{eqnarray}

\vspace{0.2cm}
\subsection{The Photon Number Rate Spectrum}
\label{pnr_spec}

The first condition in equation (\ref{vareps_lim}) is equivalent to $\beta^2\ge(\varepsilon_i/\varepsilon_{i,\obs}-1)^2$. Hence, we can rewrite equation (\ref{dcNidt_obs1a}) as
\begin{eqnarray}
  d\left(\frac{d{\cal N}_i}{dt}\right)_\obs = \frac{3N_{i,0}}{2\beta_0^3\tau_i}\exp\left(-\alpha^\prime_i\varepsilon_{i,\obs}/\varepsilon_i\right)\vartheta\left[\beta^2-\left(\frac{\varepsilon_i}{\varepsilon_{i,\obs}}-1\right)^2\right]\frac{d\varepsilon_{i,\obs}}{\varepsilon_{i,\obs}}\beta d\beta \;,
  \label{dcNidt_obs1b}
\end{eqnarray}
and now the integration range for $\varepsilon_{i,\obs}$ is from $0$ to $\infty$. Here $\vartheta(x)$ is the Heaviside step function defined by: $\vartheta(x)=1$ for $x\ge 0$, and $=0$ for $x<0$.

After working out the integration over $\beta$, we get
\begin{eqnarray}
  d\left(\frac{d{\cal N}_i}{dt}\right)_\obs = \frac{3N_{i,0}}{4\beta_0^3\tau_i}e^{-\alpha^\prime_i\varepsilon_{i,\obs}/\varepsilon_i}Y_i\frac{d\varepsilon_{i,\obs}}{\varepsilon_{i,\obs}} \;, \label{dNdtxx}
\end{eqnarray}
where $0<\varepsilon_{i,\obs}<\infty$, and
\begin{eqnarray}
  Y_i\equiv\left[\beta_0^2-\left(\frac{\varepsilon_i}{\varepsilon_{i,\obs}}-1\right)^2\right]\vartheta\left[\beta_0^2-\left(\frac{\varepsilon_i}{\varepsilon_{i,\obs}}-1\right)^2\right] \;.
  \label{Y}
\end{eqnarray}
Equation (\ref{dNdtxx}) gives rise to a specific photon number rate spectrum (photons per unit time per unit photon energy)
\begin{eqnarray}
  \left(\frac{d{\cal N}_{i,\varepsilon}}{dt}\right)_\obs = \frac{3N_{i,0}}{4\beta_0^3\tau_i}\,\frac{1}{\varepsilon}e^{-\alpha^\prime_i\varepsilon/\varepsilon_i}Y_i(\varepsilon) \;,
  \label{dcNeidt_obs}
\end{eqnarray}
where $0<\varepsilon<\infty$ and $\varepsilon$ is used to denote the observed photon energy.

Now let us consider an energy bin defined from $\varepsilon$ to $\varepsilon+\Delta\varepsilon$ in the observer frame and calculate the photon number rate in that energy bin. The result is given by
\begin{eqnarray}
  \Delta\left(\frac{d{\cal N}_i}{dt}\right)_\obs = \frac{3N_{i,0}}{4\beta_0^3\tau_i}\int_{y_1}^{y_2}\frac{e^{-\alpha^\prime_i y}}{y}Y_i(y)dy \;, 
  \label{dcNidt_obs2} 
\end{eqnarray}
where $y\equiv\varepsilon_{i,\obs}/\varepsilon_i$, $y_1\equiv\varepsilon/\varepsilon_i$, and $y_2=y_1+\Delta\varepsilon/\varepsilon_i$. The integral can be worked out with the exponential integral defined by $E_1(x)=-{\rm Ei}(-x)=\int_1^\infty e^{-xs}s^{-1}ds$ \cite[see, e.g.,][]{abr65}.

Let us define $y_\pm=(1\mp\beta_0)^{-1}$ and a function $I_2$ by
\begin{eqnarray}
  I_2(y) \equiv \frac{3}{4\beta_0^3}\left[\left(-\frac{4+\alpha^\prime_i}{2y}+\frac{1}{2y^2}\right)e^{-\alpha^\prime_i y}+\left(\beta_0^2-1-2\alpha^\prime_i-\frac{\alpha_i^{\prime 2}}{2}\right){\rm Ei}(-\alpha^\prime_i y)\right] \;. \label{I_int}
\end{eqnarray}
Then we get
\begin{eqnarray}
  \Delta\left(\frac{d{\cal N}_i}{dt}\right)_\obs = \frac{N_{i,0}}{\tau_i}
  \left\{\begin{array}{ll}
  I_2(y_1,y_2) \;, & \quad y_-<y_1<y_2<y_+ \;, \\
  I_2(y_-,y_2) \;, & \quad y_1<y_-<y_2<y_+ \;, \\
  I_2(y_1,y_+) \;, & \quad y_-<y_1<y_+<y_2 \;, \\
  I_2(y_-,y_+) \;, & \quad y_1<y_-<y_+<y_2 \;, \\
  0 \;, & \quad \mbox{else} \;,
  \end{array}\right.
  \label{dcNidt_obs3} 
\end{eqnarray}
where $I_2(a,b)\equiv I_2(b)-I_2(a)$.

\subsection{The Radiation Power}

Similar to the photon number rate, the energy rate defined in the observer frame can be calculated by
\begin{eqnarray}
  d\left(\frac{dE_i}{dt}\right)_\obs = \frac{3N_{i,0}}{2\beta_0^3\tau_i}\exp\left(-\alpha^\prime_i\varepsilon_{i,\obs}/\varepsilon_i\right)\vartheta\left[\beta^2-\left(\frac{\varepsilon_i}{\varepsilon_{i,\obs}}-1\right)^2\right]d\varepsilon_{i,\obs}\beta d\beta \;.
  \label{dEidt_obs1b}
\end{eqnarray}
After integration over $\beta$, we get
\begin{eqnarray}
  d\left(\frac{dE_i}{dt}\right)_\obs = \frac{3N_{i,0}}{4\beta_0^3\tau_i}e^{-\alpha^\prime_i\varepsilon_{i,\obs}/\varepsilon_i}Y_id\varepsilon_{i,\obs} \;. 
 \label{dEidt_obs1c}
\end{eqnarray}

So we have the energy rate in the observed photon energy bin $\varepsilon$---$\varepsilon+\Delta\varepsilon$ given by
\begin{eqnarray}
  \Delta\left(\frac{dE_i}{dt}\right)_\obs = \frac{3N_{i,0}\varepsilon_i}{4\beta_0^3\tau_i}\int_{y_1}^{y_2}e^{-\alpha^\prime_i y}Y_i(y)dy \;.
  \label{dEidt_obs1e} 
\end{eqnarray}
Let us define a function $I_1$ by
\begin{eqnarray}
  I_1(y) \equiv \frac{3}{4\beta_0^3}\left[\left(\frac{1}{y}+\frac{1-\beta_0^2}{\alpha^\prime_i}\right)e^{-\alpha^\prime_i y} +(\alpha^\prime_i+2){\rm Ei}\left(-\alpha^\prime_i y\right)\right] \;.
  \label{I1}
\end{eqnarray}
Then, setting $y_1=y_-$ and $y_2=y_+$ in equation (\ref{dEidt_obs1e}), we get the radiation power
\begin{eqnarray}
  \left(\frac{dE_i}{dt}\right)_\obs = \frac{N_{i,0}\varepsilon_i}{\tau_i}I_1(y_-,y_+) \;, 
  \label{dEidt_obs2a} 
\end{eqnarray}
where $I_1(a,b)\equiv I_1(b)-I_1(a)$.

\subsection{The Newtonian Limit}

Here we take the limit $\beta<\beta_0\ll 1$ and ignore all velocity effects except the line broadening due to the Doppler shift. Define $x=\varepsilon_{i,\obs}/\varepsilon_i-1$ (then $dx=d\varepsilon_{i,\obs}/\varepsilon_i$), $-1\le x<\infty$. Because of the step function in equation (\ref{dEidt_obs1b}), for the value of $x$ that contributes to the integral, we have $x^2(1+x)^{-2}\le\beta^2\le\beta_0^2\ll 1$, i.e., $x\ll 1$. Then, we have $\varepsilon_i/\varepsilon_{i,\obs}=(1+x)^{-1}\approx 1-x$ and $(\varepsilon_i/\varepsilon_{i,\obs}-1)^2\approx x^2$. In equation (\ref{dEidt_obs1b}), take $\exp\left(-\alpha^\prime_i\varepsilon_{i,\obs}/\varepsilon_i\right)=\exp[-\alpha^\prime_i(1+x)]\approx\exp(-\alpha^\prime_i)$. Then, we get
\begin{eqnarray}
  d\left(\frac{dE_i}{dt}\right)_\obs = \frac{3N_{i,0}\varepsilon_i}{2\beta_0^3\tau_i}e^{-\alpha^\prime_i}\vartheta\left(\beta^2-x^2\right)dx\beta d\beta \;.
\end{eqnarray}
That is, we have neglected all velocity effects except that in the step function.

After integration over $\beta$ from $\beta=0$ to $\beta=\beta_0$, we get
\begin{eqnarray}
  d\left(\frac{dE_i}{dt}\right)_\obs = \frac{3N_{i,0}\varepsilon_i}{4\beta_0^3\tau_i}e^{-\alpha^\prime_i}\left(\beta_0^2-x^2\right)\vartheta\left(\beta_0^2-x^2\right)dx \;.
  \label{dEidt_obs5}
\end{eqnarray}
After integration over $x$ from $x=-\beta_0$ to $x=\beta_0$, we get the radiation power in the Newtonian limit
\begin{eqnarray}
  \left(\frac{dE_i}{dt}\right)_\obs = \frac{N_{i,0}\varepsilon_i}{\tau_i}e^{-\alpha^\prime_i} = \frac{N_{i,0}\varepsilon_i}{\tau_i}e^{-t/\tau_i} \;,
  \label{dEidt_obs7}
\end{eqnarray}
where $t=t_\obs-D/c$. Equation~(\ref{dEidt_obs7}) is identical to equation~(\ref{dEidt}).

Similar to the case of Doppler broadening by atomic thermal motion, we can define a line profile function
\begin{eqnarray}
  \phi(\beta_0,\varepsilon_i,\varepsilon_{i,\obs})=\frac{3}{4\beta_0^3\varepsilon_{i}}\left(\beta_0^2-x^2\right)\vartheta\left(\beta_0^2-x^2\right) \;,
  \label{p_function}
\end{eqnarray}
which satisfies the normalization condition
\begin{eqnarray}
  \int_0^\infty\phi(\beta_0,\varepsilon_i,\varepsilon_{i,\obs})d\varepsilon_{i,\obs}=1 \;. \label{p_norm}
\end{eqnarray}
Then, equation (\ref{dEidt_obs5}) can be rewritten as
\begin{eqnarray}
  d\left(\frac{dE_i}{dt}\right)_\obs = \frac{\varepsilon_iN_{i,0}}{\tau_i}e^{-t/\tau_i}\phi(\beta_0,\varepsilon_i,\varepsilon_{i,\obs}) d\varepsilon_{i,\obs} \;.
  \label{dEidt_obs6}
\end{eqnarray}
The profile function $\phi$ describes the Doppler broadening of an emission line by the homogeneous expansion of the merger ejecta in the Newtonian limit.

\begin{figure}[ht!]
\centering{\includegraphics[angle=0,scale=0.65]{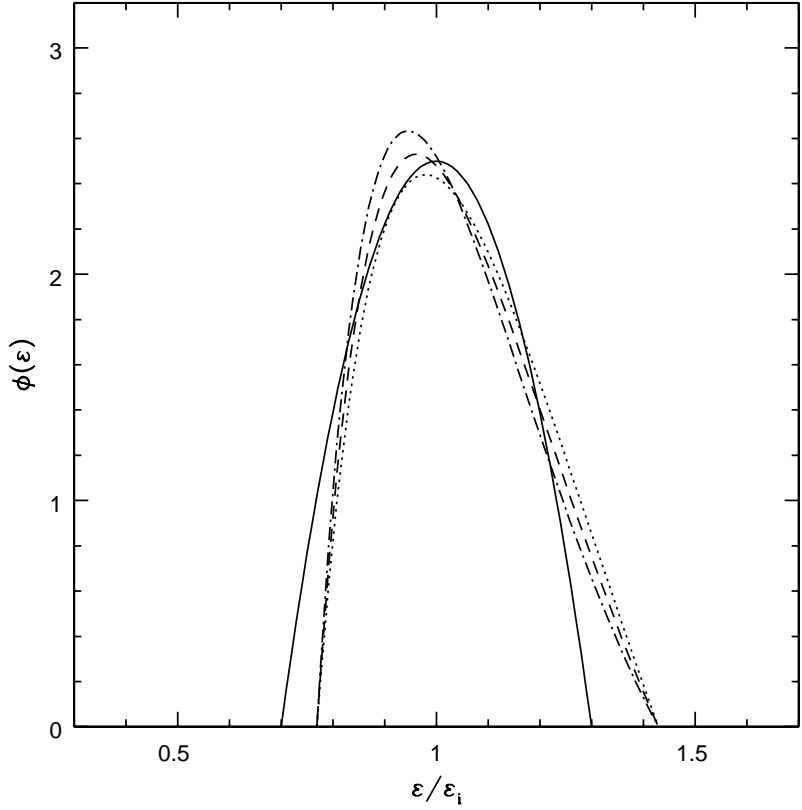}}
\caption{Line profile function due to the Doppler effect arising from the uniform expansion of a sphere, where $\varepsilon$ is the photon energy in the observer frame, and $\varepsilon_i$ is the photon energy in the frame of the line emitter. The expansion velocity of the sphere is assumed to be $V=0.3c$. The solid curve is the Newtonian solution, where all velocity effects are ignored except that on line broadening. The dashed, dotted, and dot-dashed curves are the solutions in the linear velocity approximation, where the effect of special relativity is included up to the linear order of velocity. They are the line profile functions at different moments: $t=0.5\tau_i$ (dotted curve), $t=\tau_i$ (dashed curve), and $t=1.5\tau_i$ (dot-dashed curve), where $\tau_i$ is the mean lifetime of the nuclide that emits the line.
}
\label{profile}
\end{figure}

By comparison to equation (\ref{dEidt_obs1c}), we find that in the linear velocity approximation model, the normalized line profile function should be 
\begin{eqnarray}
  \phi\left(\beta_0,\varepsilon_i,\varepsilon_{i,\obs},\alpha^\prime_i\right) = \frac{3}{4\beta_0^3\varepsilon_{i}}\,\frac{e^{-\alpha^\prime_i\varepsilon_{i,\obs}/\varepsilon_i}}{I_1(y_-,y_+)}\,Y_i \;. 
\end{eqnarray}
Because of the factor $e^{-\alpha^\prime_i\varepsilon_{i,\obs}/\varepsilon_i}$, the line profile function varies with $\alpha^\prime_i =t/\tau_i$. Note, inclusion of the linear velocity effect of relativity causes the line profile function to be asymmetric about the photon energy in the rest frame, unlike in the case of the Newtonian limit (see Figure~\ref{profile}).

Similarly, in equation (\ref{dcNidt_obs1b}), we take the approximation $\exp\left(-\alpha^\prime_i\varepsilon_{i,\obs}/\varepsilon_i\right)\approx e^{-\alpha^\prime_i}$ and $(\varepsilon_i/\varepsilon_{i,\obs}-1)^2\approx x^2$, and $d\varepsilon_{i,\obs}/\varepsilon_{i,\obs}=dx$, we get
\begin{eqnarray}
  d\left(\frac{d{\cal N}_i}{dt}\right)_\obs = \frac{3N_{i,0}}{2\beta_0^3\tau_i}e^{-\alpha^\prime_i}\vartheta\left(\beta^2-x^2\right)dx\beta d\beta \;.
\end{eqnarray}
After integration over $\beta$, we get
\begin{eqnarray}
  d\left(\frac{d{\cal N}_i}{dt}\right)_\obs = \frac{3N_{i,0}}{4\beta_0^3\tau_i}e^{-\alpha^\prime_i}\left(\beta_0^2-x^2\right)\vartheta\left(\beta_0^2-x^2\right)dx \;.
  \label{dcNidt_obs5}
\end{eqnarray}

The photon number rate in a photon energy bin defined by $\varepsilon_1=\varepsilon_i(1+x_1)$ and $\varepsilon_2=\varepsilon_i(1+x_2)$ is evaluated to be
\begin{eqnarray}
  \Delta\left(\frac{d{\cal N}_i}{dt}\right)_\obs(\varepsilon_1,\varepsilon_2) = \frac{N_{i,0}}{\tau_i}e^{-\alpha^\prime_i}
  \left\{\begin{array}{ll}
  f_1 \;, & \quad -\beta_0<x_1<x_2<\beta_0 \;, \\
  f_2 \;, & \quad x_1<-\beta_0<x_2<\beta_0 \;, \\
  f_3 \;, & \quad -\beta_0<x_1<\beta_0<x_2 \;, \\
  1 \;,   & \quad x_1<-\beta_0<\beta_0<x_2 \;, \\
  0 \;,   & \quad \mbox{else} \;,
  \end{array}\right.
  \label{dcNidt_obs5b}
\end{eqnarray}
where
\begin{eqnarray}
  f_1 &\equiv& \frac{x_2-x_1}{4\beta_0}\left[3-\beta_0^{-2}\left(x_1^2+x_1x_2+x_2^2\right)\right] \;, \\
  f_2 &\equiv& \frac{x_2+\beta_0}{4\beta_0}\left[2+\beta_0^{-2}\left(\beta_0x_2-x_2^2\right)\right] \;, \\
  f_3 &\equiv& \frac{\beta_0-x_1}{4\beta_0}\left[2-\beta_0^{-2}\left(x_1^2+\beta_0x_1\right)\right] \;.
\end{eqnarray}

\section{Mathematical Treatment of Decay Chains}
\label{tdc}

Consider a decay chain defined by $X_0\xrightarrow[]{\lambda_0} X_1\xrightarrow[]{\lambda_1} X_2\xrightarrow[]{\lambda_2} ... \xrightarrow[]{\lambda_{k-1}} X_k$, where $X_0$ is the parent nuclide, $X_1$, ..., $X_k$ are daughter nuclides, and $X_k$ is stable (i.e., it has a decay constant $\lambda_k=0$). At any time, the number of $X_i$ is denoted by $N_i$. At time $t=0$, we have $N_0=N_{0,0}$, and $N_i=0$ for all $i\ge 1$. Then, we have
\begin{eqnarray}
  \frac{dN_0}{dt} &=& -\lambda_0N_0 \;, \label{dN0dt} \\
  \frac{dN_i}{dt} &=& \lambda_{i-1}N_{i-1}-\lambda_i N_i \;, \hspace{1cm}\mbox{for $i\ge 1$} \;. \label{dNidt}
\end{eqnarray}

With the specified initial condition at $t=0$, we have the following solutions to equations (\ref{dN0dt}) and (\ref{dNidt}) \citep{bat10}
\begin{eqnarray}
  N_0 &=& N_{0,0}e^{-\lambda_0 t} \;, \label{N0(t)}\\
  N_i &=& N_{0,0}\sum_{j=0}^ih_{i,j}e^{-\lambda_j t} \;, \hspace{1cm}\mbox{for $i\ge 1$}\;, \label{Ni(t)}
\end{eqnarray}
where the coefficient $h_{i,j} (j\le i)$ is defined by
\begin{eqnarray}
  h_{0,0}=1 \;, \label{h00}
\end{eqnarray}
and
\begin{eqnarray}
  h_{i,j} = \frac{\lambda_0\lambda_1...\lambda_{i-1}}{(\lambda_0-\lambda_j)(\lambda_1-\lambda_j)...(\lambda_i-\lambda_j)} \label{hij}
\end{eqnarray}
for $i\ge 1$ and $j=0$, $1$, ..., $i-1$, and
\begin{eqnarray}
  h_{i,i} = \frac{\lambda_0\lambda_1...\lambda_{i-1}}{(\lambda_0-\lambda_i)(\lambda_1-\lambda_i)...(\lambda_{i-1}-\lambda_i)} \label{hii}
\end{eqnarray}
for $i\ge 1$.

Since $N_i(t=0)=0$ for $i\ge 1$, we have the identity
\begin{eqnarray}
  \sum_{j=0}^ih_{i,j}=0 \;, \hspace{1cm} \mbox{for $i\ge 1$} \;.  \label{sum_hij}
\end{eqnarray}

Equations (\ref{N0(t)})--(\ref{hii}) determine the number of each nuclide on the decay chain at any time $t>0$. With the solutions of $N_0$, $N_1$,...$N_{k-1}$ as a function of time, the energy generation rate of the decay chain can be calculated by
\begin{eqnarray}
  \dot{E} = \sum_{i=0}^{k-1}\varepsilon_i\lambda_iN_i=N_{0,0}\sum_{i=0}^{k-1}\varepsilon_i\lambda_i\sum_{j=0}^ih_{i,j}e^{-\lambda_j t} \;. \label{dEdt_chain}
\end{eqnarray}

Integrating $\dot{E}$ over time from $t=0$ to $t=\infty$, we get the integrated energy generation
\begin{eqnarray}
  \Delta E=\int_0^\infty\dot{E}dt =N_{0,0}\sum_{i=0}^{k-1}\varepsilon_i\lambda_i\sum_{j=0}^ih_{i,j}\lambda_j^{-1} \;. \label{delE_chain}
\end{eqnarray}

Now consider the case of a decay chain with branching or bifurcation at some position, e.g., at $i=l$. Starting from the parent nuclide $X_0$, the decay continues to $X_l$, then branching occurs: $X_l$ continues to decay along a chain $a$ to a stable nuclide $X_{k_a,a}$, and along a chain $b$ to another stable nuclide $X_{k_b,b}$. That is, we have the following processes: $X_0\xrightarrow[]{\lambda_0} X_1\xrightarrow[]{\lambda_1} ... X_{l-1}\xrightarrow[]{\lambda_{l-1}} X_l$, then $X_l\xrightarrow[]{\lambda_{l,a}} X_{l+1,a}\xrightarrow[]{\lambda_{l+1,a}} ... \xrightarrow[]{\lambda_{k_a-1,a}}X_{k_a,a}$ along one decay path, and $X_l\xrightarrow[]{\lambda_{l,b}} X_{l+1,b}\xrightarrow[]{\lambda_{l+1,b}} ... \xrightarrow[]{\lambda_{k_b-1,b}}X_{k_b,b}$ along another decay path. At $i=l$, $X_l$ decays to $X_{l+1,a}$ with a decay constant $\lambda_{l,a}$, and to $X_{l+1,b}$ with a decay constant $\lambda_{l,b}$.

According to the above results, the number of $X_{l-1}$ at time $t$ is given by
\begin{eqnarray}
  N_{l-1}=N_{0,0}\sum_{j=0}^{l-1}h_{l-1,j}e^{-\lambda_jt} \;.
\end{eqnarray}
The number of $X_l$ is determined by
\begin{eqnarray}
  \frac{dN_l}{dt}=\lambda_{l-1}N_{l-1}-\lambda_lN_l \;,
\end{eqnarray}
where $\lambda_l=\lambda_{l,a}+\lambda_{l,b}$ is the total decay constant of $X_l$. Hence, we get
\begin{eqnarray}
  \frac{d}{dt}\left(e^{\lambda_l t} N_l\right)=N_{0,0}\sum_{j=0}^{l-1}h_{l-1,j}\lambda_{l-1}e^{(\lambda_l-\lambda_j)t} \;.
\end{eqnarray}
By integration we get the solution for $N_l$
\begin{eqnarray}
  N_l=N_{0,0}\sum_{j=0}^{l-1}h_{l-1,j}\frac{\lambda_{l-1}}{\lambda_l-\lambda_j}\left(e^{-\lambda_jt}-e^{-\lambda_lt}\right) \;,
\end{eqnarray}
with the initial condition $N_l(t=0)=0$.

Since
\begin{eqnarray}
  h_{l,j}=h_{l-1,j}\frac{\lambda_{l-1}}{\lambda_l-j} \;,
\end{eqnarray}
we get
\begin{eqnarray}
  N_l = N_{0,0}\sum_{j=0}^{l-1}h_{l,j}\left(e^{-\lambda_jt}-e^{-\lambda_lt}\right)= N_{0,0}\sum_{j=0}^lh_{l,j}e^{-\lambda_jt} \;,
\end{eqnarray}
where we have used the identity (\ref{sum_hij}).

Hence, the number of the nuclide at the branching position, $N_l$, is still given by equation (\ref{Ni(t)}), but we should use the total decay constant $\lambda_l=\lambda_{l,a}+\lambda_{l,b}$ in the expression.

Similarly, for the solution of $N_{l+1,a}$, we can derive that
\begin{eqnarray}
  N_{l+1,a}=N_{0,0}\sum_{j=0}^lh_{l,j}\frac{\lambda_{l,a}}{\lambda_{l+1,a}-\lambda_j}\left(e^{-\lambda_jt}-e^{-\lambda_{l+1,a}t}\right) \;.
\end{eqnarray}
By the definition of $h_{l,j}$, we have
\begin{eqnarray}
  h_{l,j}\frac{\lambda_{l,a}}{\lambda_{l+1,a}-\lambda_j} = B_{l,a}h_{l+1,j}^a\;,
\end{eqnarray}
where $B_{l,a}=\lambda_{l,a}/\lambda_l$ is the branching ratio of $X_l$ for decaying to $X_{l+1,a}$, and
\begin{eqnarray}
  h_{l+1,j}^a=\frac{\lambda_0\lambda_1...\lambda_{l-1}\lambda_l}{(\lambda_0-\lambda_j)(\lambda_1-\lambda_j)...(\lambda_l-\lambda_j)(\lambda_{l+1,a}-\lambda_j)} 
\end{eqnarray}
is the $h_{i,j}$ parameter defined along the decay chain $X_0\xrightarrow[]{\lambda_0} X_1\xrightarrow[]{\lambda_1} ... X_l\xrightarrow[]{\lambda_{l,a}} X_{l+1,a}\xrightarrow[]{\lambda_{l+1,a}} ... \xrightarrow[]{\lambda_{k_a-1,a}}X_{k_a,a}$.

Hence, we have the number of $X_{l+1,a}$ given by
\begin{eqnarray}
  N_{l+1,a} = B_{l,a} N_{0,0}\sum_{j=0}^lh^a_{l+1,j}\left(e^{-\lambda_jt}-e^{-\lambda_{l+1,a}t}\right)= B_{l,a} N_{0,0}\sum_{j=0}^{l+1}h^a_{l+1,j}e^{-\lambda_jt} \;.
\end{eqnarray}

Similarly, along the decay chain $X_0\xrightarrow[]{\lambda_0} X_1\xrightarrow[]{\lambda_1} ... X_l\xrightarrow[]{\lambda_{l,b}} X_{l+1,b}\xrightarrow[]{\lambda_{l+1,b}} ... \xrightarrow[]{\lambda_{k_b-1,b}}X_{k_b,b}$, we have
\begin{eqnarray}
  N_{l+1,b}=B_{l,b} N_{0,0}\sum_{j=0}^{l+1}h^b_{l+1,j}e^{-\lambda_jt} \;,
\end{eqnarray}
where $B_{l,b}=\lambda_{l,b}/\lambda_l$, and
\begin{eqnarray}
  h_{l+1,j}^b=\frac{\lambda_0\lambda_1...\lambda_{l-1}\lambda_l}{(\lambda_0-\lambda_j)(\lambda_1-\lambda_j)...(\lambda_l-\lambda_j)(\lambda_{l+1,b}-\lambda_j)} \;.
\end{eqnarray}

For any $l+m$-th nuclide on the decay path $a$, we have
\begin{eqnarray}
  N_{l+m,a}=B_{l,a} N_{0,0}\sum_{j=0}^{l+m}h^a_{l+m,j}e^{-\lambda_jt} \;, \hspace{0.6cm} m\ge 1 \;,
\end{eqnarray}
where
\begin{eqnarray}
  h_{l+m,j}^a=\frac{\lambda_0...\lambda_l\lambda_{l+1,a}...\lambda_{l+m-1,a}}{(\lambda_0-\lambda_j)...(\lambda_l-\lambda_j)(\lambda_{l+1,a}-\lambda_j)...(\lambda_{l+m,a}-\lambda_j)} \;, 
\end{eqnarray}
and similarly for any $l+m$-th nuclide on the decay-path $b$.

In a brief summary, for the nuclide with branching decays, $X_l$, the total decay constant should be used in the calculation of $N_l$. For nuclides after $X_l$, e.g., the $X_{l+m}$ with $m\ge 1$, the calculation of its number can be done with the same formula for a decay chain without branching where the total decay constant is used for the $X_l$; then multiplying the result by the branching ratio of $X_l$ along the decay path to get the final result. 

If branching occurs at the beginning of a decay chain, i.e., at $i=0$, we should have $N_0=N_{0,0}e^{-\lambda_0t}$, where $\lambda_0$ is the total decay constant of $X_0$. For $i\ge 1$, we have
\begin{eqnarray}
  N_{i,a}=B_{0,a}N_{0,0}\sum_{j=0}^ih^a_{i,j}e^{-\lambda_jt} \;,
\end{eqnarray}
etc, similar to the branching case discussed above. Here the branching ratio $B_{0,a}=\lambda_{0,a}/\lambda_0$, etc.

\end{document}